\documentclass[twocolumn,astrosymb,tighten]{aastex631}
\makeatletter
\@ifundefined{SI}{\DeclareRobustCommand{\SI}[2]{\ensuremath{#1\,\mathrm{#2}}}}{}
\@ifundefined{si}{}{}
\makeatother

\providecommand{\meter}{m}
\providecommand{\hour}{h}
\providecommand{\ampere}{A}
\providecommand{\micro}{\mu}    
\providecommand{\GeV}{GeV}
\providecommand{\TeV}{TeV}
\providecommand{\per}{\!/\!}               
\providecommand{\cm}{\mathrm{cm}}
\providecommand{\second}{\mathrm{s}}
\providecommand{\s}{\mathrm{s}}            
\providecommand{\squared}{^{2}}            

\makeatletter
\DeclareRobustCommand{\ang}[1]{\ensuremath{\ang@parse#1;;;;\@nil}}
\def\ang@parse#1;#2;#3;#4;#5\@nil{%
  #1^\circ
  \ifx\relax#2\relax\else\,#2'\fi
  \ifx\relax#3\relax\else\,#3''\fi
}
\makeatother

\newcommand{\noop}[1]{}

\submitjournal{ApJ}

\usepackage{tabularx}
\usepackage{amsmath}
\usepackage{graphicx}
\usepackage[caption=false]{subfig}
\usepackage{float}
\usepackage{hyperref}

\makeatletter
\def\@fnsymbol#1{\ensuremath{%
   \ifcase#1 \or *\or \dagger\or \ddagger\or
   \mathsection\or \mathparagraph\or \|\or **\or
   \dagger\dagger \or \ddagger\ddagger \or \mathsection\mathsection \or \mathparagraph\mathparagraph \or \star
   \else \@arabic{#1}
   \fi}}
\makeatother
\shorttitle{IACT neutrino ToOs} 
\shortauthors{FACT, H.E.S.S., MAGIC, VERITAS, Fermi-LAT, and IceCube }

\graphicspath{{./}{figures/}}

\begin{document}



\title{Prompt Searches for Very-High-Energy $\gamma$-Ray Counterparts to IceCube Astrophysical Neutrino Alerts}

\author[0000-0001-8215-4377]{J.~Abhir}
\affiliation{ETH Z\"urich, CH-8093 Z\"urich, Switzerland}
\author[0000-0002-1288-833X]{A.~Biland}
\affiliation{ETH Z\"urich, CH-8093 Z\"urich, Switzerland}
\author{K.~Brand}
\affiliation{Universit\"at W\"urzburg, D-97074 W\"urzburg, Germany}
\author{T.~Bretz}
\altaffiliation{also at GSI Darmstadt, Germany}
\affiliation{ETH Z\"urich, CH-8093 Z\"urich, Switzerland}
\author[0000-0001-8823-479X]{D.~Dorner}
\affiliation{Universit\"at W\"urzburg, D-97074 W\"urzburg, Germany}

\author{L.~Eisenberger}
\affiliation{Universit\"at W\"urzburg, D-97074 W\"urzburg, Germany}
\author[0000-0001-6796-3205]{D.~Elsaesser}
\affiliation{Technische Universit\"at Dortmund, D-44221 Dortmund, Germany}
\author{P.~G\"unther}
\affiliation{Universit\"at W\"urzburg, D-97074 W\"urzburg, Germany}
\author{S.~Hasan}
\affiliation{ETH Z\"urich, CH-8093 Z\"urich, Switzerland}
\author{D.~Hildebrand}
\affiliation{ETH Z\"urich, CH-8093 Z\"urich, Switzerland}
\author{K.~Mannheim}
\affiliation{Universit\"at W\"urzburg, D-97074 W\"urzburg, Germany}
\author{M.~Linhoff}
\affiliation{Technische Universit\"at Dortmund, D-44221 Dortmund, Germany}
\author{F.~Pfeifle}
\affiliation{Universit\"at W\"urzburg, D-97074 W\"urzburg, Germany}
\author{W.~Rhode}
\affiliation{Technische Universit\"at Dortmund, D-44221 Dortmund, Germany}
\author{B.~Schleicher}
\altaffiliation{corresponding author, \href{mailto:contact@fact-project.org}{contact@fact-project.org}}
\affiliation{Universit\"at W\"urzburg, D-97074 W\"urzburg, Germany}
\author{V.~Sliusar}
\affiliation{University of Geneva, Chemin d'Ecogia 16, CH-1290 Versoix, Switzerland}
\author{M.~Vorbrugg}
\affiliation{Universit\"at W\"urzburg, D-97074 W\"urzburg, Germany}
\author{R.~Walter}
\affiliation{University of Geneva, Chemin d'Ecogia 16, CH-1290 Versoix, Switzerland}

\collaboration{1000}{(FACT Collaboration)}

\author{F.~Aharonian}
\affiliation{Dublin Institute for Advanced Studies, 31 Fitzwilliam Place, Dublin 2, Ireland}
\affiliation{Max-Planck-Institut f\"ur Kernphysik, Saupfercheckweg 1, 69117 Heidelberg, Germany}
\affiliation{Yerevan State University,  1 Alek Manukyan St, Yerevan 0025, Armenia}

\author{F.~Ait~Benkhali}
\affiliation{Landessternwarte, Universit\"at Heidelberg, K\"onigstuhl, D 69117 Heidelberg, Germany}

\author{J.~Aschersleben}
\affiliation{Kapteyn Astronomical Institute, University of Groningen, Landleven 12, 9747 AD Groningen, The Netherlands}

\author[0000-0002-2153-1818]{H.~Ashkar}
\affiliation{Laboratoire Leprince-Ringuet, École Polytechnique, CNRS, Institut Polytechnique de Paris, F-91128 Palaiseau, France}

\author[0000-0002-9326-6400]{M.~Backes}
\affiliation{University of Namibia, Department of Physics, Private Bag 13301, Windhoek 10005, Namibia}
\affiliation{Centre for Space Research, North-West University, Potchefstroom 2520, South Africa}

\author[0000-0002-5085-8828]{V.~Barbosa~Martins}
\affiliation{Deutsches Elektronen-Synchrotron DESY, Platanenallee 6, 15738 Zeuthen, Germany}

\author[0000-0002-5797-3386]{R.~Batzofin}
\affiliation{Institut f\"ur Physik und Astronomie, Universit\"at Potsdam,  Karl-Liebknecht-Strasse 24/25, D 14476 Potsdam, Germany}

\author[0000-0002-2115-2930]{Y.~Becherini}
\affiliation{Université Paris Cité, CNRS, Astroparticule et Cosmologie, F-75013 Paris, France}
\affiliation{Department of Physics and Electrical Engineering, Linnaeus University,  351 95 V\"axj\"o, Sweden}

\author[0000-0002-2918-1824]{D.~Berge}
\affiliation{Institut f\"ur Physik, Humboldt-Universit\"at zu Berlin, Newtonstr. 15, D 12489 Berlin, Germany}
\affiliation{Deutsches Elektronen-Synchrotron DESY, Platanenallee 6, 15738 Zeuthen, Germany}

\author[0000-0002-8434-5692]{M.~B\"ottcher}
\affiliation{Centre for Space Research, North-West University, Potchefstroom 2520, South Africa}

% ...

\collaboration{1000}{(H.E.S.S. Collaboration)}

\author[0000-0001-7250-3596]{S.~Abe}
\affiliation{Japanese MAGIC Group: Institute for Cosmic Ray Research (ICRR), The University of Tokyo, Kashiwa, 277-8582 Chiba, Japan}
\author[0000-0001-8215-4377]{J.~Abhir}
\affiliation{ETH Z\"urich, CH-8093 Z\"urich, Switzerland}
\author{A.~Abhishek}
\affiliation{Universit\`a di Siena and INFN Pisa, I-53100 Siena, Italy}
\author[0000-0001-8816-4920]{A.~Aguasca-Cabot}
\affiliation{Universitat de Barcelona, ICCUB, IEEC-UB, E-08028 Barcelona, Spain}
\author[0000-0002-3777-6182]{I.~Agudo}
\affiliation{Instituto de Astrof\'isica de Andaluc\'ia-CSIC, Glorieta de la Astronom\'ia s/n, 18008, Granada, Spain}
\author{T.~Aniello}
\affiliation{National Institute for Astrophysics (INAF), I-00136 Rome, Italy}
\author[0000-0002-5613-7693]{S.~Ansoldi}
\affiliation{Universit\`a di Udine and INFN Trieste, I-33100 Udine, Italy}\affiliation{also at International Center for Relativistic Astrophysics (ICRA), Rome, Italy}
\author[0000-0002-5037-9034]{L.~A.~Antonelli}
\affiliation{National Institute for Astrophysics (INAF), I-00136 Rome, Italy}
\author[0000-0001-9076-9582]{A.~Arbet Engels}
\affiliation{Max-Planck-Institut f\"ur Physik, D-85748 Garching, Germany}
\author[0000-0002-1998-9707]{C.~Arcaro}
\affiliation{Universit\`a di Padova and INFN, I-35131 Padova, Italy}

% ...


\collaboration{1000}{(MAGIC Collaboration)}


\author[0000-0002-2028-9230]{A.~Acharyya} \affiliation{CP3-Origins, University of Southern Denmark, Campusvej 55, 5230 Odense M, Denmark}
\author{A.~Archer} \affiliation{Department of Physics and Astronomy, DePauw University, Greencastle, IN 46135-0037, USA}
\author[0000-0002-3886-3739]{P.~Bangale} \affiliation{Department of Physics, Temple University, Philadelphia, PA 19122, USA}
\author[0000-0002-9675-7328]{J.~T.~Bartkoske} \affiliation{Department of Physics and Astronomy, University of Utah, Salt Lake City, UT 84112, USA}
\author[0000-0003-2098-170X]{W.~Benbow} \affiliation{Center for Astrophysics $|$ Harvard \& Smithsonian, Cambridge, MA 02138, USA}
\author[0000-0001-6391-9661]{J.~H.~Buckley} \affiliation{Department of Physics, Washington University, St. Louis, MO 63130, USA}
\author[0009-0001-5719-936X]{Y.~Chen} \affiliation{Department of Physics and Astronomy, University of California, Los Angeles, CA 90095, USA}
\author{J.~L.~Christiansen} \affiliation{Physics Department, California Polytechnic State University, San Luis Obispo, CA 94307, USA}
\author{A.~J.~Chromey} \affiliation{Center for Astrophysics $|$ Harvard \& Smithsonian, Cambridge, MA 02138, USA}
\author[0000-0002-1853-863X]{M.~Errando} \affiliation{Department of Physics, Washington University, St. Louis, MO 63130, USA}
\author{S.~Feldman} \affiliation{Department of Physics and Astronomy, University of California, Los Angeles, CA 90095, USA}
\author[0000-0001-6674-4238]{Q.~Feng} \affiliation{Department of Physics and Astronomy, University of Utah, Salt Lake City, UT 84112, USA}
\author[0000-0002-2636-4756]{S.~Filbert} \affiliation{Department of Physics and Astronomy, University of Utah, Salt Lake City, UT 84112, USA}
\author[0000-0002-1067-8558]{L.~Fortson} \affiliation{School of Physics and Astronomy, University of Minnesota, Minneapolis, MN 55455, USA}
\author[0000-0003-1614-1273]{A.~Furniss} \affiliation{Santa Cruz Institute for Particle Physics and Department of Physics, University of California, Santa Cruz, CA 95064, USA}
\author[0000-0002-0109-4737]{W.~Hanlon} \affiliation{Center for Astrophysics $|$ Harvard \& Smithsonian, Cambridge, MA 02138, USA}
\author[0000-0003-3878-1677]{O.~Hervet} \affiliation{Santa Cruz Institute for Particle Physics and Department of Physics, University of California, Santa Cruz, CA 95064, USA}
\author[0000-0001-6951-2299]{C.~E.~Hinrichs} \affiliation{Center for Astrophysics $|$ Harvard \& Smithsonian, Cambridge, MA 02138, USA and Department of Physics and Astronomy, Dartmouth College, 6127 Wilder Laboratory, Hanover, NH 03755 USA}
\author[0000-0002-6833-0474]{J.~Holder} \affiliation{Department of Physics and Astronomy and the Bartol Research Institute, University of Delaware, Newark, DE 19716, USA}
\author{Z.~Hughes} \affiliation{Department of Physics, Washington University, St. Louis, MO 63130, USA}
\author[0000-0002-1432-7771]{T.~B.~Humensky} \affiliation{Department of Physics, University of Maryland, College Park, MD, USA and NASA GSFC, Greenbelt, MD 20771, USA}
\author[0000-0002-1089-1754]{W.~Jin}\altaffiliation{corresponding author, \href{wjin@astro.ucla.edu}{wjin@astro.ucla.edu}} \affiliation{Department of Physics and Astronomy, University of California, Los Angeles, CA 90095, USA}
\author[0009-0008-2688-0815]{M.~N.~Johnson} \affiliation{Santa Cruz Institute for Particle Physics and Department of Physics, University of California, Santa Cruz, CA 95064, USA}
\author[0000-0002-3638-0637]{P.~Kaaret} \affiliation{Department of Physics and Astronomy, University of Iowa, Van Allen Hall, Iowa City, IA 52242, USA}
\author{M.~Kertzman} \affiliation{Department of Physics and Astronomy, DePauw University, Greencastle, IN 46135-0037, USA}
\author{M.~Kherlakian} \affiliation{Fakult\"at f\"ur Physik \& Astronomie, Ruhr-Universit\"at Bochum, D-44780 Bochum, Germany}
\author[0000-0003-4785-0101]{D.~Kieda} \affiliation{Department of Physics and Astronomy, University of Utah, Salt Lake City, UT 84112, USA}
\author[0000-0002-4260-9186]{T.~K.~Kleiner} \affiliation{DESY, Platanenallee 6, 15738 Zeuthen, Germany}
\author[0000-0002-4289-7106]{N.~Korzoun} \affiliation{Department of Physics and Astronomy and the Bartol Research Institute, University of Delaware, Newark, DE 19716, USA}
\author[0000-0003-4641-4201]{M.~J.~Lang} \affiliation{School of Natural Sciences, University of Galway, University Road, Galway, H91 TK33, Ireland}
\author[0000-0003-3802-1619]{M.~Lundy} \affiliation{Physics Department, McGill University, Montreal, QC H3A 2T8, Canada}
\author[0000-0001-9868-4700]{G.~Maier} \affiliation{DESY, Platanenallee 6, 15738 Zeuthen, Germany}
\author[0000-0001-7106-8502]{M.~J.~Millard} \affiliation{Department of Physics and Astronomy, University of Iowa, Van Allen Hall, Iowa City, IA 52242, USA}
\author{J.~Millis} \affiliation{Department of Physics and Astronomy, Ball State University, Muncie, IN 47306, USA and Department of Physics, Anderson University, 1100 East 5th Street, Anderson, IN 46012}
\author[0000-0002-1499-2667]{P.~Moriarty} \affiliation{School of Natural Sciences, University of Galway, University Road, Galway, H91 TK33, Ireland}
\author[0000-0002-3223-0754]{R.~Mukherjee} \affiliation{Department of Physics and Astronomy, Barnard College, Columbia University, NY 10027, USA}
\author[0000-0002-6121-3443]{W.~Ning} \affiliation{Department of Physics and Astronomy, University of California, Los Angeles, CA 90095, USA}
\author[0000-0002-4837-5253]{R.~A.~Ong} \affiliation{Department of Physics and Astronomy, University of California, Los Angeles, CA 90095, USA}
\author[0000-0003-3820-0887]{A.~Pandey} \affiliation{Department of Physics and Astronomy, University of Utah, Salt Lake City, UT 84112, USA}
\author[0000-0001-7861-1707]{M.~Pohl} \affiliation{Institute of Physics and Astronomy, University of Potsdam, 14476 Potsdam-Golm, Germany and DESY, Platanenallee 6, 15738 Zeuthen, Germany}
\author[0000-0002-4855-2694]{J.~Quinn} \affiliation{School of Physics, University College Dublin, Belfield, Dublin 4, Ireland}
\author{P.~L.~Rabinowitz} \affiliation{Department of Physics, Washington University, St. Louis, MO 63130, USA}
\author[0000-0002-5351-3323]{K.~Ragan} \affiliation{Physics Department, McGill University, Montreal, QC H3A 2T8, Canada}
\author{P.~T.~Reynolds} \affiliation{Department of Physical Sciences, Munster Technological University, Bishopstown, Cork, T12 P928, Ireland}
\author[0000-0002-7523-7366]{D.~Ribeiro} \affiliation{School of Physics and Astronomy, University of Minnesota, Minneapolis, MN 55455, USA}
\author{E.~Roache} \affiliation{Center for Astrophysics $|$ Harvard \& Smithsonian, Cambridge, MA 02138, USA}
\author[0000-0003-1387-8915]{I.~Sadeh} \affiliation{DESY, Platanenallee 6, 15738 Zeuthen, Germany}
\author{A.~C.~Sadun} \affiliation{Department of Physics, University of Colorado Denver, Campus Box 157, P.O. Box 173364, Denver CO 80217, USA}
\author[0000-0002-3171-5039]{L.~Saha} \affiliation{Center for Astrophysics $|$ Harvard \& Smithsonian, Cambridge, MA 02138, USA}
\author{M.~Santander} \affiliation{Department of Physics and Astronomy, University of Alabama, Tuscaloosa, AL 35487, USA}
\author{G.~H.~Sembroski} \affiliation{Department of Physics and Astronomy, Purdue University, West Lafayette, IN 47907, USA}
\author[0000-0002-9856-989X]{R.~Shang} \affiliation{Department of Physics and Astronomy, Barnard College, Columbia University, NY 10027, USA}
\author[0000-0002-9852-2469]{D.~Tak} \affiliation{SNU Astronomy Research Center, Seoul National University, Seoul 08826, Republic of Korea.}
\author{A.~K.~Talluri} \affiliation{School of Physics and Astronomy, University of Minnesota, Minneapolis, MN 55455, USA}
\author{J.~V.~Tucci} \affiliation{Department of Physics, Indiana University Indianapolis, Indianapolis, Indiana 46202, USA}
\author[0000-0002-8090-6528]{J.~Valverde} \affiliation{Department of Physics, University of Maryland, Baltimore County, Baltimore MD 21250, USA and NASA GSFC, Greenbelt, MD 20771, USA}
\author{V.~V.~Vassiliev} \affiliation{Department of Physics and Astronomy, University of California, Los Angeles, CA 90095, USA}
\author[0000-0003-2740-9714]{D.~A.~Williams} \affiliation{Santa Cruz Institute for Particle Physics and Department of Physics, University of California, Santa Cruz, CA 95064, USA}
\author[0000-0002-2730-2733]{S.~L.~Wong} \affiliation{Physics Department, McGill University, Montreal, QC H3A 2T8, Canada}\collaboration{1000}{(VERITAS Collaboration)}

\author[0000-0002-3308-324X]{S. Buson}
\affiliation{Deutsches Elektronen-Synchrotron DESY, Platanenallee 6, 15738 Zeuthen, Germany}
\affiliation{Julius-Maximilians-Universität Würzburg, Fakultät für Physik und Astronomie, Institut für Theoretische Physik und Astrophysik, Lehrstuhl für Astronomie, Emil-Fischer-Str. 31, D-97074 Würzburg, Germany}
\collaboration{1000}{({\em Fermi}-LAT Collaboration)}

\affiliation{III. Physikalisches Institut, RWTH Aachen University, D-52056 Aachen, Germany}
\affiliation{Department of Physics, University of Adelaide, Adelaide, 5005, Australia}
\affiliation{Dept. of Physics and Astronomy, University of Alaska Anchorage, 3211 Providence Dr., Anchorage, AK 99508, USA}
\affiliation{School of Physics and Center for Relativistic Astrophysics, Georgia Institute of Technology, Atlanta, GA 30332, USA}
\affiliation{Dept. of Physics, Southern University, Baton Rouge, LA 70813, USA}
\affiliation{Dept. of Physics, University of California, Berkeley, CA 94720, USA}
\affiliation{Lawrence Berkeley National Laboratory, Berkeley, CA 94720, USA}
\affiliation{Institut f{\"u}r Physik, Humboldt-Universit{\"a}t zu Berlin, D-12489 Berlin, Germany}
\affiliation{Fakult{\"a}t f{\"u}r Physik {\&} Astronomie, Ruhr-Universit{\"a}t Bochum, D-44780 Bochum, Germany}
\affiliation{Universit{\'e} Libre de Bruxelles, Science Faculty CP230, B-1050 Brussels, Belgium}
\affiliation{Vrije Universiteit Brussel (VUB), Dienst ELEM, B-1050 Brussels, Belgium}
\affiliation{Dept. of Physics, Simon Fraser University, Burnaby, BC V5A 1S6, Canada}
\affiliation{Department of Physics and Laboratory for Particle Physics and Cosmology, Harvard University, Cambridge, MA 02138, USA}
\affiliation{Dept. of Physics, Massachusetts Institute of Technology, Cambridge, MA 02139, USA}
\affiliation{Dept. of Physics and The International Center for Hadron Astrophysics, Chiba University, Chiba 263-8522, Japan}
\affiliation{Department of Physics, Loyola University Chicago, Chicago, IL 60660, USA}
\affiliation{Dept. of Physics and Astronomy, University of Canterbury, Private Bag 4800, Christchurch, New Zealand}
\affiliation{Dept. of Physics, University of Maryland, College Park, MD 20742, USA}
\affiliation{Dept. of Astronomy, Ohio State University, Columbus, OH 43210, USA}
\affiliation{Dept. of Physics and Center for Cosmology and Astro-Particle Physics, Ohio State University, Columbus, OH 43210, USA}
\affiliation{Niels Bohr Institute, University of Copenhagen, DK-2100 Copenhagen, Denmark}
\affiliation{Dept. of Physics, TU Dortmund University, D-44221 Dortmund, Germany}
\affiliation{Dept. of Physics and Astronomy, Michigan State University, East Lansing, MI 48824, USA}
\affiliation{Dept. of Physics, University of Alberta, Edmonton, Alberta, T6G 2E1, Canada}
\affiliation{Erlangen Centre for Astroparticle Physics, Friedrich-Alexander-Universit{\"a}t Erlangen-N{\"u}rnberg, D-91058 Erlangen, Germany}
\affiliation{Physik-department, Technische Universit{\"a}t M{\"u}nchen, D-85748 Garching, Germany}
\affiliation{D{\'e}partement de physique nucl{\'e}aire et corpusculaire, Universit{\'e} de Gen{\`e}ve, CH-1211 Gen{\`e}ve, Switzerland}
\affiliation{Dept. of Physics and Astronomy, University of Gent, B-9000 Gent, Belgium}
\affiliation{Dept. of Physics and Astronomy, University of California, Irvine, CA 92697, USA}
\affiliation{Karlsruhe Institute of Technology, Institute for Astroparticle Physics, D-76021 Karlsruhe, Germany}
\affiliation{Karlsruhe Institute of Technology, Institute of Experimental Particle Physics, D-76021 Karlsruhe, Germany}
\affiliation{Dept. of Physics, Engineering Physics, and Astronomy, Queen's University, Kingston, ON K7L 3N6, Canada}
\affiliation{Department of Physics {\&} Astronomy, University of Nevada, Las Vegas, NV 89154, USA}
\affiliation{Nevada Center for Astrophysics, University of Nevada, Las Vegas, NV 89154, USA}
\affiliation{Dept. of Physics and Astronomy, University of Kansas, Lawrence, KS 66045, USA}
\affiliation{Centre for Cosmology, Particle Physics and Phenomenology - CP3, Universit{\'e} catholique de Louvain, Louvain-la-Neuve, Belgium}
\affiliation{Department of Physics, Mercer University, Macon, GA 31207-0001, USA}
\affiliation{Dept. of Astronomy, University of Wisconsin{\textemdash}Madison, Madison, WI 53706, USA}
\affiliation{Dept. of Physics and Wisconsin IceCube Particle Astrophysics Center, University of Wisconsin{\textemdash}Madison, Madison, WI 53706, USA}
\affiliation{Institute of Physics, University of Mainz, Staudinger Weg 7, D-55099 Mainz, Germany}
\affiliation{Department of Physics, Marquette University, Milwaukee, WI 53201, USA}
\affiliation{Institut f{\"u}r Kernphysik, Universit{\"a}t M{\"u}nster, D-48149 M{\"u}nster, Germany}
\affiliation{Bartol Research Institute and Dept. of Physics and Astronomy, University of Delaware, Newark, DE 19716, USA}
\affiliation{Dept. of Physics, Yale University, New Haven, CT 06520, USA}
\affiliation{Columbia Astrophysics and Nevis Laboratories, Columbia University, New York, NY 10027, USA}
\affiliation{Dept. of Physics, University of Oxford, Parks Road, Oxford OX1 3PU, United Kingdom}
\affiliation{Dipartimento di Fisica e Astronomia Galileo Galilei, Universit{\`a} Degli Studi di Padova, I-35122 Padova PD, Italy}
\affiliation{Dept. of Physics, Drexel University, 3141 Chestnut Street, Philadelphia, PA 19104, USA}
\affiliation{Physics Department, South Dakota School of Mines and Technology, Rapid City, SD 57701, USA}
\affiliation{Dept. of Physics, University of Wisconsin, River Falls, WI 54022, USA}
\affiliation{Dept. of Physics and Astronomy, University of Rochester, Rochester, NY 14627, USA}
\affiliation{Department of Physics and Astronomy, University of Utah, Salt Lake City, UT 84112, USA}
\affiliation{Dept. of Physics, Chung-Ang University, Seoul 06974, Republic of Korea}
\affiliation{Oskar Klein Centre and Dept. of Physics, Stockholm University, SE-10691 Stockholm, Sweden}
\affiliation{Dept. of Physics and Astronomy, Stony Brook University, Stony Brook, NY 11794-3800, USA}
\affiliation{Dept. of Physics, Sungkyunkwan University, Suwon 16419, Republic of Korea}
\affiliation{Institute of Physics, Academia Sinica, Taipei, 11529, Taiwan}
\affiliation{Dept. of Physics and Astronomy, University of Alabama, Tuscaloosa, AL 35487, USA}
\affiliation{Dept. of Astronomy and Astrophysics, Pennsylvania State University, University Park, PA 16802, USA}
\affiliation{Dept. of Physics, Pennsylvania State University, University Park, PA 16802, USA}
\affiliation{Dept. of Physics and Astronomy, Uppsala University, Box 516, SE-75120 Uppsala, Sweden}
\affiliation{Dept. of Physics, University of Wuppertal, D-42119 Wuppertal, Germany}
\affiliation{Deutsches Elektronen-Synchrotron DESY, Platanenallee 6, D-15738 Zeuthen, Germany}

\author[0000-0001-6141-4205]{R. Abbasi}
\affiliation{Department of Physics, Loyola University Chicago, Chicago, IL 60660, USA}

\author[0000-0001-8952-588X]{M. Ackermann}
\affiliation{Deutsches Elektronen-Synchrotron DESY, Platanenallee 6, D-15738 Zeuthen, Germany}

\author{J. Adams}
\affiliation{Dept. of Physics and Astronomy, University of Canterbury, Private Bag 4800, Christchurch, New Zealand}

\author[0000-0002-9714-8866]{S. K. Agarwalla}
\altaffiliation{also at Institute of Physics, Sachivalaya Marg, Sainik School Post, Bhubaneswar 751005, India}
\affiliation{Dept. of Physics and Wisconsin IceCube Particle Astrophysics Center, University of Wisconsin{\textemdash}Madison, Madison, WI 53706, USA}

\author[0000-0003-2252-9514]{J. A. Aguilar}
\affiliation{Universit{\'e} Libre de Bruxelles, Science Faculty CP230, B-1050 Brussels, Belgium}

\author[0000-0003-0709-5631]{M. Ahlers}
\affiliation{Niels Bohr Institute, University of Copenhagen, DK-2100 Copenhagen, Denmark}

\author[0000-0002-9534-9189]{J.M. Alameddine}
\affiliation{Dept. of Physics, TU Dortmund University, D-44221 Dortmund, Germany}

\author[0009-0001-2444-4162]{S. Ali}
\affiliation{Dept. of Physics and Astronomy, University of Kansas, Lawrence, KS 66045, USA}

\author{N. M. Amin}
\affiliation{Bartol Research Institute and Dept. of Physics and Astronomy, University of Delaware, Newark, DE 19716, USA}

\author[0000-0001-9394-0007]{K. Andeen}
\affiliation{Department of Physics, Marquette University, Milwaukee, WI 53201, USA}

% ...
\collaboration{500}{(IceCube Collaboration)}

\author[0000-0001-7618-7527]{F. D'Ammando}
\affiliation{INAF-IRA Bologna, Via P. Gobetti 101, I-40129 Bologna, Italy}




\begin{abstract}
The search for sources of high-energy astrophysical neutrinos can be significantly advanced through a multi-messenger approach, which seeks to  detect the $\gamma$-rays that accompany neutrinos as they are produced at their sources. Multi-messenger observations have so far provided the first evidence for a neutrino source, illustrated by the joint detection of the flaring blazar TXS 0506+056 in high-energy (HE, $\mathrm{E} > 1\,\mathrm{GeV}$) and very-high-energy (VHE, $\mathrm{E} > 100\,\mathrm{GeV}$) $\gamma$ rays in coincidence with the high-energy neutrino IceCube-170922A, identified by IceCube. 
Imaging atmospheric Cherenkov telescopes (IACTs), namely FACT, H.E.S.S., MAGIC, and VERITAS, continue to conduct extensive neutrino target-of-opportunity follow-up programs. These programs have two components: follow-up observations of single astrophysical neutrino candidate events (such as IceCube-170922A), and observation of known $\gamma$-ray sources after the identification of a cluster of neutrino events by IceCube. 
Here we present a comprehensive analysis of follow-up observations of high-energy neutrino events observed by the four IACTs between September 2017 (after the IceCube-170922A event) and January 2021. Our study found no associations between $\gamma$-ray sources and the observed neutrino events. We provide a detailed overview of each neutrino event and its potential counterparts. Furthermore, a joint analysis of all IACT data is included, yielding combined upper limits on the VHE $\gamma$-ray flux.

\end{abstract}

\keywords{Neutrino astronomy (1100) --- High energy astrophysics (739) --- $\gamma$-ray observatories (632)}

\newpage
\section{Introduction} 
\label{sec:intro}

The detection of high-energy astrophysical neutrinos at hundreds of TeV is an important step towards understanding the origin of cosmic rays (CRs). 
During their acceleration and/or propagation, hadronic CRs interact with ambient matter or radiation fields, producing high-energy neutrinos through the decay of secondary particles. As such, high-energy neutrinos serve as a distinctive signature of hadronic acceleration processes~\citep{Atoyan2001, Dermer2007, Ahlers2018Sep}. 

In 2013, the  IceCube  Neutrino  Observatory~\citep{IC2017_system} announced the discovery of a diffuse flux of cosmic neutrinos~\citep{IceCube:2013low}, which is now well established through additional measurements~\citep{IceCube:2020wum}.
The astrophysical neutrino flux shows an isotropic distribution, favoring an extragalactic origin. This is also supported by recent evidence for neutrino emission from active galactic nuclei (AGNs)~\citep{TXS_2017, IceCube:2018cha, IceCube:2022der}. 

The contributions of transient and steady extragalactic sources to the diffuse astrophysical neutrino flux have been significantly constrained~\citep{Abbasi_2023}, while those of identified sources remain minimal: TXS\,0506+056 and NGC 1068 each would account for no more than $\sim$1\% of the total flux~\citep{TXS_2017, IceCube:2018cha, IceCube:2022der}, while the Galactic plane contributes approximately 10\% at 30~TeV~\citep{IC2023gal_plane}. Consequently, the sources responsible for the bulk of the diffuse neutrino flux are still to be identified ~\citep[e.g.,][]{Buson_2022}.

The challenge in identifying sources is caused by low signal statistics: only $\mathcal{O}(10)$ high-energy neutrino events with a high probability of being astrophysical are singled out by IceCube each year, and their angular localization is uncertain to $1^{\circ}$ or more~\citep{2017APh....92...30A}.

In this context, simultaneous electromagnetic (EM) observations are essential for identifying neutrino sources. A particularly important role is played by very-high-energy (VHE, $E>100$ GeV) $\gamma$-rays, which are produced together with neutrinos in the same hadronic ($pp$) or photo-hadronic ($p\gamma$) interactions. 
These interactions produce charged and neutral pions that decay into neutrinos and $\gamma$-rays, respectively.
Thus, the observation of $\gamma$-rays in coincidence with high-energy neutrinos may dramatically increase the significance of any detections and pinpoint genuine hadronic accelerators, giving us valuable insights into the CR acceleration mechanisms in the observed sources.
However, unlike neutrinos, VHE $\gamma$-rays may lose energy or be
absorbed within their source environment or during propagation over cosmological distances due to their interactions with the extragalactic background light (EBL) \citep{2013APh....43..112D, 2019ApJ...881...46R}. In such scenarios, VHE
$\gamma$-ray telescopes may not be able to detect the source.
In any case, combining multiwavelength (MWL) and VHE observations provides essential information on the emission
mechanisms occurring within sources as well as the characteristics of their emission
environments \citep{Dermer2007, Ansoldi2018, Cerruti2019, Rodrigues2019}.

The efficacy of following up neutrino events with EM observations was demonstrated with the September 22, 2017 report of a high-energy neutrino (IceCube-170922A, $\sim$~290 TeV) that was spatially and temporally coincident
with the flaring $\gamma$-ray blazar TXS\,0506+056 \citep{TXS_2017}.
This is one of the most compelling pieces of evidence for a neutrino point source so far. The significance of this correlation is estimated to be at the $3\,\sigma$ level. The MWL follow-up of this alert was key for establishing this coincidence and constraining the subsequent theoretical modeling for the emission.

Another recent example is the detection of excess astrophysical neutrino events associated with the nearby active galaxy NGC 1068. This association came out from the search for neutrinos from a list of 110 known gamma-ray sources selected \emph{a priori} from the \emph{Fermi}-LAT 4FGL-DR2 catalog.
A new analysis improves upon the previous results,
with the significance increasing from
$2.9\,\sigma$~\citep{IceCube:2020ps} to $4.2\,\sigma$~\citep{IceCube:2022der}. 

The four imaging atmospheric Cherenkov telescopes (IACTs) systems - FACT \citep[][]{2014JInst...9P0012B}, H.E.S.S.\ \citep{2006A&A...457..899A}, MAGIC \citep{2012APh....35..435A} and VERITAS \citep{Holder:2006gi} - conduct neutrino event follow-up observations in cooperation with IceCube, with the aim of identifying potential $\gamma$-ray counterparts. 
These target-of-opportunity (ToO)
programs can be broadly categorized according to two observing strategies, depending on whether they are triggered by neutrino event clusters or single high-energy neutrino events. 

The first strategy involves the follow-up observation of \emph{clusters} of candidate neutrino events with energies above $\sim$1\,TeV detected by IceCube around a hypothetical source location and within a limited time window. The cluster alerts are privately distributed by IceCube to individual IACTs under a memorandum of understanding 
in the framework of the Gamma-ray follow-up (GFU) program, in operation since 2012~\citep{2016JInst..1111009I}

The second strategy involves the follow-up observation of \emph{single high-energy neutrino events} ($>$60\,TeV), that are likely of astrophysical origin. Single neutrino alerts are publicly distributed, with a typical localization uncertainty radius of $\sim1^\circ$ (i.e., $\sim2^\circ$ diameter containment region), which matches well with current IACT fields of view (FoVs) of $3.5^{\circ}$-$5^{\circ}$.
If a promising neutrino source candidate, such as an AGN (e.g., TXS 0506+056) identified  using the Large Area Telescope (LAT) onboard the \emph{Fermi} Gamma-ray Space Telescope and IACT catalogs, or a transient source identified by one of the many EM observatories worldwide, falls within the region of interest (ROI) defined by the neutrino’s localization uncertainty, observations are usually focussed on these specific objects. If no promising source candidates can be identified \emph{a priori}, the search typically covers the whole ROI defined by the neutrino localization uncertainty.

The MAGIC and VERITAS results from the first stage of the GFU program (up to 2016) are presented in \cite{2016JInst..1111009I}, while those referring to alerts issued from 2016 up to the photon-neutrino coincidence in September 2017 are reported in \citet{Santander:20174H}, \citet{Schussler:2017mo}, and \citet[][see appendix]{TXS_2017}. 
Results from observations of high-energy neutrino event positions, under the hypothesis of steady or long-term source emission, have been presented in \cite{2017ICRC...35..618S}.

In this paper, we present a retrospective analysis of follow-up of real-time IceCube neutrino alerts using both strategies during the period from September 2017 (i.e., after the IceCube-170922A and TXS\,0506+056 detection), to January 2021. The paper is organized as follows: in Section~\ref{sec:icecube}, we give a brief overview of the alert channels distributed by IceCube. 
Section~\ref{sec:iacts} provides a description of the neutrino follow-up programs conducted by the IACTs, including an overview of the VHE $\gamma$-ray observations that have been performed (Section~\ref{vhe-obs}) and a description of the methods used to combine IACT data (Section~\ref{sec:methods}).
In Section~\ref{sec:observations}, we describe the complementary observations performed by MWL instruments and the respective analysis techniques.
Sections \ref{sec:results_flares} and \ref{sec:results_tracks} respectively present the results from the IACT and MWL observations for each neutrino alert,  whether cluster or single neutrino, that was followed up.
We conclude with a discussion in Section~\ref{sec:discussion}. The Appendix summarizes additional details of the observations and results. 

\section{IceCube neutrino alerts} \label{sec:icecube}

The IceCube Neutrino Observatory~\citep{IC2017_system} is a cubic-kilometer neutrino telescope located at the South Pole and designed to search for astrophysical sources of neutrinos and study fundamental neutrino physics. It is currently the most sensitive neutrino telescope in the TeV-PeV range. It consists of 5160 spherically shaped optical sensors called Digital Optical Modules (DOMs) that are deployed below the surface of the ice between 1450 m and 2450 m. Cherenkov light is produced by secondary charged particles when neutrinos interact in or near the active detector volume. This light can be detected by the DOMs. The signals are digitized and sent to the IceCube Laboratory at the surface of the ice sheet. Here, key information is reconstructed in real-time such as the energy, direction and time sequence of arriving neutrinos.

The construction of the IceCube detector was completed in 2010, and the discovery of an astrophysical neutrino flux in the TeV-PeV energy range was first announced in 2013~\citep{IceCube:2013low}. The properties of the diffuse neutrino flux are measured in different analyses, sensitive to various event topologies (i.e., the characteristic light patterns observed by the detector’s optical modules), flavor composition and sky regions~\citep[see e.g.,][]{IceCube:2020acn, IceCube:2020wum, IceCube:2021uhz}. 
We note that, as IceCube cannot distinguish between neutrinos and antineutrinos, we refer to both as \emph{neutrinos} in the following.

In the search for astrophysical neutrinos, the primary backgrounds are atmospheric muon neutrinos and atmospheric muons that are misidentified as neutrinos, both originating from cosmic-ray interactions in the Earth’s atmosphere. For event geometries corresponding to downgoing trajectories—as observed by IceCube—the muon background dominates, particularly affecting observations of the Southern sky. In contrast, observations of the Northern sky are primarily limited by atmospheric neutrinos alone, enabling more sensitive searches for astrophysical neutrinos in that direction.

The atmospheric neutrino background can be discriminated thanks to its soft spectrum ($\propto E^{-3.7}$) and known zenith-angle dependence. 

On the one hand, taking advantage of the isotropy of the atmospheric neutrino background, an astrophysical neutrino emission can be probed by searching for clusters of neutrino events.
This search can be performed over the entire sky, or at the location of potential neutrino emitters, and different timescales are tested. This is the basis for most point-source searches in IceCube and for the real-time GFU program which produces the \emph{GFU-cluster} alert stream, described in Section~\ref{subsec:multiplet}.

On the other hand, at energies above~150 TeV, the harder spectrum of the astrophysical neutrino flux dominates the background of muons and neutrinos arising from CR interactions in the atmosphere. This allows  selection, in near real-time, of single-neutrino events that are likely astrophysical, based on their energy. This selection feeds the 
\emph{single high-energy neutrino} alert stream described in Section~\ref{subsec:singlet}.

\subsection{GFU-cluster alert stream}\label{subsec:multiplet}

The real-time GFU event reconstruction and selection chain is designed to yield a sample of  well-pointing candidate neutrino events, suitable for searching for point sources in both space and time by  identifying localised event excesses above the expected background over time (i.e., neutrino flares)~\citep{Kintscher2020Rapid, GFUpaper}.

This scheme follows an earlier work implemented in 2006 for the AMANDA-II and MAGIC experiments, which later evolved into the GFU program between the IceCube, MAGIC and VERITAS instruments~\citep{2016JInst..1111009I, 2017APh....92...30A}.
In its most recent version, developed in 2019, a time-dependent point-source analysis is applied in real time to the GFU event sample to monitor the locations of known $\gamma$-ray emitters as well as the whole sky \citep{Kintshcer2019townhall, BoscoloMeneguolo:20233e}. The sources monitored by the GFU program are mostly BL-Lac objects and FSRQs. They are selected from the \emph{Fermi}-LAT 3FGL~\citep{2015ApJS..218...23A} and 3FHL~\citep{2017ApJS..232...18A} catalogs, with selection criteria on redshift ($z\leq$1), flux variability, spectral index and a high $\gamma$-ray flux with energy up to 100 GeV (see \cite{2016JInst..1111009I} for the detailed definition of these quantities). These criteria maximize the probability for VHE gamma-ray detection by IACTs. 



The search for neutrino flares from hypothetical sources is conducted by running a likelihood analysis across multiple sliding time windows. Different timescales are tested with increasingly larger time windows: starting from the latest recorded event (the trigger), a set of significant preceding events (defined in~\citep{Kintscher2020Rapid}) are picked for building retroactive time windows to be tested, up to a maximum duration of 180 days, with the trigger serving as the endpoint of each window.
Within each possible time window, a likelihood ratio test is performed. The likelihood accounts for the angular distance between observed neutrinos and the hypothetical source direction, estimates of the angular reconstruction uncertainty, the energy spectra of both potential sources and background, and the detector live-time during the time window being tested. 
Two parameters are fitted: the spectral index and the most likely number of signal events within the given time window.

A test statistic (TS) is calculated for each tested time window by comparing the best-fit result versus the null hypothesis (i.e., the ratio between the likelihood value corresponding to the best-fit spectral index and number of signal events, and the likelihood assuming no signal events).
The time window yielding the overall largest TS is selected as the candidate cluster. The pre-trials significance is evaluated by comparing the observed TS with the distribution of TS obtained from repeated background-only pseudo-experiments.

The outcome of the likelihood scan for every triggering event therefore includes the duration of the most likely cluster time window, the corresponding fitted spectral index and number of signal events, and the significance of the resulting candidate neutrino cluster.

When pre-trials significant excess above a predetermined threshold is registered, an alert is privately sent to partner IACTs, allowing them to rapidly re-point their telescopes to acquire $\gamma$-ray observations from the direction of the candidate source. 
Alerts are issued only once the significance threshold is exceeded and are not updated afterwards. New alerts for the same source can only be generated after the significance level falls back below the threshold.  

It is likely that some alerts are given by background fluctuations. Hence, we calculate the false-alert rate (FAR)~\citep{BoscoloMeneguolo:20233e} in order to quantify the frequency at which a cluster is found from some particular direction with a significance surpassing the pre-defined threshold, being produced by chance in the background only scenario. For each source, the FAR is estimated by applying the cluster search algorithm to one year of time-scrambled data at the same source declination. For the all-sky scan, the FAR is estimated from one year of time-scrambled data at sampled declinations.

The significance threshold for alert issuing was pre-set to 3\,$\sigma$ for individual sources and 4.2\,$\sigma$ for the whole sky, in order to obtain a total FAR of ~10 alerts/year per each IACT, from monitoring their respective full source catalogs, and one all-sky alert/year to be shared with all partners. Indeed, a higher threshold was set for all-sky alerts to take into account the larger trials arising from testing multiple sky pixels.

No post-trials significance is reported for alerts, as the real-time analyses are continuously updated with the arrival of new events. Trial corrections for neutrino alerts are only applied retrospectively in \textit{a posteriori} checks to the best pre-trials $p$-value recorded. This calculation of the trial factors accounts for the total exposure time at the time of evaluation since the program was started and, in the case of predefined source hypotheses, the total number of monitored sources.

The complete information on the GFU-cluster alerts used in this paper is provided in Table~\ref{tab_gfu_alerts}.

\subsection{Single High-Energy Neutrino alert streams}\label{subsec:singlet}

Since 2016, IceCube has been promptly broadcasting the positions of single potential astrophysical neutrino events to the wider astronomical community after detecting them 
at the South Pole, encouraging timely follow-up observations~\citep{2017APh....92...30A}. Two alert streams were first defined to select astrophysical neutrino candidates based on their energy and the geometry of the Cherenkov light deposition in the IceCube detector. 

The High-Energy Starting Events (HESE) stream uses a vetoing technique to reject atmospheric background and select only events with an interaction vertex inside the detector. This guarantees a very good signal purity at the expense of a low event rate, due to a reduced sensitive volume, and a moderate angular resolution (1.89$^{\circ}$ for 90\% containment), due to limited track lengths. 

The Extreme High-Energy (EHE) event selection features a two-dimensional cut on the quality of the directional reconstruction and the total amount of measured Cherenkov light (a proxy for the energy of the incoming muon). A better angular resolution can be achieved (0.81$^{\circ}$ for 90\% containment), thanks to the long selected muon tracks.

In May 2019, the two separate streams were replaced by a unified algorithm selecting candidate astrophysical neutrinos from three channels: the existing EHE selection, an improved version of the HESE selection, and the GFU-GOLD/BRONZE selection (i.e., high-energy muon track events selected from the GFU event sample based on their reconstructed energy and signal purity) constituting the majority (86\%) of the issued alerts. More details about the above alert  streams can be found in~\cite{Blaufuss:20199c}.

For each event that passes these selections, a quantity called ``signalness'' is calculated to quantify the probability that it is of astrophysical origin. Two alert streams, dubbed ``GOLD'' and ``BRONZE'', are defined with different event rates and an average astrophysical signalness value of 50\% and 30\%, respectively.

The sky coordinates, angular uncertainty, energy, signalness and FAR values of selected events are publicly distributed through the NASA General Coordinates Network (GCN) as a ``GCN Notice''.\footnote{An example GCN Notice for the Gold alert IceCube-190730A is available in \href{https://gcn.gsfc.nasa.gov/notices_amon_g_b/132910_57145925.amon}{\nolinkurl{https://gcn.gsfc.nasa.gov/notices\_amon\_g\_b/132910\_57145925.amon.}}} The median delay between the event detection at the South Pole and the successful dissemination via a GCN Notice was 41.8 s for GOLD and BRONZE alerts during the period under consideration in this work. 

More advanced reconstruction algorithms are applied to the event once the data have been transferred to computing clusters in the Northern Hemisphere, and updated position and angular uncertainty estimates are circulated in a revised GCN Notice accompanied by a GCN Circular\footnote{An example GCN Circular for the Gold alert IceCube-190730A is available in \href{https://gcn.nasa.gov/circulars/25225}{https://gcn.nasa.gov/circulars/25225.}}. This updated position can be used by follow-up instruments to revise their observing strategy with respect to the first notice or identify alternative potential counterparts. The median delay between the initial event detection and the dissemination of the second notice was  3.35 hours over the period covered by this work.  

For more details on the alert streams and a recent, refined reconstruction of the single high-energy neutrino events triggering alerts, see \emph{IceCat-1}: The IceCube Event Catalog of Alert Tracks~\citep{2023ApJS..269...25A}. 

For completeness, we mention that since 2020, IceCube has been issuing alerts also for cascade-like events,\footnote{\url{https://gcn.gsfc.nasa.gov/amon_icecube_cascade_events.html}} showing a different topology compared to the track-like GFU event sample~\citep{abbasi2021convolutional}. However, they are not part of the neutrino target-of-opportunity programs reported in this work.

\section{IACT neutrino follow-up programs} \label{sec:iacts}
The MAGIC, FACT, H.E.S.S. and VERITAS IACT instruments operate ToO programs designed to search for the VHE $\gamma$-rays that are expected to be emitted in association with astrophysical neutrinos. We present here a brief overview of the main characteristics of these observatories and describe the evolution of their neutrino ToO programs.

\subsection{MAGIC}
MAGIC is a system of two IACTs located at the Observatorio del Roque de Los Muchachos (\ang{28;45.70;} N, \ang{17;53.42;} W; \SI{2200}{\meter} above sea level) in La Palma, Canary Islands, Spain.
The two telescopes each have a reflector with a diameter of \SI{17}{\meter} and are situated \SI{85}{\meter} apart, comprising a system capable of achieving an energy threshold as low as \SI{50}{\GeV} ($\sim$\SI{20}{\GeV} using the Sum-Trigger-II analog trigger \citep{magic-sumt}). 
MAGIC can detect $\gamma$-rays up to $\sim$\SI{50}{\TeV} with an integral sensitivity of $\sim$0.7\% of the Crab Nebula flux above \SI{220}{\GeV} in \SI{50}{\hour} of observation \citep{aleksic_2016}.

Together with other IACTs, MAGIC has
conducted follow-up observations of neutrino triggers from IceCube since 2012. As soon as IceCube started delivering single high-energy neutrino alerts via the GCN in 2016, they were added to the MAGIC transient alert system and automatic reaction was implemented in December 2017. The MAGIC neutrino ToO follow-up program was upgraded after the IceCube transition from the HESE/EHE to the GOLD/BRONZE channels. Currently, MAGIC automatically repoints to any 
GOLD alert position that is visible during the night, with a zenith angle smaller than 60$^{\circ}$ and with an angular distance to the moon larger than 30$^\circ$.
BRONZE and GFU-cluster alerts are scheduled manually. 
The final decision to perform follow-up observations typically relies on a combination of several factors like the intrinsic parameters of the neutrino alert (e.g., signalness, FAR, duration of the flare, etc.), the available visibility window, weather conditions, or the presence of a candidate electromagnetic counterpart. 
MAGIC allocates approximately 40\,hours of dark time and 20\,hours of moon time per year for the follow-up of neutrino events and their potential counterparts. 

The majority of observations presented here were performed using the standard stereo trigger \citep{Paoletti2007, Aleksic2016}, except for the observations of the AGNs OP\,313 and PKS\,1502+106. Those two distant sources ($z \gtrsim 1.0$) were observed using the Sum-Trigger-II analog trigger which is optimized for a low-energy threshold and delivers improved sensitivity in the $<200\,$GeV energy range \citep{magic_sumt}. All observations were performed in wobble mode with an offset of \ang{0.4} from the source, allowing a simultaneous background measurement in the telescope's FoV \citep{FOMIN1994137, IACTs_bkgmodel}. MAGIC data were analysed using the MARS (\textit{MAGIC Analysis and Reconstruction Software}) proprietary package \citep{mars_2013}. When necessary, the flux values were corrected for atmospheric extinction due to clouds and aerosols, using the LIDAR system at the MAGIC site \citep{Schmuckermaier2022, Schmuckermaier2023May}. For the observations performed during moonlight, the analysis method described in \cite{magicperf_moon:2017} was applied. 

\subsection{FACT}
The first Geiger-mode avalanche photodiode (G-APD) Cherenkov telescope (FACT) is an imaging atmospheric Cherenkov telescope located at the Observatorio del Roque de los Muchachos next to the two MAGIC telescopes.
It has a mirror surface of 9.51\,m$^2$ and a camera with a FoV of \ang{4.5} \citep{2013JInst...8P6008A}. 
While the main focus of the project is the monitoring of bright TeV blazars, a follow-up program for multi-messenger and multi-wavelength alerts has been set up. Thanks to the use of silicon-based photosensors \citep{2014JInst...9P0012B}, the duty cycle of the instrument is maximized. The ability to operate the telescope even in bright moonlight \citep{2013ICRC...33.1132K} not only minimizes gaps in the long-term light curves of monitored sources, but also allows for follow-up observations during light conditions in which other IACTs are not available. From the beginning, the project was targeted towards robotic operation, and automating the operations further increased the duty cycle of the instrument \citep{2019ICRC...36..665D}. 

Up to May 2019, FACT followed up multi-wavelength and multi-messenger alerts on a best-effort basis, adding the observations manually to the schedule. 
Then, an automatic follow-up mode was introduced for alerts received via GCN. Additional alerts received by email and extensions of interesting follow-ups are handled manually. 
For automatic follow-ups, the following strategy is applied: alerts for $\gamma$-ray bursts (from e.g.,\ {\it INTEGRAL}, {\em Swift}/BAT, {\it Agile}, {\em Fermi}-GBM, HAWC), neutrino alerts (e.g.,\ IceCube single high-energy events and multi-messenger coincidences) and transient {\em Fermi}-LAT sources are followed-up for one hour. 
Automatic follow-up observations are carried out if the following conditions are satisfied: (1) the observation occurs during astronomical night (Sun elevation $<$\ang{-18}); (2) the zenith distance of the source’s position is less than \ang{45}; (3) the angular distance between the source and the Moon is between \ang{10} and \ang{170}; (4) the currents in the photosensors are predicted from moonlight to be less than \SI{110}{\micro\ampere} \citep[see][]{2013ICRC...33.3024B}; and (5) the predicted energy threshold is not more than 10 times higher than for the best observing conditions~\citep[see][]{2013ICRC...33.3024B}.
In case these conditions are not met when the alert arrives, but at any time within the ongoing or following observing night (for alerts arriving during day), an automatic call to the alert expert is issued, who decides on how to proceed and schedules a follow-up manually if needed. 

So far, FACT has followed up 62 multi-wavelength and multi-messenger alerts, of which 24 were observed in automatic mode since May 2019.
Of the latter, four were following up neutrino alerts, three of which occurred between May and November 2019.
Before May 2019, six neutrino alerts were followed up, of which two were after September 2017. For this paper, we therefore study five neutrino alerts followed up by FACT. 

All observations were performed in wobble mode with an offset of 0.6$^\circ$ from the source location. The data were analyzed using the open-source Modular Analysis and Reconstruction Software \citep[MARS,][]{2010apsp.conf..681B}. The details of the analysis are described in \cite{2015arXiv150202582D} and the background suppression cuts can be found in \cite{2019ICRC...36..630B}. 
Integral-flux upper limits are determined using the ``light curve cuts'', yielding an energy threshold of 810\,GeV, assuming a power-law spectral index of $\Gamma = 2.5$. For the differential-flux upper limits, the ``spectrum cuts'' were used with an energy threshold of 490\,GeV, using the same assumption for the spectrum as for the integral-flux upper limits.
The dependence of the $\gamma$-ray rate on the zenith angle  and the trigger threshold (which changes according to the level of ambient light) was determined and corrected for using the $\gamma$-ray rate measured from the Crab Nebula, a standard source at TeV energies. 
More details on the correction and spectral analysis can be found in \cite{2021A&A...647A..88A} and \cite{2022icrc.confE.760S} respectively.

To reject data taken with bad atmospheric conditions, the observed cosmic-ray rate ($R750_{\mathrm{cor}}$) is extracted by applying an artificial trigger of 750 DAQ counts to the data and correcting it for the effect of zenith angle~\citep{2017ICRC...35..612M, 2019APh...111...72B}. Seasonal variations of the cosmic-ray rate due to changes in the Earth's atmosphere are considered by determining a monthly reference value $R750_{\mathrm{ref}}$. Good-quality data were selected using a cut of $0.93 < R750_{\mathrm{cor}}/R750_{\mathrm{ref}} < 1.3$.

\subsection{H.E.S.S.}
The High Energy Stereoscopic System (H.E.S.S.) 
consists of four 12-m IACTs that have been operating since 2004, and one 28-m telescope that was added in 2012. The telescopes are located in the Khomas Highlands, about 100\,km south-west of Windhoek, Namibia (\ang{23;16.18;} S, \ang{16;30.00;} E; 1800\,m above sea level).
Currently, H.E.S.S. is the only IACT-array in the
Southern Hemisphere.
The original four telescopes are placed in the corners of a square of side length 120\,m, and the fifth telescope is placed at the center of this square. 
The initial four-telescope configuration is sensitive to $\gamma$-rays with energies between $\sim$100\,GeV and $\sim$100\,TeV and has a FoV equivalent to \ang{5.0} in the sky \citep{2006A&A...457..899A}.
The fifth telescope has a mirror area of $614\,\mathrm{m^2}$, making it the largest optical instrument in the world. Its mirror is segmented into 875 hexagonal facets. The addition of this telescope enables the energy threshold of H.E.S.S.\ to be lowered to a few tens of GeV. 

Since 2012, H.E.S.S.\ has conducted a neutrino ToO program, searching for spatial and temporal correlation between neutrinos and VHE $\gamma$-ray emission. 
While observations were initially triggered by real-time alerts from the ANTARES neutrino telescope~\citep{ANTARES_TAToO, ANTARES_followup_res}, observations of archival and real-time neutrino detections by IceCube have been an integral part of the H.E.S.S.\ multi-messenger program since 2015.

The H.E.S.S.\ Collaboration joined the GFU program in December 2018 \citep{schussler2019hess} when IceCube extended the search for neutrino clusters to an \emph{a priori} defined list of known $\gamma$-ray sources in the Southern Hemisphere. H.E.S.S.\ observation time is awarded by an Observations Committee in a competitive process every year. 
Over the past few years, the typical amount of time allocated for neutrino-related observations was about 20 hours per year. 
This time is mostly devoted to deep observations of particularly interesting events that enable not only the detection but also the characterization of possible VHE $\gamma$-ray emission. ToO observations are usually triggered fully automatically via the H.E.S.S.\ Transient alert system~\citep{2022A&A...666A.119H}, which is connected to a large variety of brokers and alert channels, including a dedicated link to the IceCube computing center at the University of Wisconsin–Madison.

Focussing primarily on high-energy sensitivity, the H.E.S.S.\ results presented in this work make use of data collected by the four 12-m telescopes, except for the GFU-cluster alert associated with the source 1ES\,1312-423 where data from all five telescopes are included. All observations are performed in wobble mode, where the telescopes are pointed with an offset of \ang{0.5} from the source position. 
The results are derived using the semi-analytical {\it Model
Analysis}~\citep{de_Naurois_2009} with loose cuts and cross-checked by the independent calibration and analysis procedure described in \cite{PARSONS201426} for all the neutrino ROIs.

\subsection{VERITAS}

VERITAS \citep{Holder2006} consists of an array of four 12-m IACTs located at the Fred Lawrence Whipple Observatory (FLWO) in Southern Arizona, USA (\ang{31;40.50;} N,
\ang{110;57.11;} W, 1300\,m above sea level). Each telescope is equipped with a camera containing 499 photomultiplier tubes covering a FoV of 3.5$^{\circ}$. 
It reaches its best sensitivity in the 100\,GeV to 30\,TeV energy range.
The angular resolution of VERITAS is $\sim0.1^{\circ}$ at 1\,TeV (for 68\% containment), and the energy resolution is 15-25\% at the same energy. In its current configuration, VERITAS can detect a source with a VHE flux of 1\% of the Crab Nebula above an energy threshold of 240 GeV in less than 25\,hours at a statistical significance of $5\,\sigma$  \citep{2015ICRC...34..771P}.

First efforts within VERITAS to study potential correlations between VHE $\gamma$ rays and IceCube neutrinos date back to 2007 and involved a neutrino search associated with flares of the known VHE blazar Markarian\,421 ~\citep{Bayer:2007pm}. 
Subsequent efforts concentrated on follow-ups of GFU-cluster alerts~\citep{2016JInst..1111009I} which, following the IceCube announcement of the discovery of an astrophysical neutrino flux in 2013, were complemented by a broad program of deep exposures on archival IceCube neutrino positions of likely astrophysical origin~\citep{Santander:20167z, Santander:2016chw, Santander:2017zkl}. 
This program later transitioned to the real-time efforts described in this paper where, as part of its long-term scientific program, VERITAS allocates approximately 45\,hours of dark time observations per year to follow up neutrino alerts from IceCube or to perform deep exposures of potential VHE $\gamma$-ray counterparts of neutrino events. 

Through its ToO program, VERITAS initiates automatic follow-up observations after receiving an alert from IceCube through GCN and will accumulate a three-hour initial exposure as long as the target's  zenith angle is smaller than 50$^{\circ}$. Assuming that VERITAS is operating at the time of the alert, the total delay between the detection of the neutrino at the South Pole and the start of pointed observations by VERITAS is typically about a few minutes. For alerts that occur during local daytime, longer delays of up to 24\,hours until the start of observations may be considered if the neutrino event has high astrophysical probability ($>90\%$) and is well localized ($<1^{\circ}$), or if a promising $\gamma$-ray counterpart (such as a source in the \emph{Fermi}-LAT 3FHL or 4FGL catalogs) is identified within the neutrino uncertainty region. 
In the latter case, a long exposure (up to tens of hours) is initiated, targeting a sensitivity level of a few percent of the Crab Nebula flux.

The analysis of VERITAS data presented here was performed using the standard VERITAS analysis tools \citep[Eventdisplay;][]{2017arXiv170804048M} with background-rejection cuts optimized for soft-spectrum sources ($\Gamma \geq 3.5$). The background was estimated through the standard “reflected regions” technique \citep{Fomin1994}. A cross-check analysis was performed with an independent software package \citep[VEGAS;][]{2008ICRC....3.1385C}. 

\begin{figure}[ht]
\begin{center}
\includegraphics[width= 0.5\textwidth]{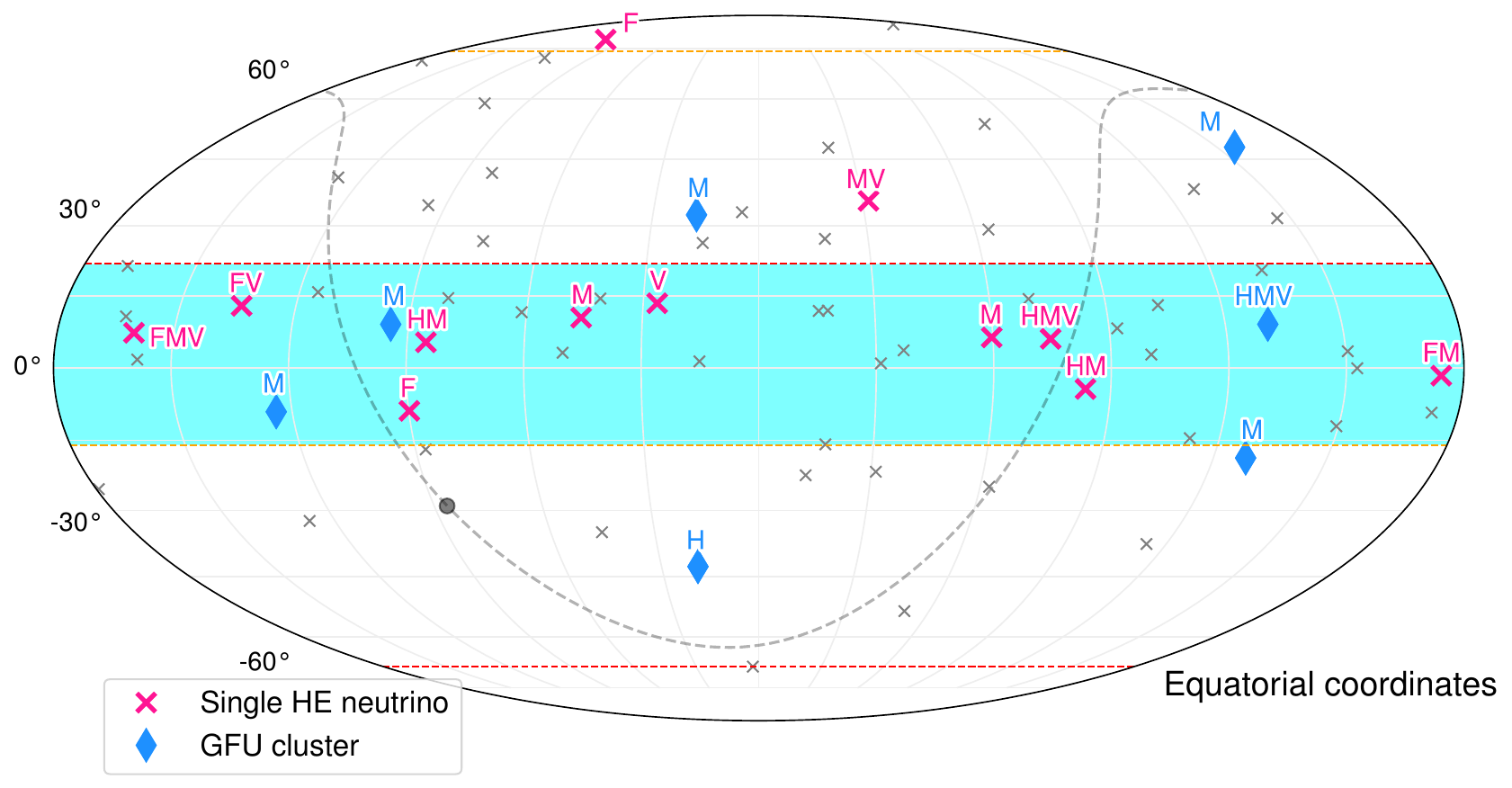}
\caption{Skymap in equatorial coordinates showing IceCube alert positions in the period from September 2017 to January 2021. Alerts followed up by IACTs are shown in color (according
to the alert type), and those not followed up are shown in gray.
Letters indicate which IACTs participated in the observations (F - FACT, H - H.E.S.S., M - MAGIC, V - VERITAS). 
The latitude band between two dashed orange lines and two dashed red lines indicate regions of the sky that are potentially observable at zenith angles less than 45$^{\circ}$ from the northern (FACT, MAGIC, VERITAS) and southern (H.E.S.S.) IACTs, respectively. The light cyan band represents the overlapping visibility window for instruments in both hemispheres around the celestial equator, where the IceCube sensitivity to neutrinos in the $\sim 100$ TeV energy range is at its best.}
\label{fig:alert_skymap}
\end{center}
\end{figure}

\subsection{Overview of VHE $\gamma$-ray observations}
\label{vhe-obs}

\begin{figure}[th]
\begin{center}
\includegraphics[width= 0.5\textwidth]{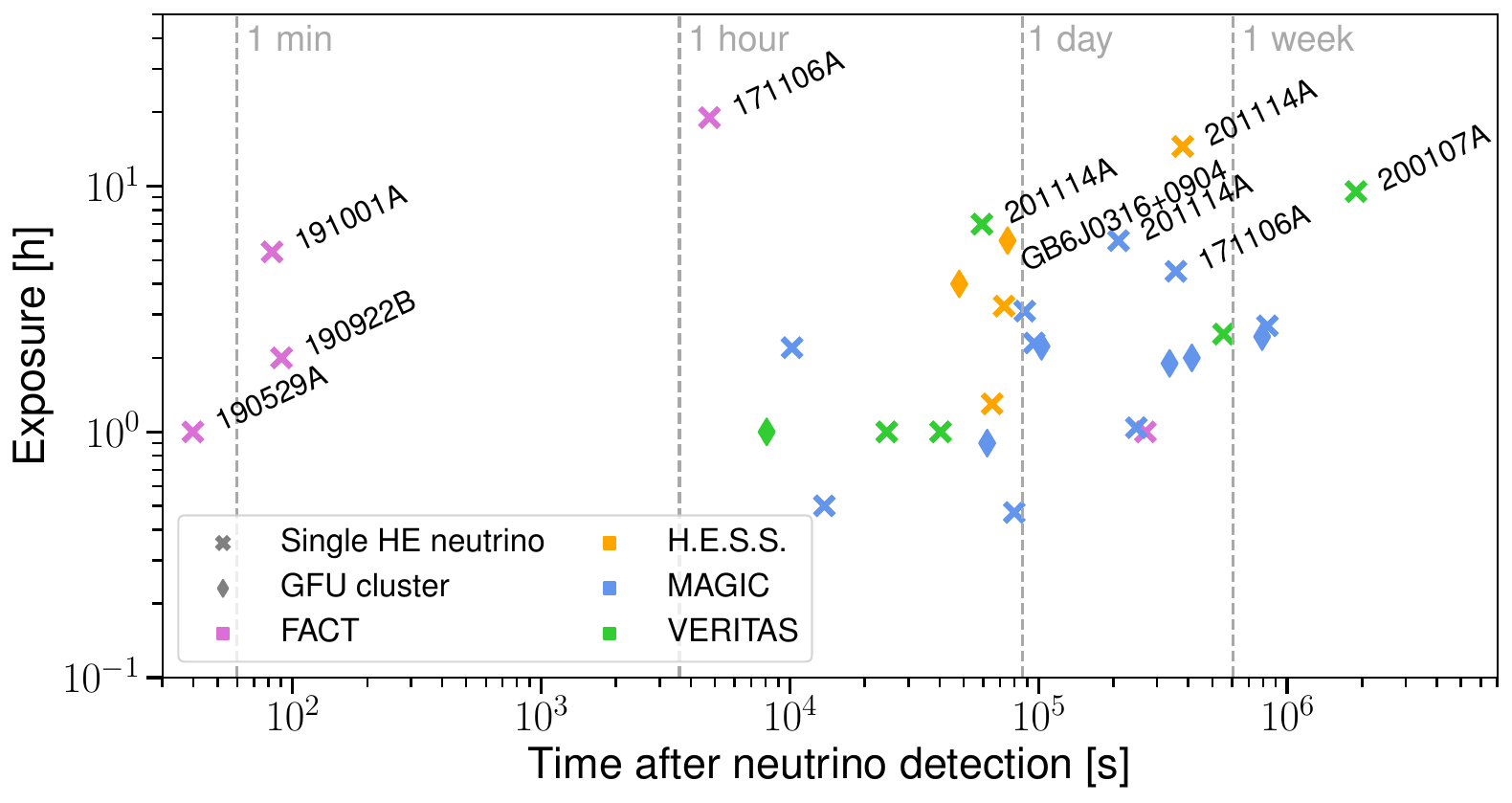}
\caption{
Delay times plotted against exposure times for IACT follow-up observations of neutrino alerts in the period from September 2017 to
January 2021.
The delay is calculated from the neutrino event arrival time (single events) or the flare threshold-crossing time (clusters) up to the start of the IACT observation. 
Observations performed with delays of less than 100\,s or total exposures longer than 4 hours are labeled by alert names.
The marker color represents the IACT performing the observation while the marker type represents the alert type.}
\label{fig:delay}
\end{center}
\end{figure}

Fig.~\ref{fig:alert_skymap} provides a skymap of the direction of the alerts
sent by IceCube as single high-energy neutrino events and GFU-clusters.
The follow-up observations of different IACTs are indicated using letters (see figure caption). 
Table~\ref{tab_GFU} and Table~\ref{tab_tracks} explicitly list the observed alerts and 
provide information on MWL data collected.
The IACT delay and exposure times for all single events and GFU-clusters discussed here are presented in Fig.~\ref{fig:delay}. 
The delay is calculated from the neutrino event arrival time (single events) or the time at which the significance threshold is exceeded (clusters) up to the start of the IACT observation. 

Already from this broad overview, we can deduce some general trends in the follow-up strategies. 
Reaction times of less than a day are achieved in 50\% of the cases, and observations that began more than a week after the trigger are rare. The total time spent on following up public IceCube alerts is very similar for all collaborations ($\sim$20\,hours).

\begin{table*}[ht]
\small
\caption{List of GFU-cluster alerts followed up by IACTs in the years 2019-2020. The all-sky alert is shown separately. IACT-named columns give the total exposure time (in hours) for each instrument. 
Information on the available MWL data is also provided.
Complete information on the GFU-cluster alerts is shown in Table~\ref{tab_gfu_alerts}. 
}
\label{tab_GFU}
\centering
\begin{tabular}{l|cccccc}
\hline
Source & Duration [days]& Pre-trials significance  & H.E.S.S & MAGIC & VERITAS & MWL\\
 \hline

1ES\,1312-423 & 0.26 & 3.4\,$\sigma$ & 2.6\,h & --- & --- & {\em Swift} \\

MG1\,J181841+0903 
& Multiple alerts & $>$ 3.3\,$\sigma$ & --- & 1.6\,h & --- & {\em Swift} \\

PMN\,J2016-0903  & 0.01 & 3.6\,$\sigma$  & --- & 0.9\,h & --- &  \\
OP\,313  
& Multiple alerts & $>$ 3.0\,$\sigma$ & --- & 3.2\,h & --- & {\em Swift} \\
OC\,457  & 0.30 & 3.3\,$\sigma$ & --- & 2.5\,h & --- & {\em Swift}\\
GB6\,J0316+0904 & 2.25 & 3.1\,$\sigma$ & 6.0\,h  & 1.9\,h  & 1.0\,h & \\
\hline
All-sky alert (PMN\,J0325-1843)   & 3.67 & 5.1\,$\sigma$ & ---  & 2.0\,h & --- & \\
 \end{tabular}
\end{table*}

\begin{table*}[ht]
\small
\caption{List of IceCube single high-energy neutrino alerts followed up by at least one IACT. A link to the corresponding GCN Circular with updated coordinates (including information regarding the initial localization) released by IceCube is provided in the alert name. The alert nature is provided: E = EHE, H = HESE, G = GOLD. Energy and signalness estimates are not available for 200107A and 190529A.
IACT-named columns give the total exposure time (in hours) for each instrument. The name of the potential counterpart and the available MWL data discussed in the text are also given. 
Complete information on the single high-energy neutrino alerts is shown in Table~\ref{tab_tracks_extended}. 
}
\label{tab_tracks}
\centering
\begin{tabular}{lc|cccccccc}
\hline
Name & & Energy  & Signalness & FACT & H.E.S.S.\ & MAGIC & VERITAS & Pot. Count. & MWL\\
\multicolumn{2}{l|}{(GCN Circular)} &[TeV] &  &  & &  &  &  & \\
 \hline
\href{https://gcn.nasa.gov/circulars/22105}{IC-171106A} &E & 230  & 0.75 & 4.0\,h  & --- & 4.5\,h  & 2.5\,h  & 87GB 223537.9+070825 & \\
\href{https://gcn.nasa.gov/circulars/23375}{IC-181023A}&E & 120 & 0.28 & 1.0\,h & --- & --- & --- & & \\
\href{https://gcn.nasa.gov/circulars/24378}{IC-190503A}&E & 100 & 0.36 & --- & --- & 0.5\,h & --- & & \\
\href{https://gcn.nasa.gov/circulars/24674}{IC-190529A$^\dagger$}&H & --- & --- & 1.0\,h & --- & --- & --- & & \\
\href{https://gcn.nasa.gov/circulars/25225}{IC-190730A}&G & 299 & 0.67 & ---  & --- & 3.1\,h & --- & PKS 1502+106 & {\em Swift}\\ 
\href{https://gcn.nasa.gov/circulars/25806}{IC-190922B}&G & 187 & 0.50 & 2.0\,h & --- & 2.2\,h & --- & AT2019pqh & \\
\href{https://gcn.nasa.gov/circulars/25913}{IC-191001A}&G & 217 & 0.59 & 5.4\,h& ---  & --- & 1.0\,h & AT2019dsg &  \\ 
\href{https://gcn.nasa.gov/circulars/26655}{IC-200107A$^\ddag$}&--- & --- & --- & ---  & --- & 2.7\,h & 9.5\,h & 4FGL~J0955.1+3551 & {\em Swift} \\ 
\href{https://gcn.nasa.gov/circulars/28504}{IC-200926A}&G & 670 & 0.44 & --- & 1.3\,h & 1.0\,h& & \\
\href{https://gcn.nasa.gov/circulars/28575}{IC-201007A}&G & 683 & 0.88 & --- & 3.0\,h & 0.5\,h & ---  &  & \\
\href{https://gcn.nasa.gov/circulars/28887}{IC-201114A}&G & 214 & 0.56 & ---  & 14.3\,h & 6.0\,h & 7.0\,h & 4FGL~J0658.6+0636 & {\em Swift}\\
\href{https://gcn.nasa.gov/circulars/29120}{IC-201222A}&G & 186 & 0.53 & ---  & --- & --- & 1.0\,h & & \\
\hline
\multicolumn{10}{l}{%
  \begin{minipage}{15cm}~\\
$\dag$ Retracted \\
$\ddag$ The high energy starting track was not identified as either GOLD or BRONZE
  \end{minipage}%
}\\
\end{tabular}
\end{table*}

In detail, the IACTs followed 12 out of the 62 single high-energy event alerts sent by IceCube in the period between September 2017 and January 2021.\footnote{Full list available at: \url{https://gcn.gsfc.nasa.gov/amon_icecube_gold_bronze_events.html} and following links. 
IceCube-200107A was announced several hours after its detection, through a GCN Circular (see GCN \#26655).}

In the framework of the GFU program, for each IACT experiment a separate list of possible follow-up sources was prepared.
The numbers of sources and the corresponding total FAR (computed according to~\citep{BoscoloMeneguolo:20233e}) was 139 sources, 6.2 alerts/year for H.E.S.S.; 179 sources, 9.9 alerts/year for MAGIC and 190 sources, 11.4 alerts/year for VERITAS. No private alert was sent to FACT.

From March 2019 till January 2021 (1.9 years), IceCube issued 27 GFU-cluster alerts (17 for H.E.S.S., 12 for MAGIC and 8 for VERITAS) from 17 sources, and a single all-sky alert.
Some of the alerts have been issued simultaneously to multiple IACTs due to partial overlaps in the source catalogues, and in some cases when the source repeatedly went below and above the threshold, multiple correlated alerts were sent.
During the first months of the online system’s operation, it experienced inconsistent performance due to the development and testing phase, which caused the effective number of issued alerts to be lower than the expected FAR.
Of the cluster alerts issued, 7 were followed by at least one IACT (2 H.E.S.S., 6 MAGIC, 1 VERITAS).

Seven of the single-event alerts and one GFU-cluster were followed by more than one IACT. We used the combined exposure to calculate joint upper limits, which are more constraining than single-instrument limits. For example, 4FGL~J0955.1+3551, a possible counterpart to IceCube-200107, was observed by MAGIC and VERITAS, while the GFU-cluster alert on GB6~J0316+0904 was followed by all three large instruments. In the case of IceCube-201114A, a dedicated multi-wavelength follow-up campaign (including H.E.S.S., MAGIC and VERITAS) was organized for its potential counterpart 4FGL~J0658.6+0636. The observation campaign and its results are discussed in more detail in a separate publication~\citep{201114A}.

\subsection{Calculation of individual and combined upper limits} 
\label{sec:methods}

For the upper-limits (ULs) calculation, we used the Rolke method \citep{Rolke2005Oct, Lundberg_2010} with a confidence level set to 95\% and including a 30\% global systematic uncertainty in the efficiency of the applied cuts. The upper limits were calculated considering an observed spectrum modeled by a power-law function, d$N/$d$E = K E^{-\Gamma}$, $K$ being the normalization constant of the flux, and $\Gamma$ the index. Following the slope of the IceCube spectrum for the astrophysical neutrino flux, $\Gamma$ was set to 2.5~\citep{PhysRevD.110.022001}–as the gamma-ray and neutrino emissions are expected to show the same spectral shape~\citep{Ahlers2018Sep}.

If the cluster alerts arrived from known sources or potential counterparts to the high-energy single-event alerts were found, both differential- and integral-flux upper limits above a given energy threshold (which varies for each alert and observatory) were calculated at the source or counterpart position. 

Furthermore, the single-event public IceCube alerts are characterized by large localization uncertainties ($\sim$ \ang{0.5} to a few degrees at 50\% C.L.).  
As a consequence, the exact localization of the neutrino origin cannot be pinpointed to any point source with a high
degree of accuracy. Therefore, we provide skymaps containing the integral $\gamma$-ray flux upper limits and covering the neutrino arrival localization error. The upper limits for those maps are calculated following the description above.

Whenever a cluster alert or a potential counterpart is observed by multiple IACTs, we use a method that combines individual likelihoods to produce combined flux upper limits, as described in \cite{MAGIC:2022jop}. 
All IACTs adopted the same energy binning with four bins per decade in the energy range of 71\,GeV to 71\,TeV. 
Therefore, the resulting spectra cover a subset of this range, depending on the energy threshold and maximum energy of the single instruments during each observation. 
The latter quantity, i.e. the maximum energy, is defined in a different way in the single experiments: H.E.S.S. and VERITAS  use Gaussian statistic for the estimation of the background, while FACT and MAGIC use Poissonian approximation. 
This results in spectra whose maximum energy is computed only for $N_\mathrm{off}>10$ in the first case, while in the second case also $N_\mathrm{off}<10$ is allowed.
The combined-upper-limit calculation uses a profile maximum likelihood method to estimate the 95\% containment for each experiment separately. For each energy bin the likelihood function is calculated as
\begin{equation}
    \label{eq:L_estimate}
    \begin{aligned}
    L = \frac{(\epsilon \mu + b)^{N_{\rm on}}e^{-(\epsilon \mu + b)} }{N_{\rm on}!}
    \cdot 
    \frac{ (\tau b)^{N_{\rm off}} e^{-\tau b}}{N_{\rm off}!}
    \cdot \\
    \frac{1}{\sigma_{\epsilon} \sqrt{2\pi}} \exp{\left[-\frac{1}{2} \left(\frac{\epsilon - \epsilon_0}{\sigma_{\epsilon}}\right)^2\right]},
    \end{aligned}
\end{equation}\noindent where $N_{\rm on}$ and $N_{\rm off}$ are corresponding IACT-measured events in the signal and background regions, i.e., the $\rm ON$ and $\rm OFF$ regions, $\mu$ and $b$ are the expected gamma-ray signal and background events in the ON region, $\tau$ is the ratio between OFF and ON exposure, $\epsilon$ is the expected detector efficiency, $\epsilon_0$ is the common efficiency of the detectors, which is set to 1. The efficiency systematic uncertainty, denoted by $\sigma_\epsilon$, is conservatively taken as 0.3 in this study for all the detectors. Thus, the derived upper limits based on this estimate are also conservative. The first term in Equation \ref{eq:L_estimate} describes the Poissonian signal, the second term describes the Poissonian background, and the third term describes the Gaussian detection efficiency.

The likelihood can be converted into a likelihood ratio test statistic $\lambda$ after determining the maximum likelihood estimators $\hat{b}$ and  ${\hat{\epsilon}}$: 

\begin{equation}\label{profile likelihood}
\lambda_{\rm i}\left(\mu \mid N_{\rm {on}},N_{\rm{off }}\right)=\frac{L\left(\mu, \hat{\varepsilon}, \hat{b} \mid N_{\rm{on}}, N_{\rm {off}}\right)}{L\left(\hat{\mu}, \hat{\varepsilon}, \hat{b} \mid N_{\rm{on}}, N_{\rm {off}}\right)},
\end{equation}

\noindent where i denotes different experiments. As the observations from different experiments are independent, the test statistics of individual experiments $-2\ln \lambda_{\rm i}$ can be added to extract a combined-upper-limit value:
\begin{equation}
    -2\ln \lambda_{\rm comb} = \displaystyle\sum_{i = 1}^{N} -2\ln \lambda_{\rm i},
\end{equation}

\noindent where, $N$ is the number of experiments performing follow-up observations of the same event. The expected number of ON-region events, $N_{\rm{on}}$, is related to the differential flux $\Phi(E)$ through the instrument response functions (i.e., effective area, observation time, and energy dispersion) of each IACT, assuming a simple power-law spectral model with flux normalization $K$.

We note that FACT, H.E.S.S. and VERITAS use an upper limit calculation that gives a particularly constraining limit when the number of events in the signal region is below the estimated background. The effect of these fluctuations is reduced by the combination of data sets from different instruments. The resulting, combined upper limits on the $\gamma$-ray flux can therefore be higher, i.e., less constraining, than the ones derived by the individual instruments.

\section{Complementary observations with satellite facilities} \label{sec:observations}
In this section we describe complementary observations carried out by space-based observatories, i.e., the $Fermi$-LAT and the Neil-Gehrels $Swift$ Observatory.


\subsection{{\it Fermi} Large Area Telescope observations}

The \emph{Fermi} Large Area Telescope (LAT) is a pair-conversion telescope sensitive to $\gamma$-rays with energies from $20$\,MeV to greater than $300$\,GeV~\citep{2009ApJ...697.1071A}. It has a large FoV ($>2$\,sr) and scans the entire sky every three hours during standard operation, making it well suited to monitor variable and transient $\gamma$-ray phenomena on different timescales, from seconds to years.

The real-time neutrino monitoring program carried out by the {\em Fermi}-LAT team started in mid-2016, i.e.,\  when IceCube began distributing public alerts to the community~\citep{garrappa:2022}. 
The potential and importance of the program was demonstrated with the detection of spatial and temporal coincidence of IceCube-170922A with the $\gamma$-ray flaring blazar TXS~0506$+$056. The prompt observations by the {\em Fermi}-LAT triggered a rich campaign, including follow-up by several multi-wavelength ground- and space-based facilities, which led to the first  identification of a compelling astrophysical counterpart to a source of IceCube neutrinos.

In this study, the LAT data analysis is performed using the Python package \texttt{fermipy}~\citep{Wood:2018}, adopting the same procedure for each source in the sample. We select photons of the \texttt{Pass 8 SOURCE} class, in a ROI of 15$^\circ$ $\times$ 15$^\circ$ square, centered at the target. To minimize contamination from $\gamma$-rays produced in the Earth’s upper atmosphere, a zenith-angle cut of $\theta < 90^{\circ}$ is applied. The standard data-quality cut $\textnormal{(DATA\_QUAL $>$ 0) \&\& (LAT\_CONFIG == 1)}$ is applied, and time periods coinciding with solar flares and $\gamma$-ray bursts detected by the LAT are removed. The ROI model includes all 4FGL catalog sources~\citep[][]{ballet2020_4FGL-DR2} located within 20$^{\circ}$ from the ROI center, as well as the Galactic and isotropic diffuse backgrounds\footnote{\url{https://fermi.gsfc.nasa.gov/ssc/data/access/lat/BackgroundModels.html}} (\texttt{gll\_iem\_v07.fits} and \texttt{iso\_P8R3\_SOURCE\_V3\_v1.txt}). 

We perform a binned analysis in the energy range 0.1-800\,GeV, using 10 bins per decade in energy and 0.1$^{\circ}$-wide spatial bins and adopting the \texttt{P8R3\_SOURCE\_V3} instrument response functions. For each given target, a maximum likelihood analysis is performed over the time range of interest, 
ensuring overlap between Fermi data and IACT observations.
When a candidate astrophysical counterpart to the neutrinos is available, we select this as the target of interest, and the analysis is centered at its position. For IceCube neutrino events with no candidate astrophysical counterparts, an analysis of the region centered at the best-fit position provided by IceCube was performed.

In the fit, the sources in the ROI are modeled by adopting the spectral shapes and parameters reported in 4FGL. The fit is performed using the ``optimize'' function, implemented in \texttt{fermipy}. The method performs an automatic optimization of the ROI by fitting all sources using an iterative strategy.

Since our data span a different integration time with respect to 4FGL, the results are checked for potential newly-detected sources using the iterative procedure implemented in the \texttt{fermipy} function ``find$\_$sources.'' To this end, a TS map is produced. Following~\citep{1979ApJ...228..939C}, the TS is defined as $2\log(L/L_0)$, where
\textit{$L$} is the likelihood of the model with a point source at a given position and \textit{$L_0$} is the likelihood without the source. A TS value of 25 corresponds to a statistical significance of $\gtrsim\,4.0\,\sigma$~\citep[according to the prescription adopted in][]{4fgl:2020}. 
A TS map is produced by including a putative point source at each pixel of the map and evaluating its significance over the current best-fit model. The test source is modeled with a power-law spectrum where only the normalization is allowed to vary in the fit, whereas the photon index is fixed at 2. We test whether there are significant peaks (TS $>$ 25) in the TS map, with a minimum separation of 0.5$^{\circ}$  from existing sources in the model, and add a new point source to the model at the position of the most significant peak found. Then, the ROI is fitted again, and a new TS map is produced. This process is iterated until no more significant excesses are found.
Flux upper limits are computed if the TS of the target of interest is lower than 4 ($\sim2\sigma$).

\subsection{\textit{Swift}}

The {\em Neil Gehrels {\em Swift} Observatory} satellite \citep{2004ApJ...611.1005G} carried out observations of a few sources involved in the follow-up program described in the present paper (See Section \ref{sec:results_tracks}).
NVSS\,J065844+063711 (counterpart of IceCube-201114A). 
The observations were performed with all three instruments onboard: the X-ray Telescope \citep[XRT;][0.2--10.0\,keV]{2005SSRv..120..165B}, the Ultraviolet/Optical Telescope \citep[UVOT;][170--600\,nm]{2005SSRv..120...95R}, and the Burst Alert Telescope \citep[BAT;][15--150\,keV]{2005SSRv..120..143B}.

All XRT observations were performed in photon-counting mode \citep[for a description of the XRT read-out modes, see][]{hill04}. The XRT spectra were generated with the {\em Swift}-XRT data product generator tool at the UK {\em Swift} Science Data Center\footnote{\url{http://www.swift.ac.uk/user\_objects}}
\citep[Version 1.10 of the product generator module was released as part of swifttools v3.0; for details, see][]{2009MNRAS.397.1177E}. Spectra having count rates higher than 0.5\,counts\,s$^{-1}$ may be affected by pile-up. To correct for this effect, the central region of the image has been excluded, and the source image has been extracted with an annular extraction region with an inner radius that depends on the level of pile-up \citep[see e.g.,][]{2005SPIE.5898..360M}. We used the spectral redistribution matrices in the Calibration database maintained by HEASARC. The X-ray spectral analysis was performed using the \texttt{XSPEC 12.9.1} software package \citep{1996ASPC..101...17A}. Data were grouped by single photons with \texttt{grppha} and the Cash statistic was used \citep{1979ApJ...228..939C}. All XRT spectra are fitted with an absorbed power-law model \texttt{tbabs * pow} and an HI-column density set to the Galactic value in the direction of the source \citep[$N_{H}$;  taken from][]{2016A&A...594A.116H}.

The hard X-ray flux of these sources is usually below the sensitivity of the BAT instrument for daily short exposures. Moreover, none of the sources were bright enough in the hard X-rays to be detected in the Swift/BAT 157-month catalog\footnote{\url{https://swift.gsfc.nasa.gov/results/bs157mon/}}. This reflects in a flux limit in the 14-195 keV energy range of $\sim$ 8.4$\times$10$^{-12}$ erg s$^{-1}$ cm$^{-2}$.

During the {\em Swift} pointings, the UVOT instrument observed the sources in its optical ($v$, $b$, and $u$) and UV ($w1$, $m2$, and $w2$) photometric bands \citep{2008MNRAS.383..627P,2010MNRAS.406.1687B}. The UVOT data in all filters were analysed with the \texttt{uvotimsum}  and \texttt{uvotmaghist} tasks and the 20201215 CALDB-UVOTA release. Source counts were extracted from a circular region of 5-arcsecond  radius centered on the source, while background counts were derived from a circular region with a 20-arcsecond radius in a nearby source-free region. All UVOT exposures were checked for possible small-scale sensitivity problems, which occur when the image of the source falls on small detector regions where the sensitivity is lower.\footnote{\url{https://swift.gsfc.nasa.gov/analysis/uvot\_digest/sss\_check.html}} 
The UVOT magnitudes are corrected for Galactic extinction using the $E(B–V)$ value from \citet{S&F2011} and the extinction laws from \citet{Cardelli1989Oct} and converted to flux densities using the conversion factors from \citet{Breeveld2010Aug}.

\section{Follow-up results on GFU-cluster alerts}
\label{sec:results_flares}
A summary of the follow-up observation campaigns of GFU-cluster alerts is listed in Table~\ref{tab_GFU} while details on the alerts including the cluster time window, duration, significance and FAR are listed in Table~\ref{tab_gfu_alerts}. 

No significant VHE emission is detected from any of the sources under study with the exception of blazar 1ES\,1312-423 (see Section~\ref{sec:1es1312} and Fig.~\ref{fig:1es1312_sed}). 
The flux upper-limits are calculated as reported in Section~\ref{sec:methods}.

Although for some sources a change of the X-ray spectrum or an enhancement of the X-ray activity has been detected during the period of the neutrino detection, X-rays alone are not sufficient to conclusively establish a correlation between neutrino emission and a specific blazar. Assuming that the neutrinos were produced in proton-photon interactions, the observed X-ray emission could come from the population of target photons in these processes. However, to better understand the mechanism(s) producing X-rays, a detailed modeling of the broad-band spectra is needed.

The spectral energy distribution (SED) of each source, which includes MWL data covering the same observation time period and archival data, is shown in Fig.~\ref{fig:sed_multiplets}, while the combined upper-limits at VHE are shown in Fig.~\ref{fig:ul_GB6J0316+0904} for the source GB6~J0316+0904, the only one observed by more than one IACT for the GFU-cluster alerts.

\subsection{1ES\,1312-423}
\label{sec:1es1312}

\begin{figure}[ht]
\begin{center}
\includegraphics[width= 0.48\textwidth]{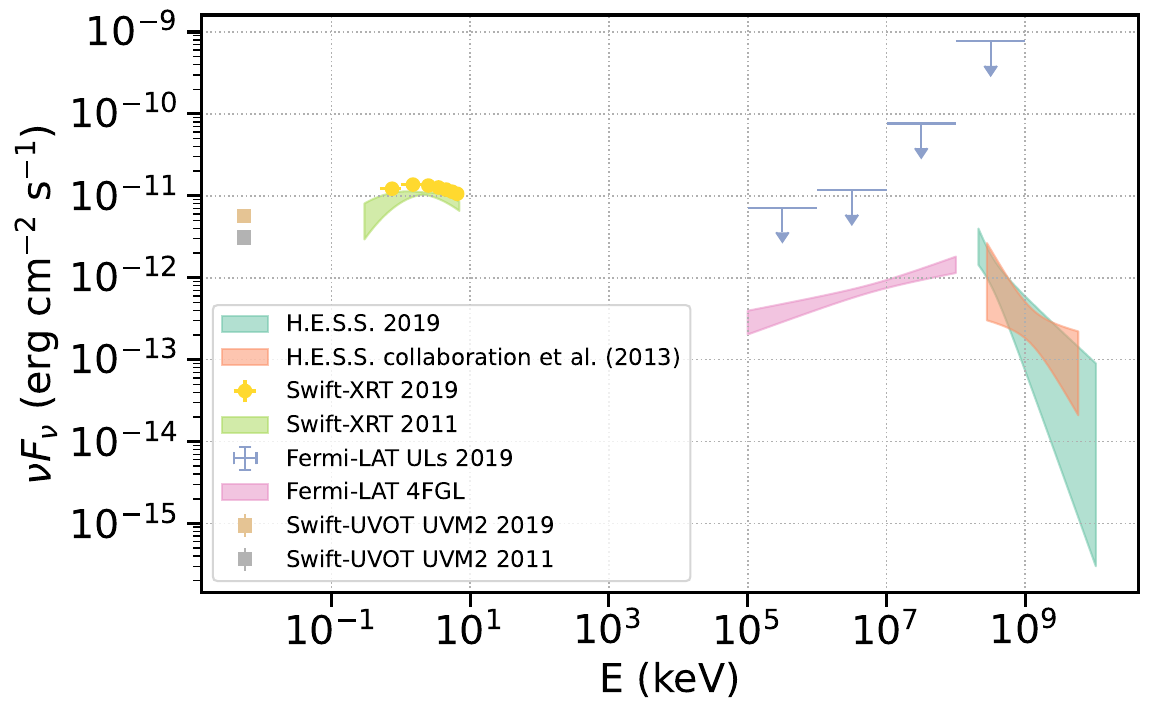} 
\caption{1ES\,1312-423 MWL SED showing both archival data and observations obtained during the period following the GFU neutrino alert (March 12th-13th, 2019). }
\label{fig:1es1312_sed}
\end{center}
\end{figure}

1ES\,1312-423 is a blazar located at RA:~13$^{\circ}$ 15' 03.39'', Dec: -42$^{\circ}$ 36' 49.75'', i.e.,\ about $2^{\circ}$ from the Centaurus\,A (Cen A) radio galaxy, with a redshift of $z = 0.105\pm0.001$~\citep{2011NewA...16..503M}. The source was detected with H.E.S.S.\ in VHE $\gamma$-rays during the period from April 2004 to July 2010 using a total exposure time of $150.6\,\mathrm{h}$ ~\citep{2013MNRAS.434.1889H}. Based on this archival dataset, fitting the differential energy spectrum $\phi(E) = \mathrm{d}N/\mathrm{d}E$ of the VHE $\gamma$-ray emission above $280\,\mathrm{GeV}$ with a power-law function  $\phi(E) = \phi(E_0) \times (E/E_0)^{-\Gamma}$ yields the best-fit parameters $\Gamma = 2.85 \pm 0.47 (\mathrm{stat}) \pm 0.20 (\mathrm{sys})$ and a differential flux at 1\,TeV of $\phi(1~\mathrm{TeV}) = (1.91 \pm 0.59 (\mathrm{stat}) \pm 0.39 (\mathrm{sys})) \times 10^{-13}~\mathrm{cm}^{-2}~\mathrm{s}^{-1}~\mathrm{TeV}^{-1}$.

On March 12th, 2019, IceCube announced the detection of a neutrino cluster from the location. H.E.S.S.\ observed the source for a total of 2.6 hours divided over two nights: March 12th and 13th, 2019. Applying a set of {\it loose} cuts~\citep{2006A&A...457..899A}, VHE emission from 1ES\,1312-423 was observed above $140\,\mathrm{GeV}$ with a significance of $4\sigma$. For these observations, the best-fit parameters are $\Gamma = 3.57 \pm 0.60 (\mathrm{stat}) \pm 0.20 (\mathrm{sys})$ and $\phi(1~\mathrm{TeV}) = (1.72 \pm 1.4 (\mathrm{stat}) \pm 0.4 (\mathrm{sys})) \times 10^{-13}\,\mathrm{cm}^{-2}\,\mathrm{s}^{-1}\,\mathrm{TeV}^{-1}$. A comparison of the SED with the archival dataset is given in Fig.~\ref{fig:1es1312_sed} together with the results obtained from dedicated ToO observations by {\em Swift} (UV + X-rays). The flux levels and energy spectra in the TeV domain are compatible, while some variations in the X-ray and UV bands can be seen in the figure. It is not clear whether this can be linked to a definite change in the state of 1ES\,1312-423.

The source was observed six times by {\em Swift} during the period from January 25, 2011 to April 19, 2019, with five of these observations performed in 2019. The 0.3-10\,keV spectrum can be fitted with an absorbed power-law model with $N_{H}$ fixed to 7.25\,$\times$10$^{20}$\,cm$^{-2}$. No significant flux increase or spectral index change was found in the X-ray (see Table~\ref{1312_XRT}). The {\em Swift}/UVOT observation results can be found in Table~\ref{1312_UVOT}.

\subsection{MG1\,J181841+0903}
\label{sec:mg1J1818}
MG1\,J181841+0903 is a flat-spectrum radio quasar (FSRQ) located at RA:~18$^\circ$ 18' 40.06'', Dec: +09$^\circ$ 03' 46.20'' at unknown redshift. On June 5th, 2019, the MAGIC Collaboration received two GFU-cluster alerts related to this source. 
On June 7th, 2019, the MAGIC telescopes observed MG1\,J181841+0903 within a zenith-angle range of $25^\circ - 46^\circ$ and collected 2.2 hours of good-quality data. 
An upper limit of $2.37\times10^{-11}\,\mathrm{cm^{-2}}\,\mathrm{s^{-1}}$
was calculated for the integral flux above 110\,GeV.

The source was observed 10 times by {\em Swift}-XRT between November 17th, 2013 and June 6th, 2019. As the the first nine observations (performed between November 17th, 2013 and November 12th, 2016) consisted of short exposures and low count rates, we summed all of them for a total exposure of 5554\,s. The summed spectrum in the $0.3-10\,$keV energy range can be fitted by an absorbed power-law model with a fixed $N_{H}$ = 1.25\,$\times$10$^{21}$\,cm$^{-2}$ and a photon index of 1.60\,$\pm$\,0.53. The corresponding unabsorbed ($0.3-10\,$keV) flux is (3.2\,$\pm$\,1.3)\,$\times$10$^{-13}$\,erg\,cm$^{-2}$\,s$^{-1}$. The X-ray spectrum collected on June 6th, 2019 can be well fitted by an absorbed power-law model with a photon index of 1.32 $\pm$ 0.35 that corresponds to an unabsorbed ($0.3-10\,$keV) flux of (1.2\,$\pm$\,0.4)\,$\times$10$^{-12}$\,erg\,cm$^{-2}$\,s$^{-1}$. Thus, there was an increase in X-ray flux for this source on June 6th, 2019, which was accompanied by a hint of hardening in the spectrum.
On the same day, the source was detected ($>$\,3\,$\sigma$) by UVOT in the $w2$ band with a magnitude $w2$ = 19.39 $\pm$ 0.11. Observations in the other bands resulted in upper limits ranging from 19.98 and 19.11. As such, the increase in X-ray activity coincided with an increase in the UV band (see Fig.~\ref{fig:MG1J181841+0903}).

\subsection{PMN\,J2016-0903}
\label{sec:pmnj2016}
PMN\,J2016-0903 is a BL Lac type object located at RA:~20$^{\circ} $16' 24.00'', Dec:~-09$^\circ$ 03' 32.70'' at a redshift of $z=0.367$~\citep{2022ApJS..263...24A}.
On November 29th, 2019, IceCube detected a 
neutrino cluster from this direction. 
MAGIC observed the source on the night of November 30th, 2019, within a zenith-angle range of $56^\circ-67^\circ$. 
The good-quality data collected cover about 0.86~hours of observation time.
An upper limit on the integral flux above an energy threshold of 450\,GeV was computed to be $7.40\times10^{-12}\,\mathrm{cm^{-2}}\,\mathrm{s^{-1}}$.

The source was only observed by {\em Swift} once, on  December 8, 2012 with an exposure time of 1581\,s. The $0.3-10\,$keV spectrum can be fitted with an absorbed power-law model with a fixed $N_{H} = 3.97 \,\times\,$10$^{21}$\,cm$^{-2}$ and a photon index of 2.69 $\pm$ 0.31. The corresponding unabsorbed flux in the $0.3-10\,$keV band is $(1.9 \pm 0.4) \times$10$^{-12}$\,erg\,cm$^{-2}$\,s$^{-1}$. The $0.1-2.4\,$keV flux reported in the ROSAT All Sky Survey (RASS)~\citep{2016A&A...588A.103B} is $6.42\times 10^{-13}$\,erg\,cm$^{-2}$\,s$^{-1}$, with an exposure of 1374 s, consistent with the \emph{Swift} one. For comparison, the $0.1-2.4\,$keV flux estimated by {\em Swift}-XRT on December 8, 2012 is 1.36 $\times$10$^{-12}$\,erg\,cm$^{-2}$\,s$^{-1}$, a factor of two higher than the flux observed by ROSAT. 

The measured UVOT magnitudes estimated on December 8, 2012 are $v$ = 17.55 $\pm$ 0.20, $b$ = 18.10 $\pm$ 0.15, $u$ = 17.02 $\pm$ 0.10, $w1$ = 17.24 $\pm$ 0.11, $m2$ = 17.25 $\pm$ 0.11, and $w2$ = 17.30 $\pm$ 0.09.

\begin{figure*}[!th]
\centering
\subfloat[MG1J181841+0903 (Section~\protect\ref{sec:mg1J1818})]{
  \includegraphics[width=0.45\textwidth]{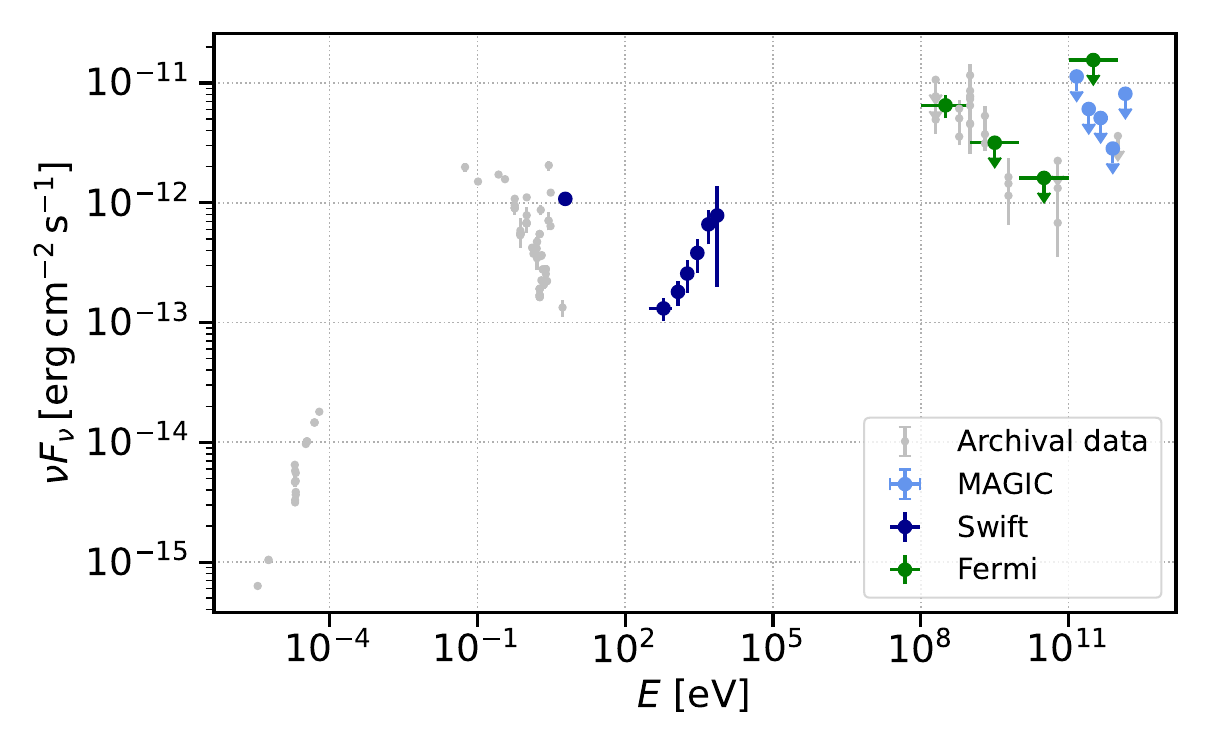}
  \label{fig:MG1J181841+0903}
  }
\subfloat[PMNJ2016-0903 (Section~\protect\ref{sec:pmnj2016})]{%
  \includegraphics[width=0.45\textwidth]{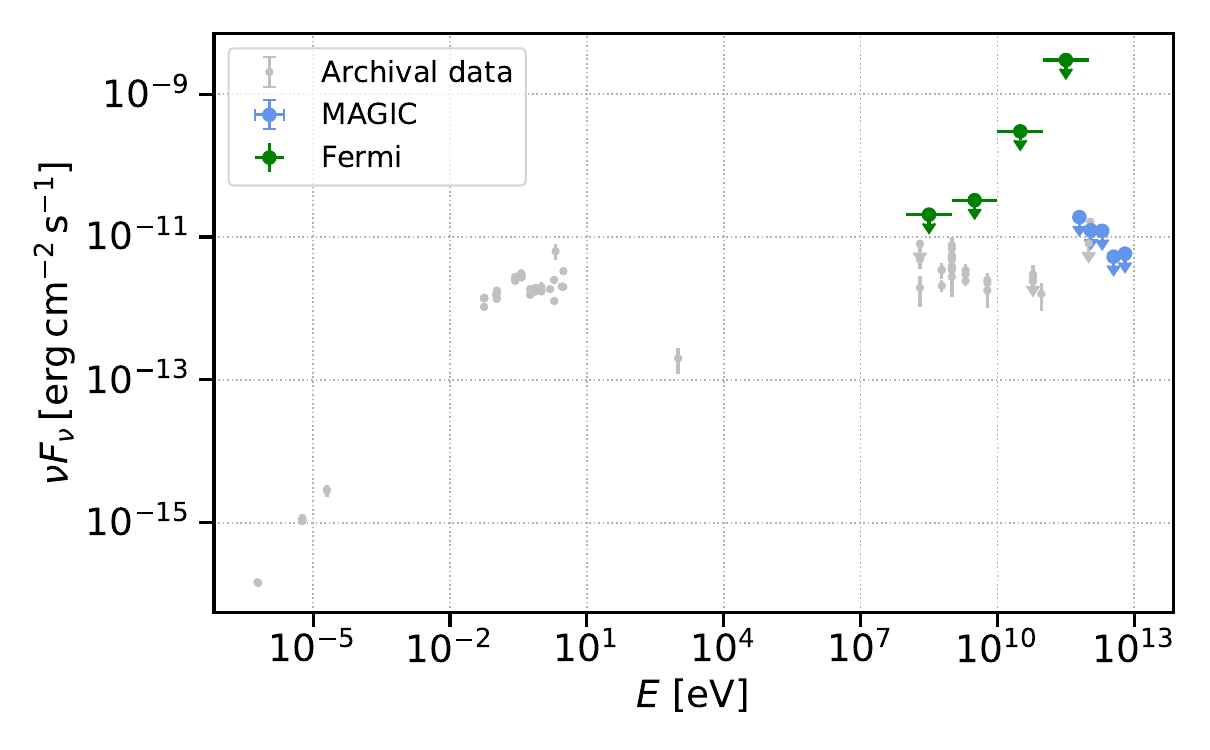}%
  \label{fig:PMNJ2016-0903}%
}\\
\subfloat[OP313 (Section~\protect\ref{sec:op313})]{%
  \includegraphics[width=0.45\textwidth]{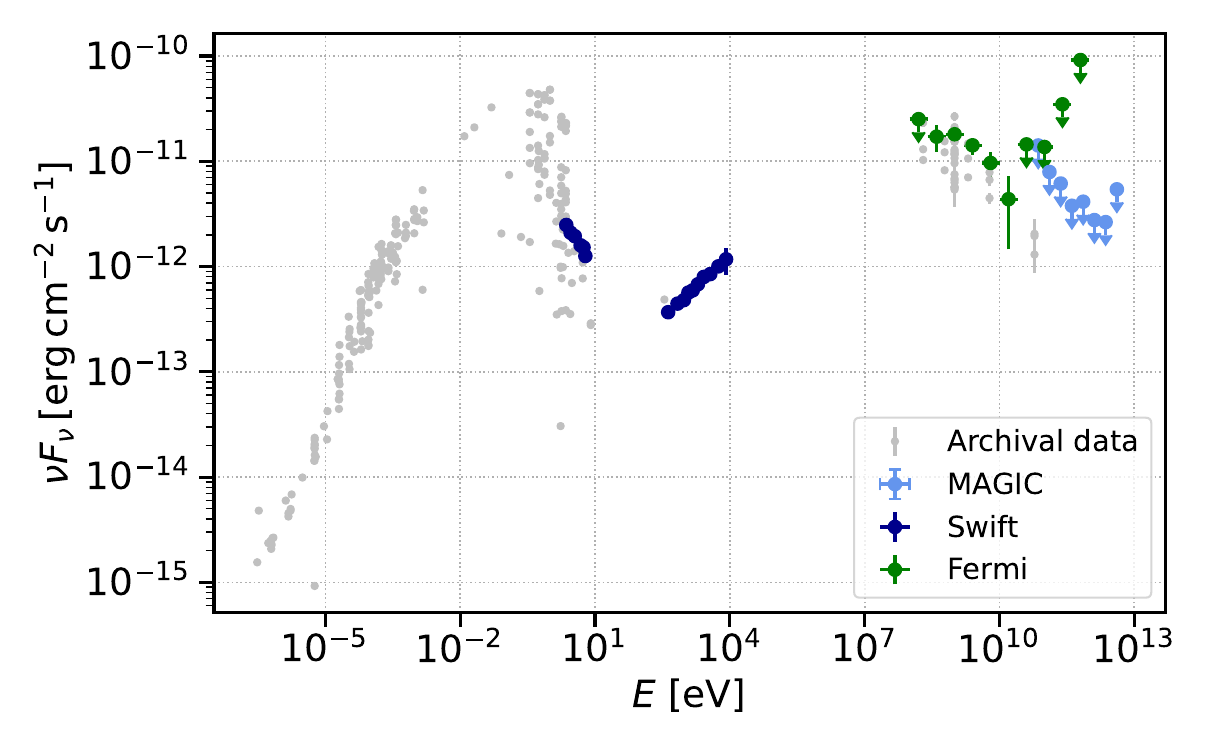}%
  \label{fig:OP313}%
}
\subfloat[OC457 (Section~\protect\ref{sec:oc457})]{%
  \includegraphics[width=0.45\textwidth]{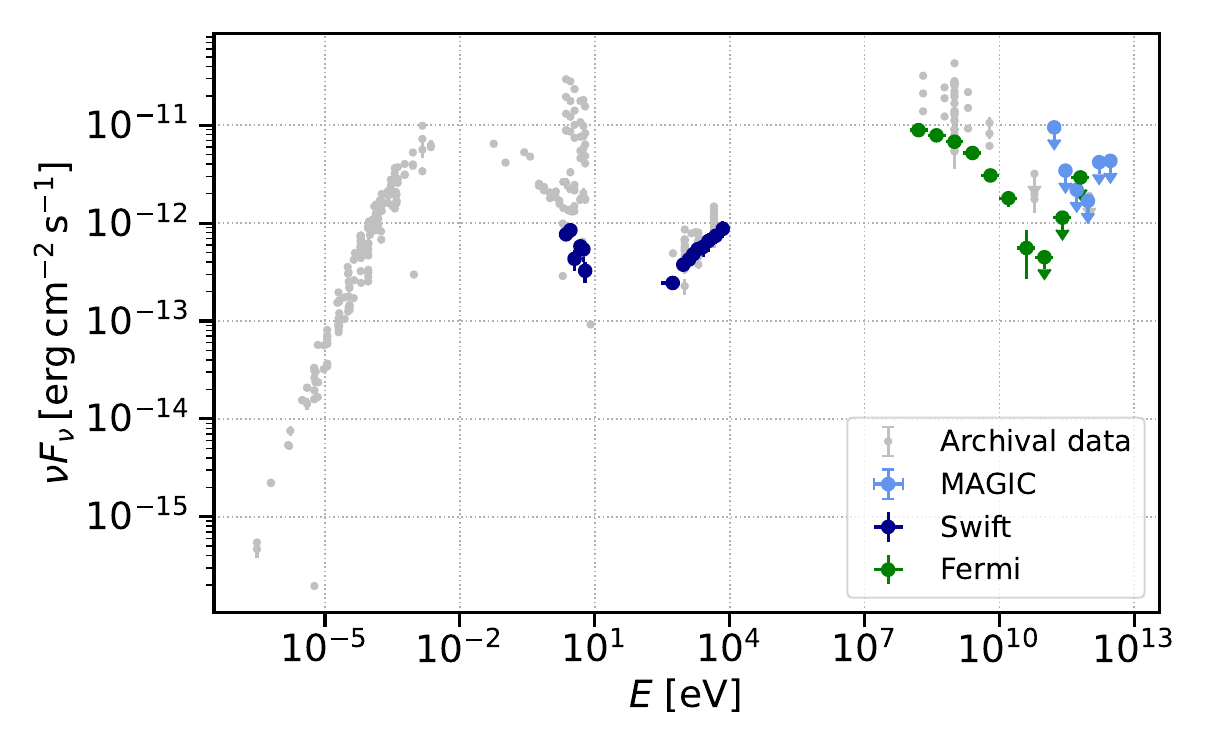}%
  \label{fig:OC457}%
}\\
\subfloat[GB6~J0316+0904 (Section~\protect\ref{sec:gb6j0316})]{%
  \includegraphics[width=0.45\textwidth]{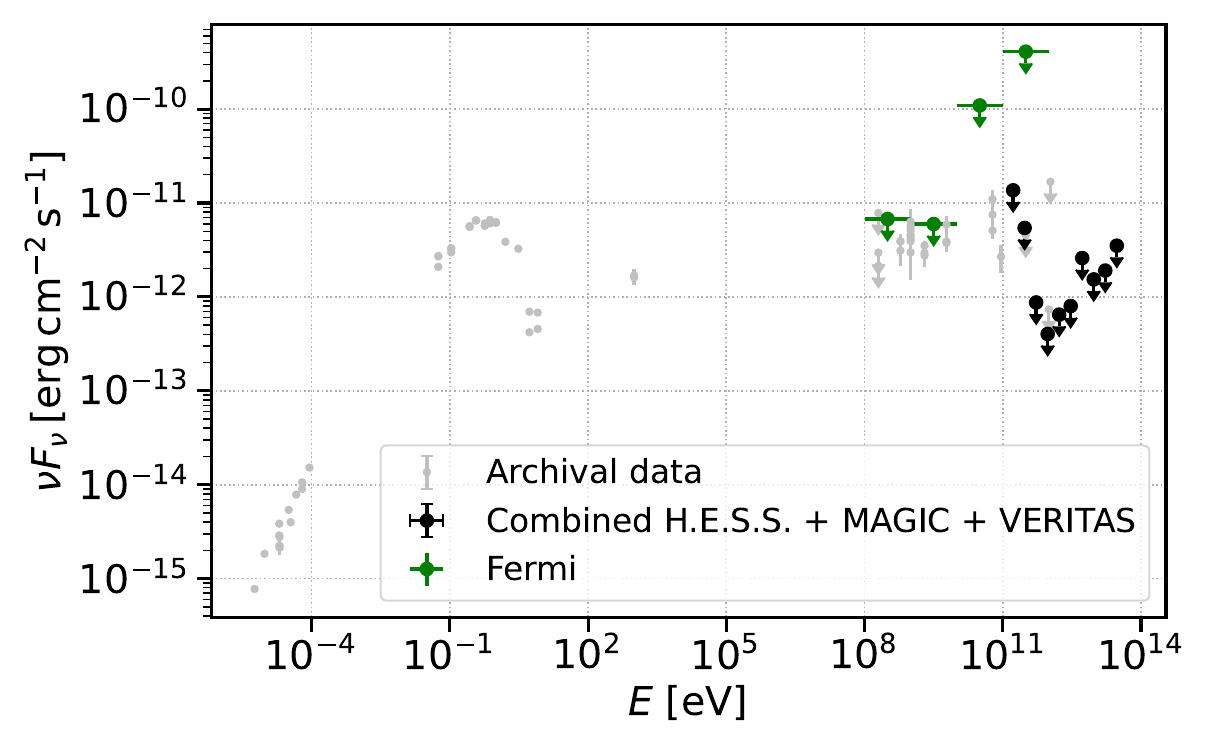}%
  \label{fig:GB6J0316+0904}%
}
\subfloat[PMN~J0325+1843 (Section~\protect\ref{sec:allskygfu})]{%
  \includegraphics[width=0.45\textwidth]{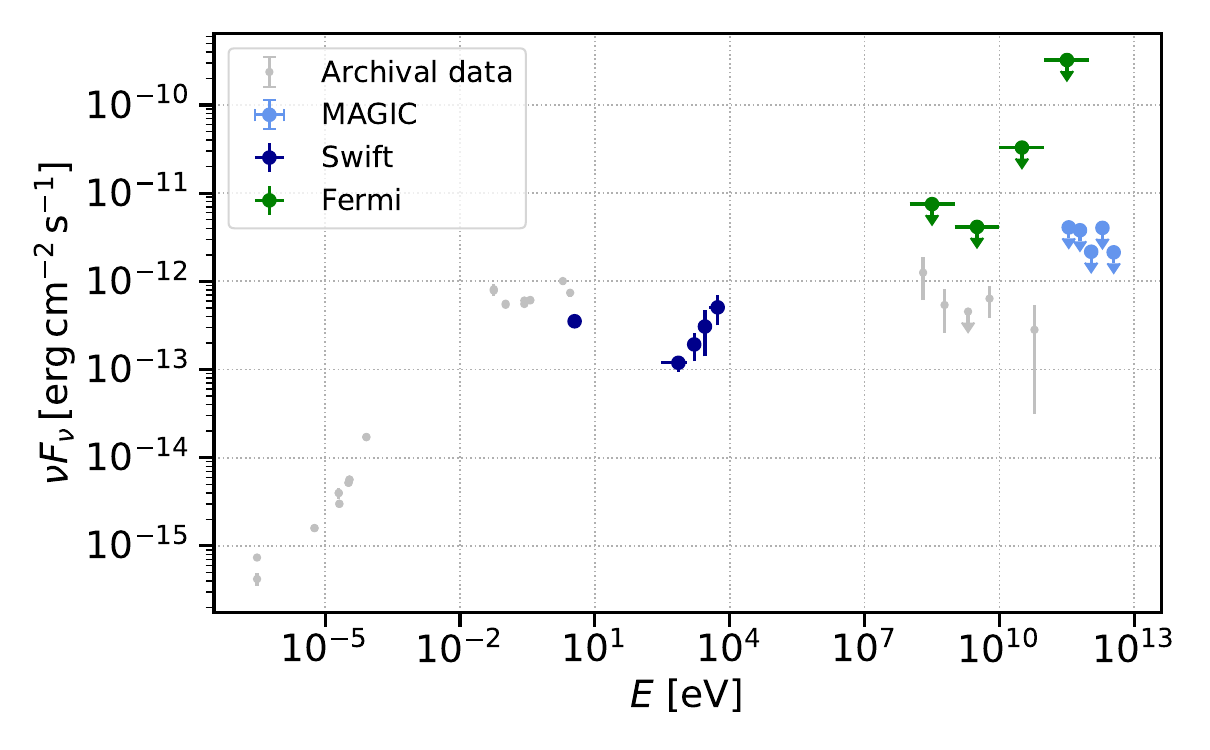}%
  \label{fig:PMNJ0325-1843}%
}
\caption{SEDs for the counterparts of the GFU-cluster alerts mentioned in the text. They comprise IACT ULs and simultaneous MWL data, together with archival data provided for comparison.}
\label{fig:sed_multiplets}
\end{figure*}

Since there are no simultaneous {\em Swift} observations at the time of the neutrino alert, we do not compare the state of activity of the source during the neutrino event with the archival one.
Thus, the SED of the source showing simultaneous MWL data (see Fig.~\ref{fig:PMNJ2016-0903}) does not include information on the {\em Swift} flux.

\begin{figure*}
\centering
\subfloat[IceCube-171106A/87GB~223537.9+070825 (Section~\protect\ref{sec:results_ehe171106A})]{%
  \includegraphics[width=0.45\textwidth]{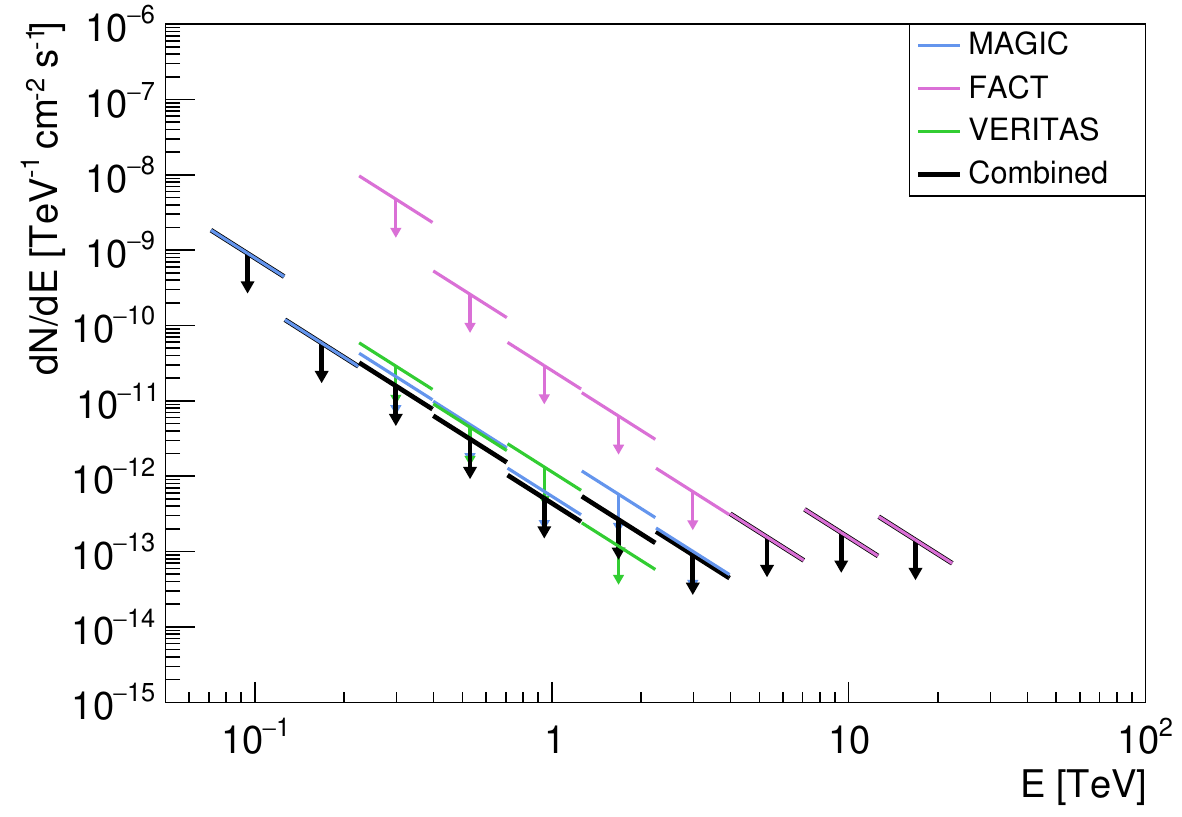}%
  \label{fig:ul_EHE171106A}%
 }
\subfloat[IceCube-191001A/AT2019dsg (Section~\protect\ref{sec:results_ic191001})]{%
  \includegraphics[width=0.45\textwidth]{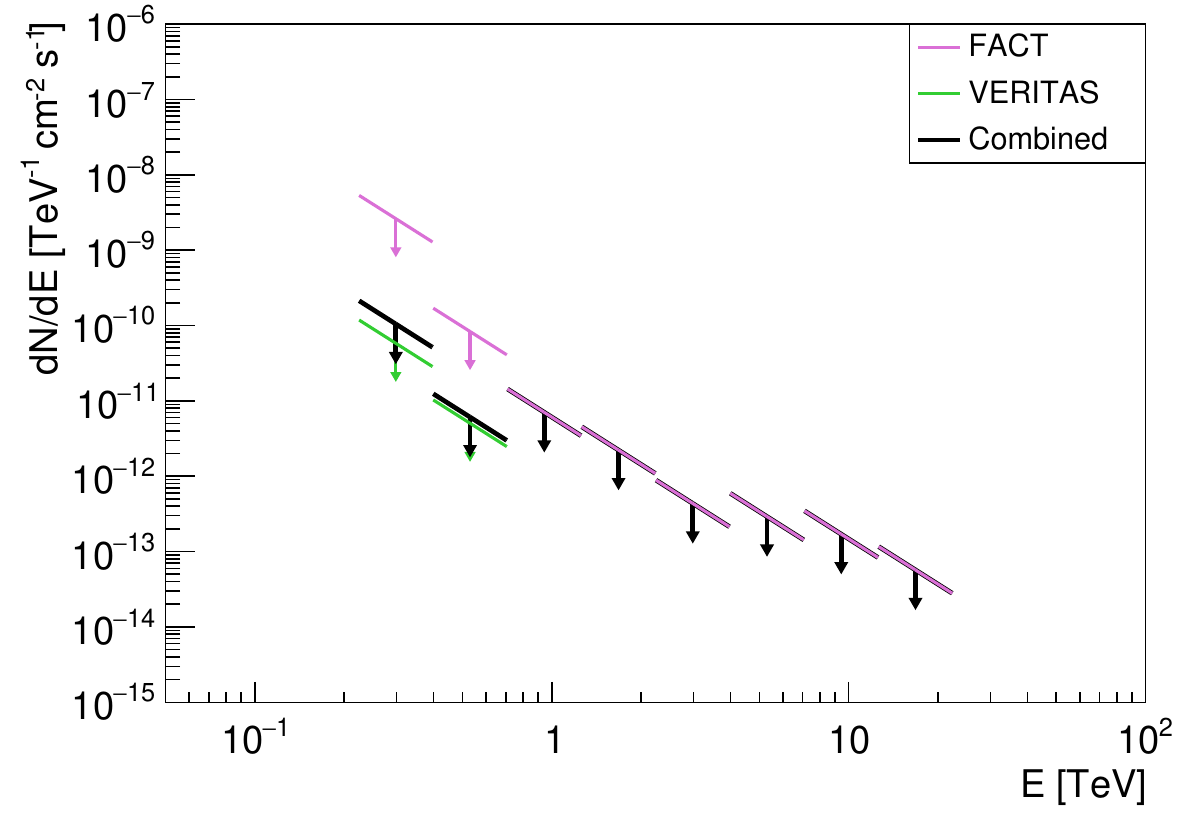}%
  \label{fig:ul_IC191001A}%
}\\
\subfloat[IceCube-190922B/AT2019pqh (Section~\protect\ref{sec:results_ic190922})]{%
  \includegraphics[width=0.45\textwidth]{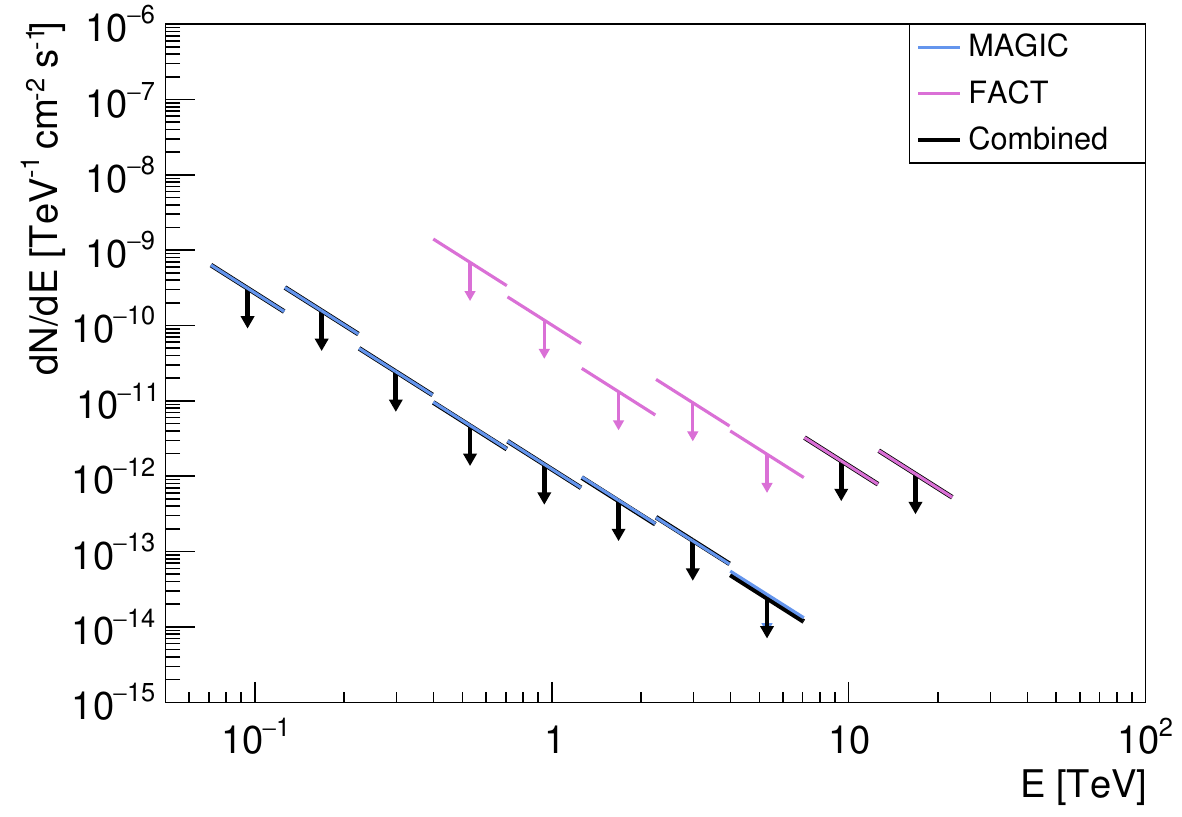}%
  \label{fig:ul_IC190922B}%
}
\subfloat[IceCube-201114A/4FGL J0658.6+0636 (Section~\protect\ref{sec:results_ic201114})]{%
  \includegraphics[width=0.45\textwidth]{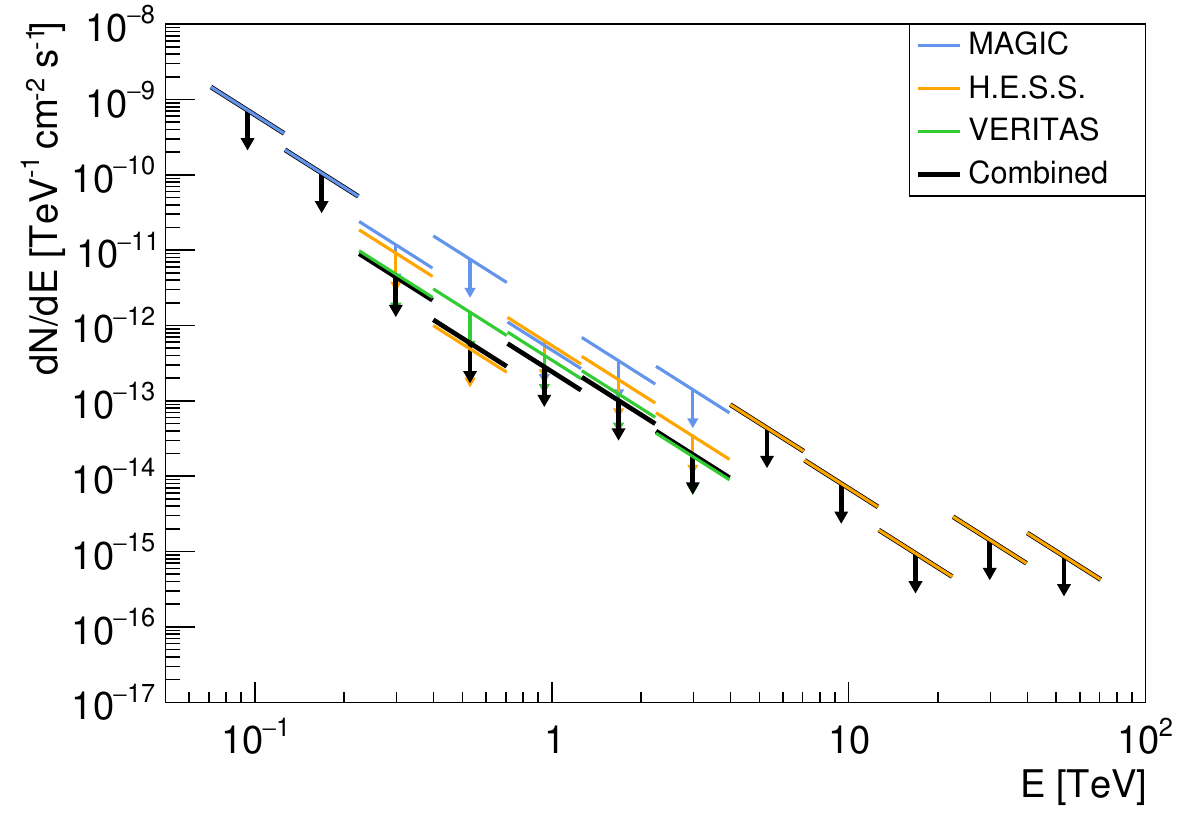}%
  \label{fig:ul_IC201114A}%
}\\
\subfloat[GB6\,J0316+0904 (Section~\protect\ref{sec:gb6j0316})]{%
  \includegraphics[width=0.45\textwidth]{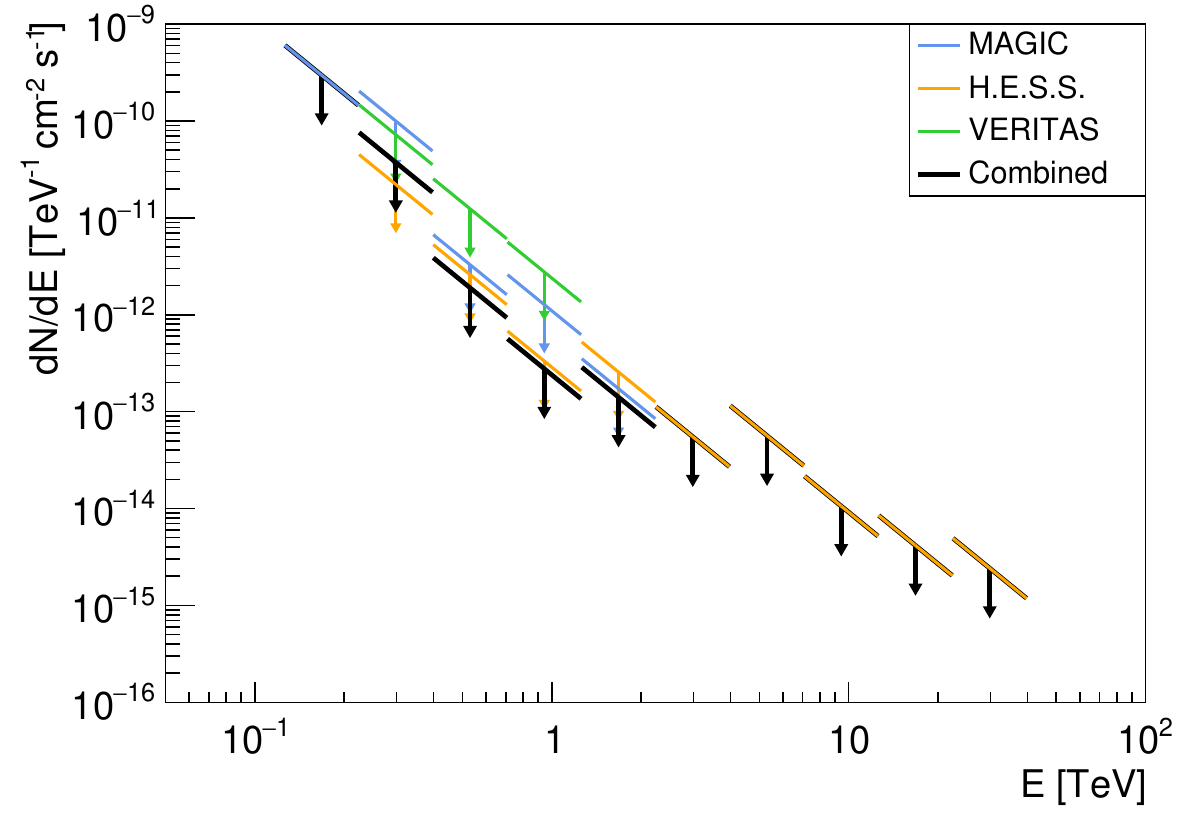}%
  \label{fig:ul_GB6J0316+0904}%
}
\subfloat[IceCube-200107A/4FGL J0955.1+3551 (Section~\protect\ref{sec:results_ic200107})]{%
  \includegraphics[width=0.45\textwidth]{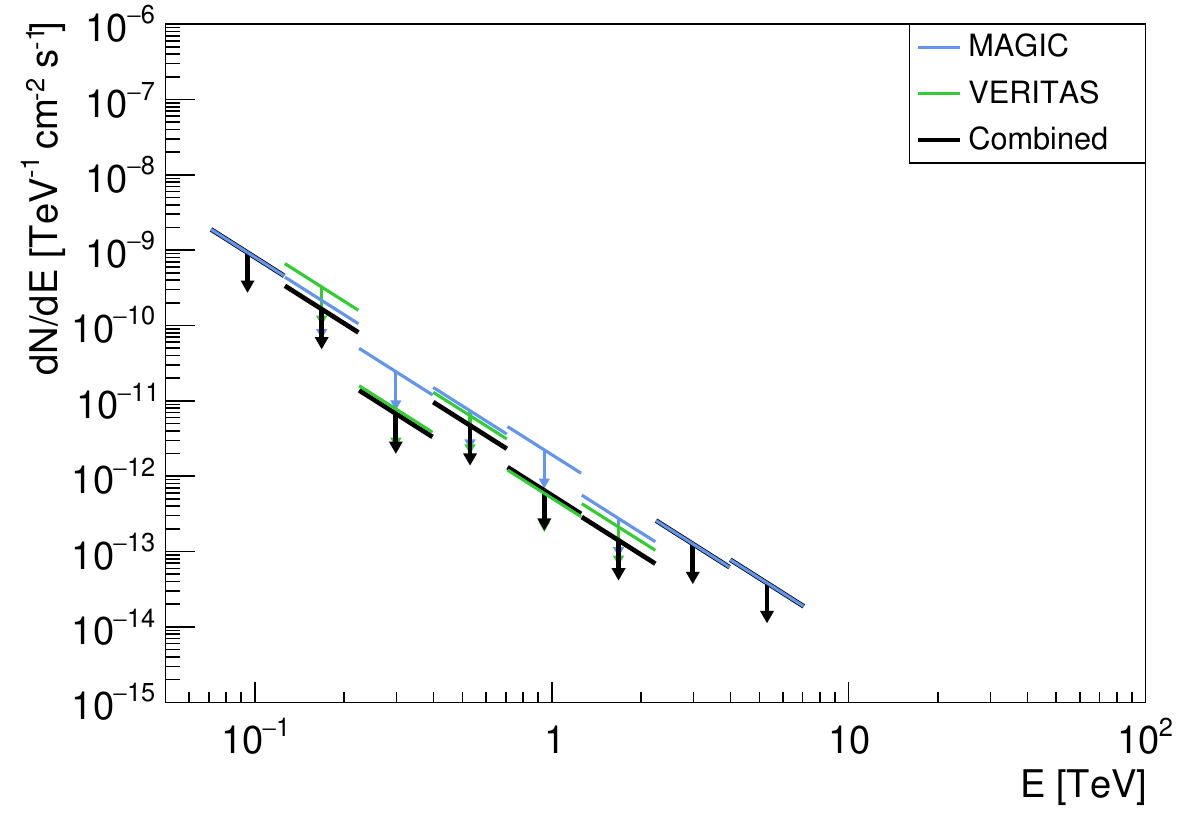}%
  \label{fig:ul_IC200107A}%
}\\
\caption{Combined differential-flux upper limits at $95\%~\mathrm{C.L.}$ for sources observed by multiple IACTs.}
\label{fig:ul_combined}
\end{figure*}

\subsection{OP~313}
\label{sec:op313}
OP~313 is an FSRQ located at RA:~13$^\circ$ 10' 28.66'', Dec:~+32$^\circ$ 20' 43.78'', with redshift $z=0.996$~\citep{Grasha2019Oct}. 
This source has been observed by MAGIC since 2014, though without detecting it \citep{fsrq_ul_paper_magic}.
In 2020, IceCube issued four GFU-cluster alerts of increasing statistical significance from this region.
The MAGIC telescopes observed the source on June 19, 20 and 23 with the Sum-Trigger-II analog trigger within a low zenith-angle range $16^\circ-34^\circ$.
The observations were performed in good weather conditions, providing 3.2~hours of good-quality data. 
Also in this case the source was not detected.
An upper limit on the integral flux above an energy threshold of 55\,GeV was computed to be $5.20\times10^{-11}\,\mathrm{cm^{-2}}\,\mathrm{s^{-1}}$.
More recently, in December 2023, the prototype Large-Sized Telescope (LST-1) achieved the first detection of OP 313 in the VHE gamma-ray band during a high state of activity of the source, marking it as the most distant AGN detected by a Cherenkov telescope \citep{2023ATel16381....1C}. 

The source was observed 32 times by {\em Swift}  between April 3, 2007 and March 17, 2021. The $0.3-10\,$keV spectrum can be fitted with an absorbed power-law model with $N_{H}$ fixed to 1.23\,$\times$10$^{20}$\,cm$^{-2}$. 
After a period of high activity in 2007, the X-ray flux of OP\,313 decreased during 2008 to 2014. 
The source exhibited renewed activity during the period from December 27, 2019 to March 21, 2020 (see Table~\ref{OP313_XRT}). 
The X-ray photon index during this period is relatively hard (i.e.,\ $1.5-1.7$) in comparison to the flat spectrum ($\Gamma$ $\sim$ 2) observed during previous low activity periods. 
During a similar period, increases in the optical and UV bands were also observed (see Table~\ref{OP313_UVOT}). The MWL SED is shown in Fig.~\ref{fig:OP313}.

\subsection{OC~457}
\label{sec:oc457}
OC~457 is an FSRQ located at RA:~01$^\circ$ 36' 58.59'', Dec:~+47$^\circ$ 51' 29.10'' at a redshift of $z=0.86$~\citep{2017ApJS..233....3T}. On August 4th, 2020, IceCube detected a neutrino cluster from this source.
The source was observed by the MAGIC telescopes more than a week after the alert, on August 14th and 15th, 2020, for a total of 2.5~hours of observation during moderate moonlight conditions. 
The observations were performed within a zenith-angle range of $22^\circ-34^\circ$.
The integral-flux upper limit above an energy threshold of 125\,GeV is $1.43\times10^{-11}\mathrm{cm^{-2}}\,\mathrm{s^{-1}}$.

The source was observed by {\em Swift} twenty times between July 16th, 2007 and August 17th, 2020 (see Table~\ref{OC457_XRT}). The 0.3 -- 10 keV spectrum can be fitted with an absorbed power-law model with $N_{H}$ = 1.02 $\times$10$^{21}$ cm$^{-2}$ s$^{-1}$. 
Between 2007 and 2020, the source was variable in X-rays, with the flux changing by a factor of three and the photon index varying between 1.1 and 1.8. Interestingly, at the time of the neutrino detection, on August 5, 2020, the X-ray flux reached the minimum value observed during the 2007-2020 period. Similarly, the optical and UV magnitudes observed by UVOT on the same day are significantly dimmer than the values observed before 2020 (although not at the dimmest value observed over the entire period, see Table~\ref{OC457_UVOT}). The SED is shown in Fig.~\ref{fig:OC457}.

\subsection{GB6~J0316+0904}
\label{sec:gb6j0316}

GB6~J0316+0904 is a BL Lac type object located at RA:~03$^{\circ}$ 16' 12.733'', Dec:~+09$^{\circ}$ 04' 43.283'' with redshift $z = 0.372$ \citep{2014ApJ...784..151S}. It was selected as a potential target for the GFU program for all three participating IACTs. On January 15, 2021, a 
neutrino cluster from that region was reported. 

The VERITAS telescopes were the first on target due to their automatic response and were able to collect 1.0\,hour of data on the night of the alert. H.E.S.S. pointed to the source within the next 24\,hours and performed 6\,hours of observations in total over three consecutive nights.
Due to high humidity on site, MAGIC was not able to follow-up the alert until three days later, at which point, 1.9 hours of observations were collected.

 Fig.~\ref{fig:GB6J0316+0904} and Fig.~\ref{fig:ul_GB6J0316+0904} show the MWL SED of the source and the combined differential-flux ULs. The VHE $\gamma$-ray upper limits were obtained by combining the data from all three IACTs (see Section~\ref{sec:methods}).

The source was observed by {\em Swift} five times from March 9th, 2009 to January 9th, 2015. The $0.3-10\,$keV spectrum can be fitted with an absorbed power-law model with $N_{H}$ = 1.27 $\times$10$^{21}$ cm$^{-2}$. 
We combined the observations that were carried out on July 3rd and 4th, 2011 in order to improve the statistics for the spectral fit. Results can be found in Table~\ref{0316_XRT}. A significant change of the X-ray flux was observed during the period from 2009 to 2015, indicating that the source is highly variable in X-rays. Similar high degrees of variability have been observed in the optical and UV bands (see Table~\ref{0316_UVOT}).
However, a comparison with the source activity at the time of the neutrino alert is not possible since there are no {\em Swift} observations in that period.

\subsection{All-sky alert/PMN~J0325-1843}
\label{sec:allskygfu}
On September 19, 2019, IceCube observed a GFU-cluster through its all-sky search for time-variable point sources.
This search exploits the same algorithm used to search for neutrino multiplets from known $\gamma$-ray emitters (see Section~\ref{subsec:multiplet}).
The location of the alert is consistent with the position of PMN~J0325-1843, a candidate blazar source. 

On September 24, 2019, the MAGIC telescopes observed the source for a total of 2.3~hours within a zenith-angle range of $47^\circ-51^\circ$. An upper limit on the integral flux above 250 GeV was computed to be $5.16\times10^{-12}\mathrm{cm^{-2}}\,\mathrm{s^{-1}}$.

The source was only observed by {\em Swift} once, on September 22nd, 2019 for an exposure time of 1631\,s. The $0.3-10\,$keV spectrum can be fitted with an absorbed power-law model with $N_{H}$ = 3.16 $\times$10$^{20}$ cm$^{-2}$ and a photon index of 1.18 $\pm$ 0.48. The corresponding unabsorbed flux in the $0.3-10\,$keV band is (1.0 $\pm$ 0.4) $\times$10$^{-12}$ erg cm$^{-2}$ s$^{-1}$. The 0.1-2.4\,keV flux reported in the RASS is 2.51 $\times$10$^{-13}$ erg cm$^{-2}$ s$^{-1}$, with an exposure of 1139 s, consistent with the \emph{Swift} one. 
The {\em Swift}-XRT observations provided a comparable flux in the $0.1-2.4\,$keV energy range, 2.62 $\times$10$^{-13}$\,erg cm$^{-2}$ s$^{-1}$ on September 22nd, 2019. The magnitude measured by UVOT on September 22, 2019 is $u$ = 19.05 $\pm$ 0.09. The MWL SED is given in Fig.~\ref{fig:PMNJ0325-1843}.

\section{Follow-up results on single high-energy neutrino alerts} \label{sec:results_tracks}
A summary of the follow-up observation campaigns of single high-energy neutrino events is listed in Table~\ref{tab_tracks} while detailed information on the alerts are given in Table~\ref{tab_tracks_extended}, including the event direction, energy, signalness and FAR as well as the refined reconstruction of the events that was performed offline and reported in~\citep{2023ApJS..269...25A}. 

No VHE emission was detected from the neutrino alert directions and from the potential counterparts identified in the uncertainty region of six events.
Fig.~\ref{fig:ul_combined} gives combined differential-flux upper limits for sources observed by multiple IACTs. The details of combined upper limits calculation can be found in Section \ref{sec:methods}. 

Changes in the X-ray spectra and flux level have been observed in some of these sources, providing useful information about the sources' behavior in a broad band context and thus for modeling their SED. However, X-rays alone are not sufficient to conclusively establish a correlation between neutrino emission and a specific source.  

Fig.~\ref{fig:sed_tracks} provides the SED for the potential counterparts of the single neutrino alerts, plotted with simultaneous MWL and archival data. While detailed SED modeling for all sources is beyond the scope of this paper, in Section~\ref{sec:results_ic190730} we use PKS~1502+106 as an example to discuss the potential effects as well as the limitation of our results on the current modeling work.

\subsection{IceCube-171106A}\label{sec:results_ehe171106A}

Follow-up observations were carried out by FACT, MAGIC and VERITAS.

FACT devoted a total of 19 hours of observations in following up the alert. FACT’s initial observations, conducted on November 6, 2017, focused on the early reported position for the neutrino and lasted for 3.3~hours. Another 15.6~hours of observations, distributed over the following five nights, focused on updated positions for the neutrino. After removing data influenced by bad weather conditions, 4~hours on the updated position remained, covering a zenith-angle range from 21$^\circ$ to 57$^\circ$. An integral-flux upper limit above an energy threshold of 810\,GeV was determined as $8.9\times10^{-12}\mathrm{cm^{-2}}\,\mathrm{s^{-1}}$. The FACT differential-flux upper limits are shown in Fig.~\ref{fig:ul_EHE171106A}. 

MAGIC observed the event in the direction of 87GB~223537.9+070825, on Nov 10th, 13th, 15th, 16th, 2017, collecting a total of 4.5~hours of good-quality data.
The observations were performed in wobble mode with a zenith-angle range of $20^\circ - 45^\circ$.
The integral-flux upper limit estimated by MAGIC at the location of the source above an energy threshold of 120\,GeV is $8.31\times10^{-12}\mathrm{cm^{-2}}\,\mathrm{s^{-1}}$.

\begin{figure*}
\centering
\subfloat[IceCube-171106A (Section~\ref{sec:results_ehe171106A})]{%
\includegraphics[width=0.35\textwidth]{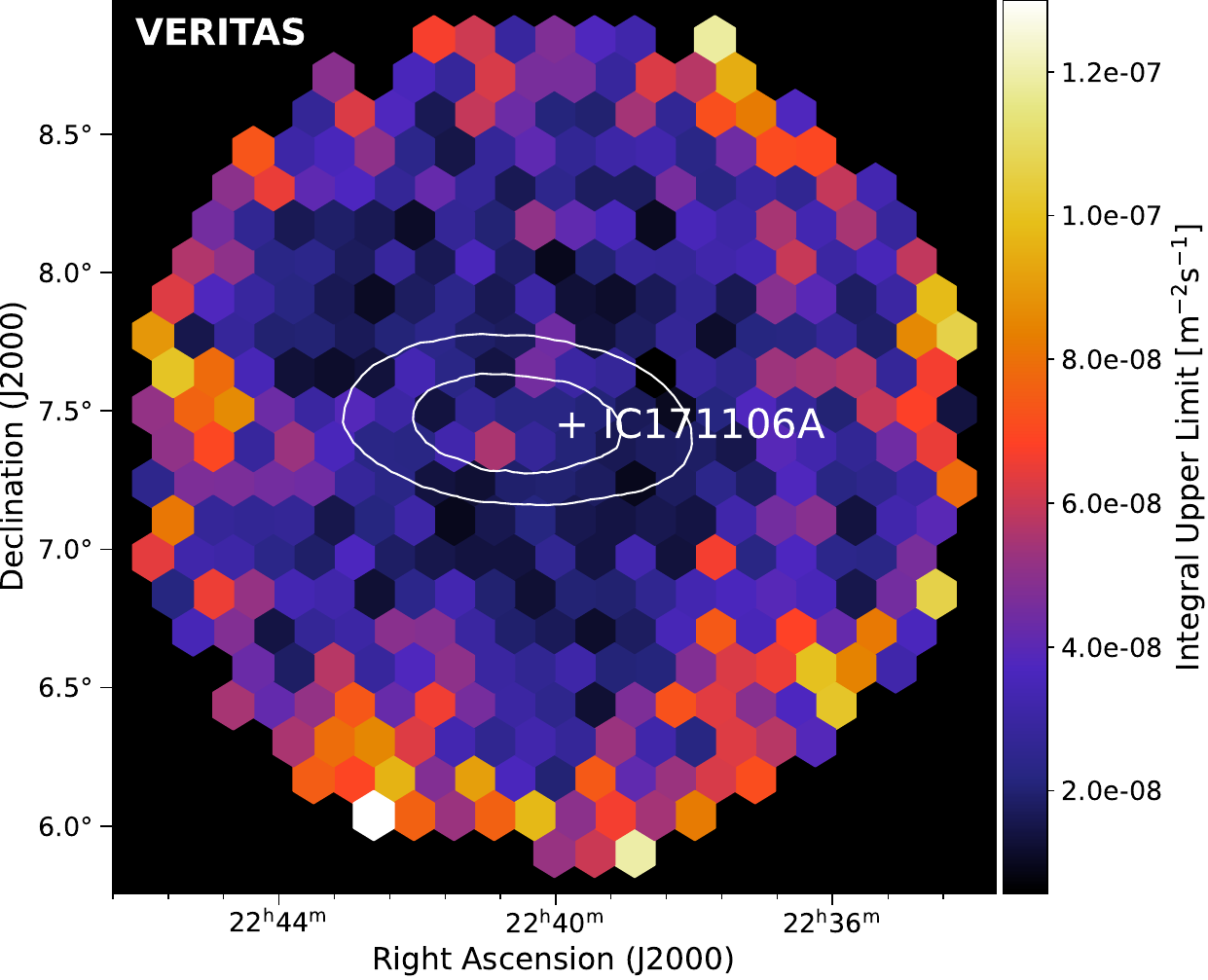}%
\label{fig:IC171106A_VERITAS}%
} 
\subfloat[IceCube-201222A (Section~\ref{sec:app_ic201222})]{%
\includegraphics[width=0.35\textwidth]{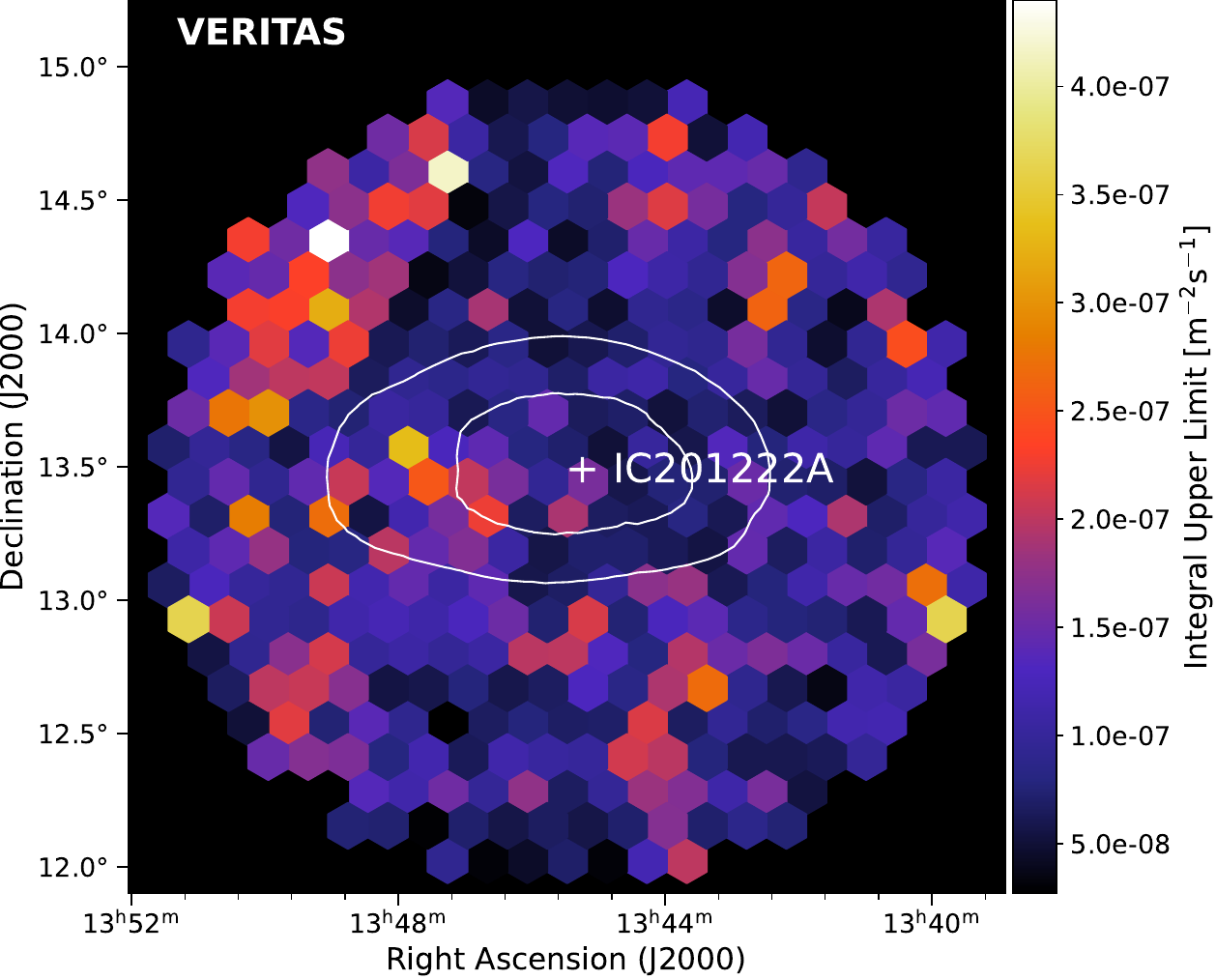}%
\label{fig:IC201222A_VERITAS}%
}\\

\subfloat[IceCube-191001A (Section~\ref{sec:results_ic191001})]{%
\includegraphics[width=0.35\textwidth]{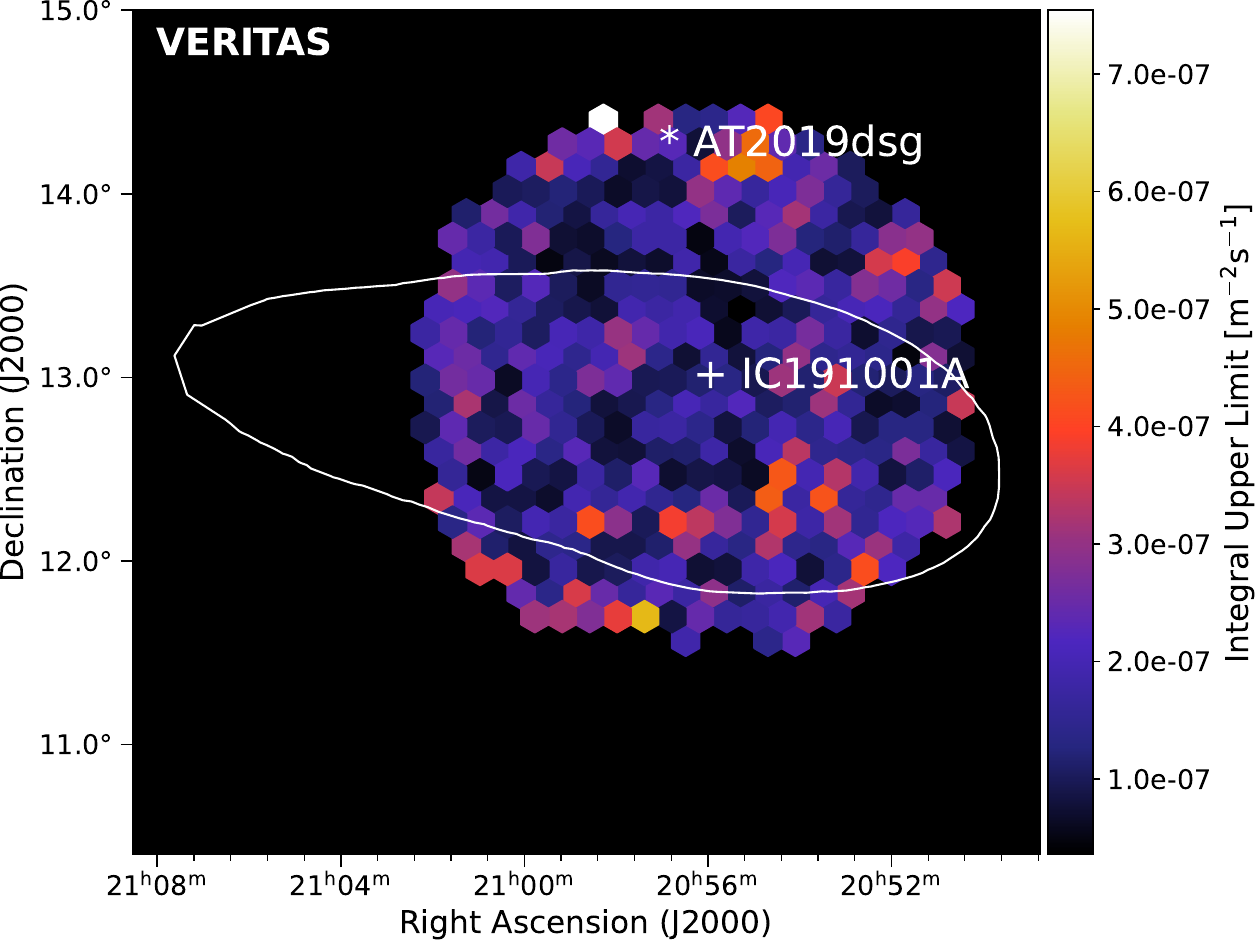}%
\label{fig:IC191001A_VERITAS}%
} 
\subfloat[IceCube-200107A (Section~\ref{sec:results_ic200107})]{%
\includegraphics[width=0.335\textwidth,trim= 0cm 0cm 0cm 1cm]{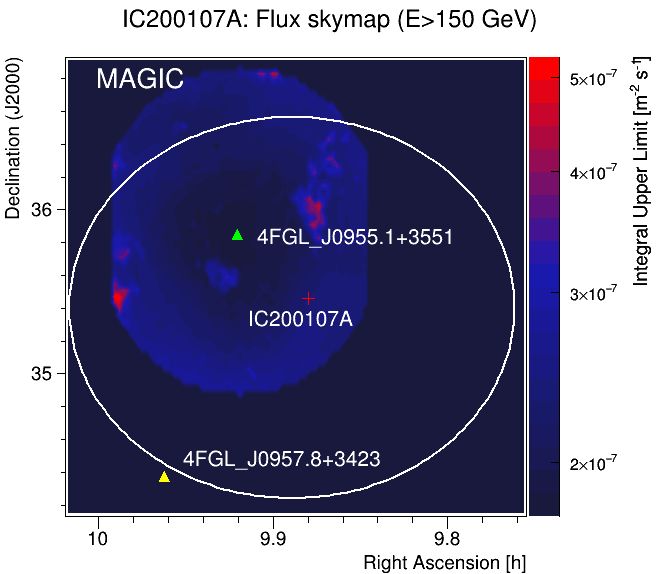}%
\label{fig:IC200107A_MAGIC}%
}\\

\subfloat[IceCube-200926A  (Section~\ref{sec:results_ic200926})]{%
\includegraphics[width=0.35\textwidth,trim= 1cm .2cm 1.5cm 1cm]{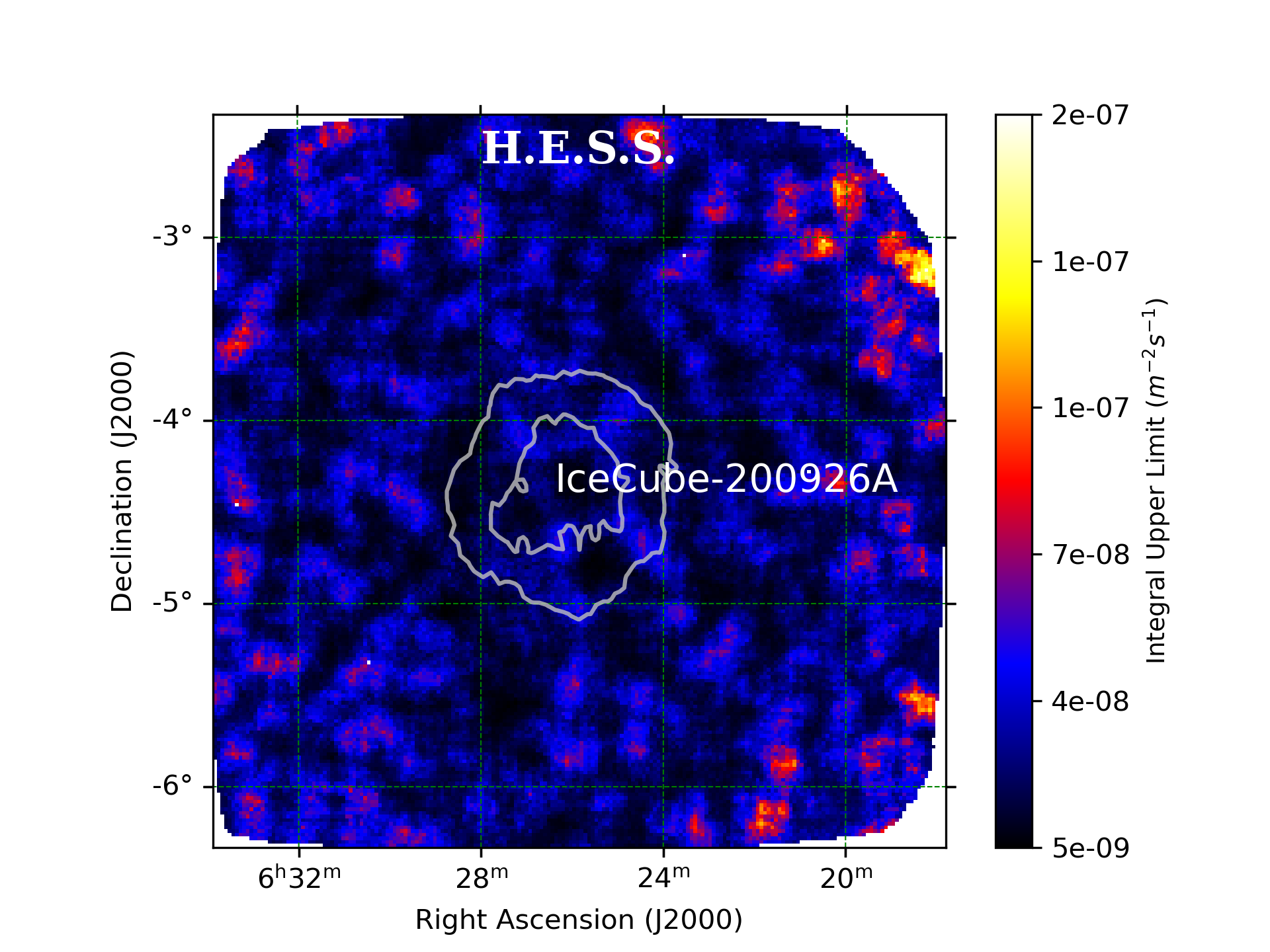}%
\label{fig:200926A_HESSUL_contours}%
} 
\subfloat[IceCube-201007A  (Section~\ref{sec:results_ic201007})]{%
\includegraphics[width=0.35\textwidth,trim= 1cm .2cm 1.5cm 1cm]{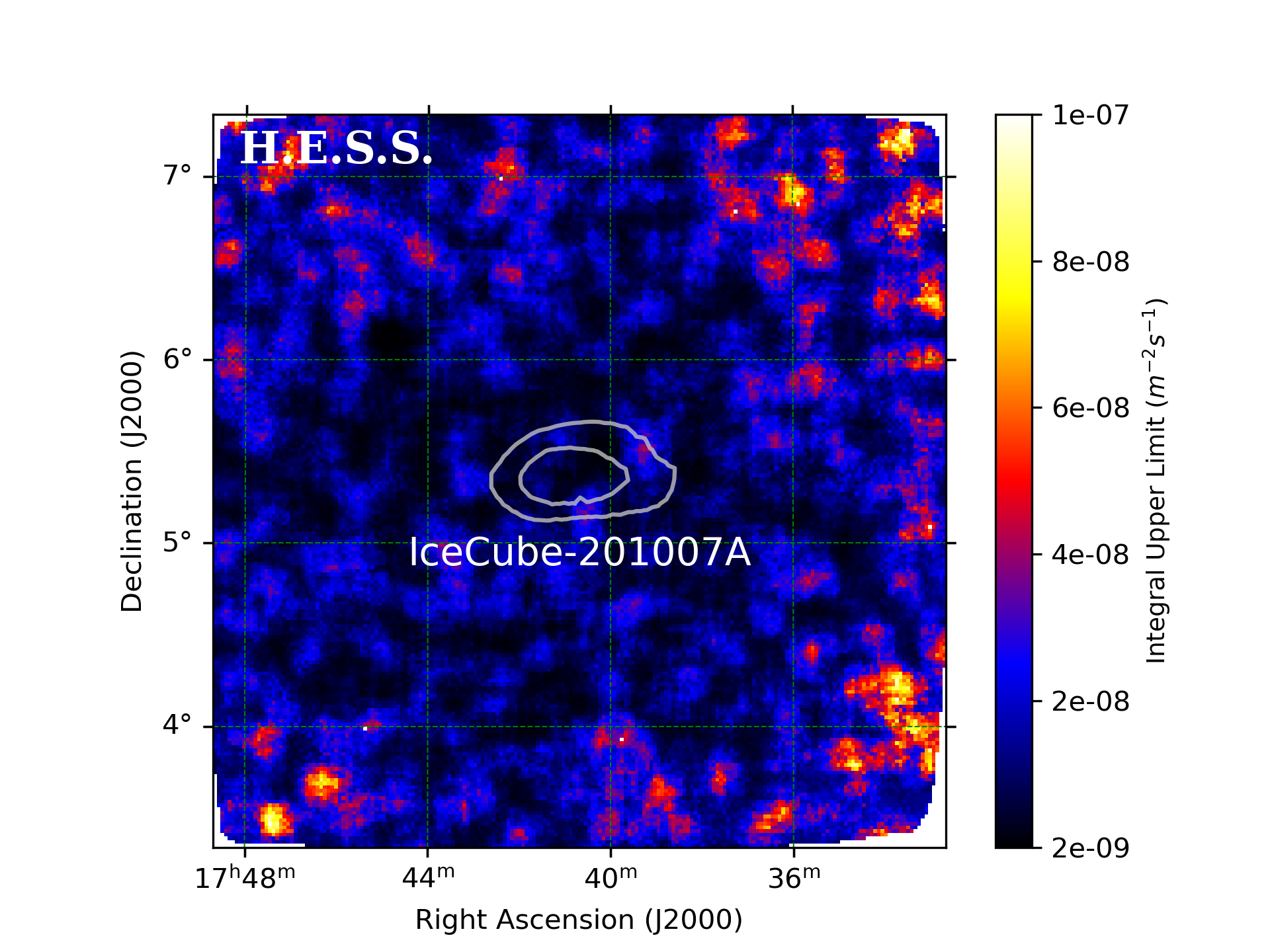}%
\label{fig:201007A_HESSUL_contours}%
}\\ %
\subfloat[IceCube-201114A  (Section~\ref{sec:results_ic201114})]{%
\includegraphics[width=0.35\textwidth,trim= 1cm .2cm 1.5cm 1cm]{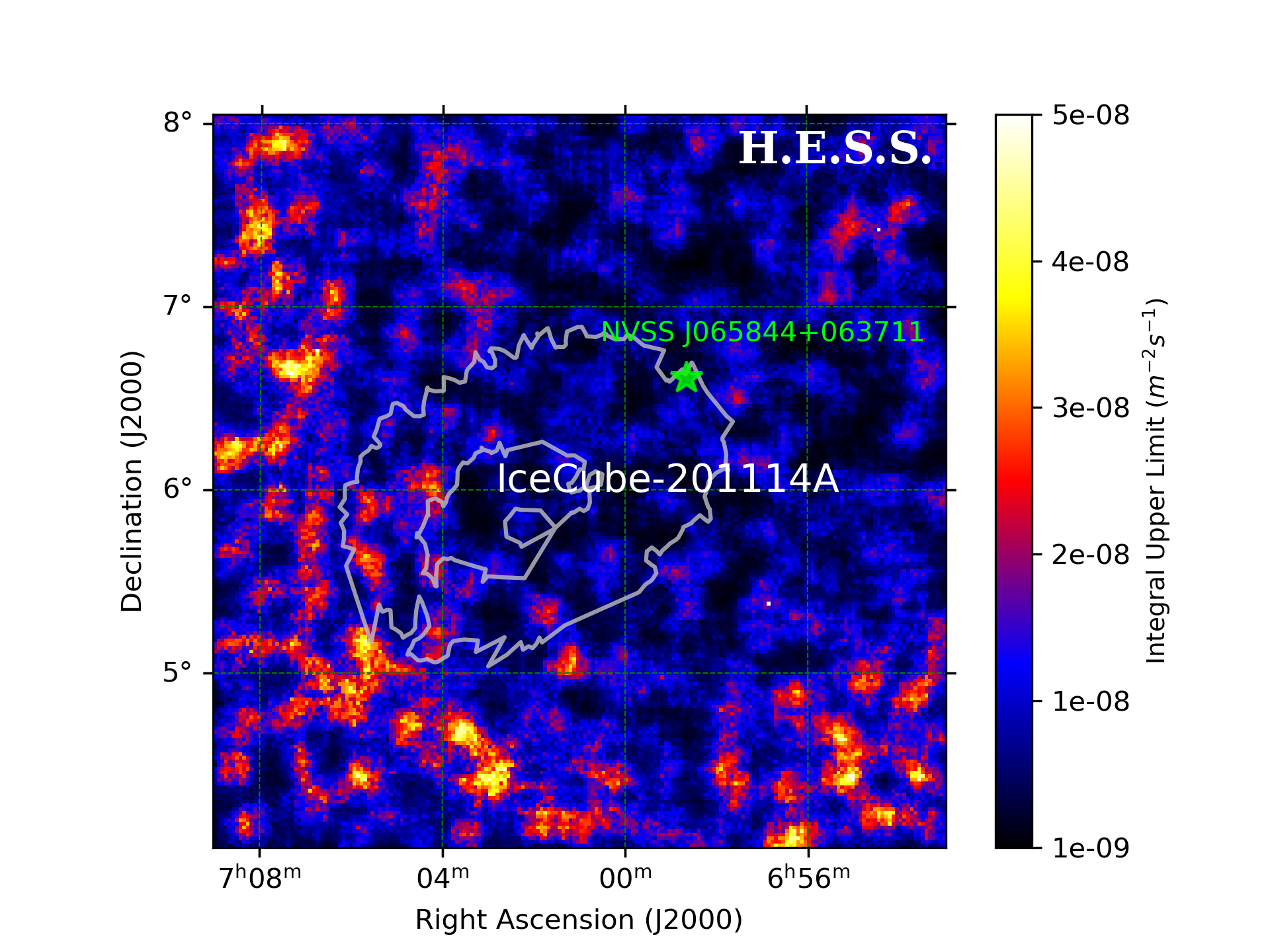}%
\label{fig:201114A_HESSUL_contours}%
}\\
\caption{Integral VHE $\gamma$-ray flux upper-limit maps derived from VERITAS (a, b, c), MAGIC (d), and H.E.S.S. (e, f, g). The two white lines denote the 50\% and 90\% containment contours of the IceCube event localizations (for panels (c) and (d), only the 50\% containment contours are shown). The energy thresholds used to derive the upper-limit maps are 350, 200, 138, 150, 307, 530, and 326 GeV for panels (a) through (g), respectively.}
\label{fig:UL_maps}
\end{figure*}

VERITAS observed the event in the direction of  the IceCube alert and collected about 2.5~hours of quality-selected observation data from Nov 13 to Nov 20, 2017 with an average zenith angle of 35.1$^{\circ}$. Observations were performed using the standard wobble observation mode with a 0.5$^{\circ}$ offset in each of the four cardinal directions. The integral-flux upper limit above an energy threshold of 350\,GeV at the location of the FSRQ 87GB\,223537.9+070825 is $1.49\times10^{-12}\mathrm{cm^{-2}}\,\mathrm{s^{-1}}$. An integral-flux upper-limit map and IceCube localization region are shown in Fig.~\ref{fig:IC171106A_VERITAS} and a MWL SED is shown in Fig.~\ref{fig:IC171106A}. 

\subsection{IceCube-181023A}\label{sec:app_ic181023}

FACT followed up on the alert and observed the position of the neutrino for one hour on the night of October 26th, 2018. Earlier observations were not possible, as during full moon, remote operations cannot be carried out for safety reasons. After data-quality selection, 0.6~hours remained, covering a zenith-angle range from 52$^\circ$ to 60$^\circ$. An integral-flux upper limit above an energy threshold of 810\,GeV was determined as $1.0\times10^{-11}\mathrm{cm^{-2}}\,\mathrm{s^{-1}}$.

\subsection{IceCube-190503A}\label{sec:app_ic190503}
($+0.76^\circ$, $-0.70^\circ$, 90\% PSF containment).
The event was observed on May 3rd, 2019, by the MAGIC telescopes, which pointed in the direction of the alert position for a total observation time of 0.5 hours.
The zenith-angle range of the observation is $44^\circ-52^\circ$.
An integral-flux upper limit was calculated above an energy threshold of 200\,GeV.
The computed value is $1.60\times10^{-11}\mathrm{cm^{-2}}\,\mathrm{s^{-1}}$.

\subsection{IceCube-190730A / PKS~1502+106}
\label{sec:results_ic190730}

The distant blazar PKS~1502+106 (\textit{z}=1.84, \citep{Hewett2010Jul}) has been proposed as a potential electromagnetic counterpart to the IceCube-190730A Gold alert in several publications \citep[e.~g.][]{2021JCAP...10..082O, Rodrigues2021May}, as it is located within the $50\%$ neutrino uncertainty region, precisely  $0.31^\circ$ from the best-fit neutrino location. 

On July 31st, 2019, a day after the alert was issued, MAGIC observed the source for 3.1 hours with a zenith-angle range $29^\circ - 50^\circ$. 
Due to the high redshift of the target, the observations were performed using the Sum-Trigger-II analog stereo trigger with the aim of achieving the lowest possible energy threshold. 
The integral-flux upper limit computed at the location of the source, above an energy threshold of 150 GeV, is $8.09\times10^{-12}\mathrm{cm^{-2}}\,\mathrm{s^{-1}}$.

{\em Swift}-XRT observations performed between July 4 and July 30, 2019 found the source in a low-activity state with respect to the average flux reported in the second {\em Swift}-XRT point source (2SXPS) catalog \citep[1.97 $\times$ 10$^{-12}$ erg\,cm$^{-2}$\,s$^{-1}$;][]{Evans20}, based on all observations carried out between January 1, 2005 and August 1, 2018 (see Table~\ref{PKS1502_XRT}). In 2020, the $0.3-10$\,keV fluxes detected with the XRT were comparable to the average 2SXPS value. The photon index measured for the source in $2019-2020$ was comparable with the average value reported in the 2SXPS catalog (i.e., 1.43) within the uncertainties. 

\begin{figure*}
\centering
\subfloat[Comparison with model from \citet{2021JCAP...10..082O} in quiescent state. The black and green shaded areas correspond to the SED model and $\nu + \bar{\nu}$ all flavor from the paper  respectively, while the red dashed line corresponds to the IC-190730A energy.]{%
  \includegraphics[width=0.45\textwidth]{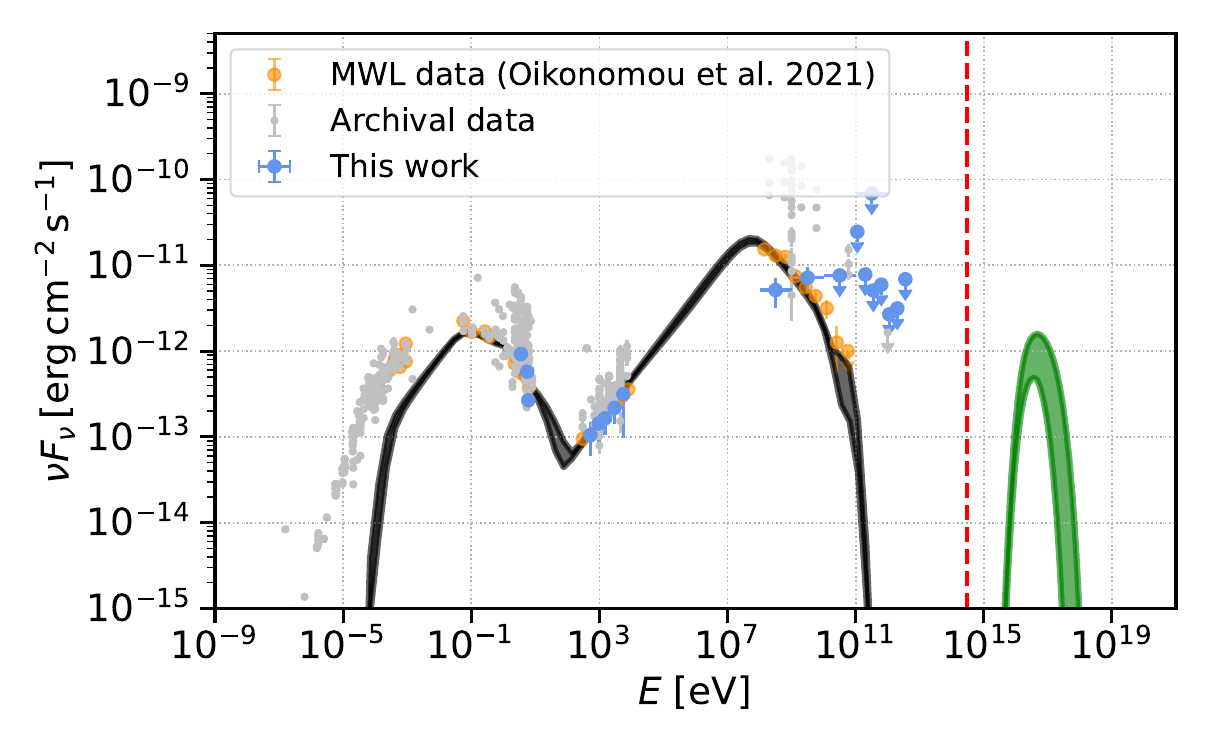}%
  \label{fig:PKS1502_Oikonomou}%
} 
\subfloat[Comparison with model from \citet{Rodrigues2021May} in quiescent state. The purple solid line corresponds to the SED model from the paper, while the red dashed line corresponds to IC-190730A energy.]{%
  \includegraphics[width=0.45\textwidth]{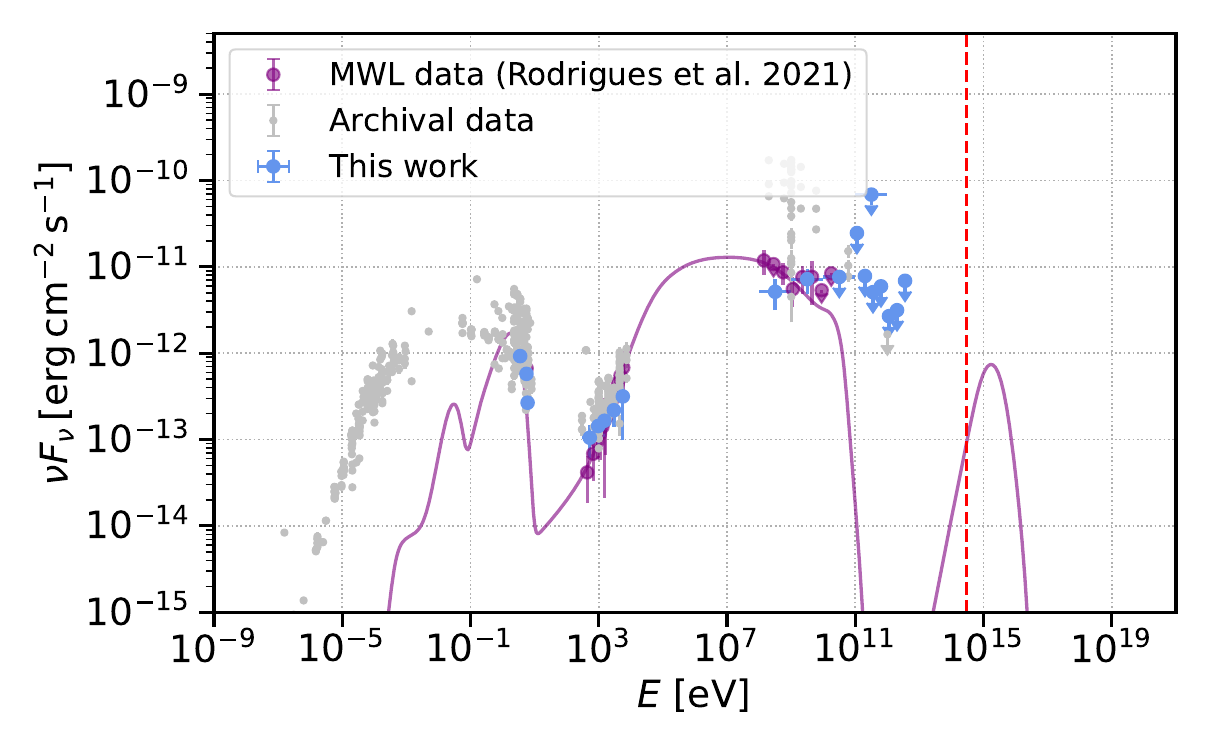}%
  \label{fig:PKS1502_Rodrigues}%
}\\
\caption{Model comparison of the SED for blazar PKS~1502+106, potentially associated the single high-energy neutrino alert IC-190730A  (Section~\ref{sec:results_ic190730}). Two different models are compared with IACT ULs and simultaneous MWL data, together with archival data~\citep[from ASI ASDC,][]{Stratta2011Mar}.}
\label{fig:sed_PKS_models}
\end{figure*}

The MWL SED resulting from this work is shown in Fig.~\ref{fig:IC190730A} and it is compared with models in Fig.~\ref{fig:sed_PKS_models}, as an example to discuss the potential effects as well as the limitation of our results on the current modeling work.

\citet{2021JCAP...10..082O} modeled the MWL SED of PKS 1502+106 using a lepto-hadronic framework. Since the source was in a low state at the time of the neutrino detection, they used MWL data from the  Wide-field Infrared Survey Explorer (\emph{WISE}), {\em Swift}/UVOT, {\em Swift}-XRT and {\em Fermi}-LAT telescopes taken between 2010 and 2014, which was a period of persistent low activity for the source. Several different locations for the emitting region were tested to explain the MWL and neutrino emission. To demonstrate how VHE $\gamma$ rays constrain the model, we choose the one that places the emitting region of PKS~1502+106 beyond the broad-line region (BLR) but inside the dust torus. In Fig.~\ref{fig:PKS1502_Oikonomou}, we present this model together with the VHE $\gamma$-ray ULs obtained from the MAGIC observations and MWL data. 

\citet{Rodrigues2021May} also modeled the emission of this source in a lepto-hadronic context, selecting three different emission states of the source starting from the {\em Fermi}-LAT 11-year light curve. Their model considers a single emitting region placed at the boundary of the BLR, resulting in substantial external inverse Compton emission from that region. This component dominates the hard X-ray and soft $\gamma$-ray emission, while soft X-rays and hard $\gamma$-rays are dominated by photons arising from inverse Compton scattering by pairs produced through the Bethe-Heitler process or the annihilation of VHE photons in the jet. In Fig.~\ref{fig:PKS1502_Rodrigues}, we show a comparison between the low state considered in \citet{Rodrigues2021May} and the results obtained in this work.

Compared with both works, the VHE $\gamma$-ray data presented here do not provide strong constraints on either of these models. One reason could be the limited observations performed on this source. Moreover, the high energy (HE) $\gamma$-ray data in Figures \ref{fig:PKS1502_Oikonomou} and \ref{fig:PKS1502_Rodrigues} deviate from both models, whereas the UV and X-ray data appear to be consistent with them. The inconsistency between the HE $\gamma$-ray data and the models could be because the MWL data were not taken simultaneously, although each of the periods considered coincides with a low state of activity of the source. We note that both \citet{2021JCAP...10..082O} and \citet{Rodrigues2021May} consider a long-term period of low-state activity for the source, whereas in this study, we only analyzed one month of {\em Fermi} data centered on the neutrino arrival time.

\subsection{IceCube-190922B}\label{sec:results_ic190922}

Both FACT and MAGIC observed the event, collecting 1.9~hours and 2.2~hours of good-quality data, respectively.

FACT’s observations started on September 22, 2019 at 23:05:26 UTC (i.e.,\ 63~seconds after the alert) and lasted for two hours. After data-quality selection, 1.9~hours remained, covering a zenith-angle range from 35$^\circ$ to 46$^\circ$. 
An integral-flux upper limit above an energy threshold of 810\,GeV was determined to be $1.1\times10^{-11}\mathrm{cm^{-2}}\,\mathrm{s^{-1}}$.

MAGIC conducted follow-up observations on September 22 and 25, 2019. The first day of observations focused on the direction of the neutrino event, the second day of observations focused on the location of the supernova candidate AT2019pqh, following its announcement by the optical telescope Zwicky Transient Facility \citep[ZTF,][]{2020PASP..132c8001D,2019PASP..131a8002B}
on September 23, 2019 as a possible counterpart of the neutrino event ~\citep{2019ATel13125....1S}.
The position of this source was inside the alert error region, about 0.5$^\circ$ shifted with respect to the neutrino direction.
The integral-flux upper limit found by the MAGIC telescopes at the location of AT2019pqh, above an energy threshold of 150\,GeV, is $1.25\times10^{-11}\mathrm{cm^{-2}}\,\mathrm{s^{-1}}$.

\subsection{IceCube-191001A /  AT2019dsg}\label{sec:results_ic191001}

Following the neutrino alert, ZTF identified radio-emitting tidal disruption event AT2019dsg as a possible counterpart. A TDE is a rare, transient event that occurs when a star comes sufficiently close to a supermassive black hole (SMBH) to be torn apart by tidal forces. Eventually, roughly 50\% of the star’s mass is captured and forms an accretion disk around the black hole.
The probability of a chance coincidence between a radio-emitting TDE that is as bright as AT2019dsg in bolometric energy flux and an astrophysical neutrino event has been estimated as $<0.2\%$ \citep{Stein2021}.

FACT observed for a total of 5.4~hours starting on October 1st, 2019 at 
20:10:41, i.e.,\ with a delay of 53~seconds after the alert. Of the total observation time, 1~hour was carried out in automatic mode the first night, and the remaining time was scheduled manually in the second night. After data-quality selection, 0.5~hours remained, covering a zenith-angle range of 16$^\circ$ to 32$^\circ$. An integral-flux upper limit above an energy threshold of 810\,GeV was determined to be $1.9\times10^{-11}\mathrm{cm^{-2}}\,\mathrm{s^{-1}}$. With slightly relaxed data-quality selection cuts ($0.8 < R750_{\rm cor}/R750_{\rm ref} < 1.3$), there are 5~hours of data, covering a zenith-angle range of 16$^\circ$ to 45$^\circ$. In this case, the integral-flux upper limit above an energy threshold of 810\,GeV is $4.6\times10^{-12}\mathrm{cm^{-2}}\,\mathrm{s^{-1}}$. 
Since this second choice may add a
small systematic effect, we believe the true upper limit to fall in
between the two values we found, possibly closer to the stricter one.
For the calculation of the differential-flux upper limits, we restrict the data sample to trigger thresholds below 560~DAQ~counts (moderate moonlight), leaving 3.7 hours of data.

On October 2nd, 2019, VERITAS performed a follow-up observation in response to the IceCube alert. The observations resulted in 40 minutes of quality-selected data taken from 02:24:00 to 04:04:48 UTC with an average zenith angle of $\sim$20$^\circ$. Observations were performed using the standard wobble observation mode with a 0.7$^{\circ}$ offset in each of four cardinal directions. Assuming a photon index of 2.5, the integral-flux upper limit above an energy threshold of 138 GeV is $8.6\times10^{-12}$ cm$^{-2}$ s$^{-1}$, which corresponds to $2\%$ of the Crab Nebula flux \citep{meagher2015years}. An integral-flux upper limit map and IceCube localization region are shown in Fig.~\ref{fig:IC191001A_VERITAS} and a MWL SED is shown in Fig.~\ref{fig:IC191001A}.

\subsection{IceCube-200107A / NVSS~J095508+355102}\label{sec:results_ic200107}
 
The event did not pass the GOLD/BRONZE classification, but it was identified as a starting track. 

This localization region contains a candidate counterpart source NVSS~J095508+355102 (4FGL~J0955.1+3551). This object is a blazar, located at redshift $z=0.557$~\citep{2020MNRAS.495L.108P}. The MWL SED of NVSS~J095508+355102 reveals a synchrotron peak at $\sim4\times10^{17}$\,Hz in the rest-frame \citep{2023MNRAS.526..661K}, indicating that this source belongs to the rare class of extreme high-synchrotron-peak blazars. Subsequent X-ray observations revealed a flaring episode just after the detection of the neutrino~\citep{ic200107a_swift_1, ic200107a_swift_2}.

The MAGIC telescopes observed the location of the source on January 17 and 18, 2020,  collecting 2.7 hours of good-quality data.
The observations were made in a zenith-angle range of $20^\circ-48^\circ$.
The computed value for the upper limit on the integral flux above an energy threshold of 120 GeV is $1.75\times10^{-11}\mathrm{cm^{-2}}\,\mathrm{s^{-1}}$. An integral-flux upper-limit map of the region observed by MAGIC is shown in Fig.~\ref{fig:IC200107A_MAGIC} together with the IceCube localization region.

VERITAS collected 9.5 hours of quality-selected data between Jan 29, 2020 and Feb 2, 2020, with an average zenith angle of 11.8$^{\circ}$. Observations were performed using the standard wobble observation mode with a 0.5$^{\circ}$ offset in each of four cardinal directions. The computed value for the upper limit on the integral flux above an energy threshold of 150 GeV at the location of the blazar NVSS J095508+355102 is $4.08\times10^{-12}\mathrm{cm^{-2}}\,\mathrm{s^{-1}}$.

In follow-up observations that were performed on January 8, 2020, {\em Swift} detected the source with the highest X-ray flux recorded in the period between 2012 and 2020. The source remained at a high X-ray flux (compared to the observations performed in 2012--2013) up to February 21, 2020 (see Table~\ref{0955_XRT} and Table~\ref{0955_UVOT}).
 
The MWL SED is shown in Fig.~\ref{fig:IC200107A}. The proposed multi-messenger SED models have difficulties with reconciling the observed neutrino and electromagnetic emissions. \cite{2020ApJ...902...29P} explore a lepto-hadronic model with an external photon field. A larger number of possible scenarios is discussed in \cite{ 2020ApJ...899..113P}, including neutrinos being produced in the vicinity of the accreting SMBH or through interactions with photons from a possibly weak BLR. Both groups argue that the X-ray flaring event is very unlikely to be directly connected with the observed neutrino event.

\subsection{IceCube-200926A}\label{sec:results_ic200926}

The event was observed by H.E.S.S.\ and MAGIC, which collected 1.3~hours and 1.0~hour of good-quality data, respectively.

H.E.S.S.\ observed at the direction of the neutrino event on September 27 from 02:00 UTC to 03:20 UTC with an average zenith angle of 47.5$^\circ$. H.E.S.S.\ did not detect any significant emission in the ROI defined by the IceCube localization uncertainty. In Fig.~\ref{fig:200926A_HESSUL_contours}, we show the integral VHE $\gamma$-ray flux upper-limit map of the region covered by the H.E.S.S.\ observations with sufficient statistics, together with the localization contours of IceCube-200926A. The upper limit on the integral flux (above an energy threshold of 307\,GeV) at the best-fit IceCube position is found to be $1.86\times10^{-12}\mathrm{cm^{-2}}\,\mathrm{s^{-1}}$.

The MAGIC telescopes observed at the neutrino direction on September 29, 2020, within a zenith-angle range of $38^\circ-48^\circ$. The upper limit on the integral flux (above an energy threshold of 200\,GeV) at the best-fit IceCube position is $9.78\times10^{-12}\mathrm{cm^{-2}}\,\mathrm{s^{-1}}$.

\subsection{IceCube-201007A}\label{sec:results_ic201007}

The event was observed with H.E.S.S. and MAGIC, which collected 3 hours and 0.5 hours of good-quality data, respectively.

From October 8, 2020, to  October 13, 2020, H.E.S.S. conducted follow-up observations of the neutrino localisation region for a total of 3 hours. The zenith angles of observation ranged from 51$^\circ$ to 60$^\circ$. In Fig.~\ref{fig:201007A_HESSUL_contours}, we show the integral VHE $\gamma$-ray flux upper limits map of the region covered by the H.E.S.S.\ observations, together with the localization contours of IceCube-200926A. The upper limit on the integral flux (above an energy threshold of 530\,GeV) at the best-fit IceCube position is found to be $8.12\times10^{-13}\mathrm{cm^{-2}}\,\mathrm{s^{-1}}$.

The MAGIC telescopes observed the location of event IceCube-201007A on October 9, 2020 in good weather conditions. The zenith-angle range of the observations is $42^\circ-50^\circ$. The upper limit on the integral flux (above an energy threshold of 200\,GeV) at the best-fit IceCube position is found to be $2.46\times10^{-11}\mathrm{cm^{-2}}\,\mathrm{s^{-1}}$.

\subsection{IceCube-201114A / NVSS~J065844+063711}\label{sec:results_ic201114}

The location of the neutrino event was found to be consistent with the position of the blazar NVSS J065844+063711, also known as the {\em Fermi} source 4FGL J0658.6+0636 (RA: 104.64$^\circ$, Dec: 6.60$^\circ$). The event was observed with H.E.S.S., MAGIC and VERITAS. 

H.E.S.S.\ observed the region for 14.3~hours. The observation campaign was carried out from November 18, 2020, to November 25, 2020, and from December 10, 2020, to December 11, 2020. The zenith angle for observation ranged from 30$^\circ$ to 42$^\circ$. 
In Fig.~\ref{fig:201114A_HESSUL_contours}, we show the VHE $\gamma$-ray integral-flux upper-limit map of the region covered by the H.E.S.S.\ observations together with the localization contours of IceCube-201114A. The upper limit on the integral flux above 326 GeV at the position of NVSS\,J065844+063711 is found to be $5.37\times10^{-13}\mathrm{cm^{-2}}\,\mathrm{s^{-1}}$.

The MAGIC telescopes observed on November 16 and 17, 2020 and during the period of November 19 to 25, 2020. They collected 6\,h in the direction of NVSS\,J065844+063711.
The observations were performed at a zenith angle in the range of $22^\circ-47^\circ$.
The upper limit on the integral flux above 120 GeV was computed to be \SI{1.09e-11}{\per\cm\squared\per\second}.

VERITAS collected about 7~hours of quality-selected data between Nov 15, 2020 and Nov 19, 2020 and with an average zenith angle of 31.1$^{\circ}$. Observations were performed using the standard wobble observation mode with a mixture of 0.5$^{\circ}$ and 0.7$^{\circ}$ offset in each of four cardinal directions. The upper limit on the integral flux above an energy threshold of 200\,GeV at the location of the blazar NVSS\,J065844+063711 is $1.37\times10^{-12}\mathrm{cm^{-2}}\,\mathrm{s^{-1}}$. 

{\em Swift}-XRT observations carried out on November 15, 2020 indicate an increase of the X-ray flux at the time of the neutrino emission compared with previous observations performed in May 2012 (see Table~\ref{0658_XRT}). The source remained at a comparable X-ray flux level up to December 11, 2020. In contrast, {\em Swift}/UVOT observations indicate that in 2020, the source was less bright in the $u$, $m2$ and $w2$ bands than the baseline values found in 2012 (see Table~\ref{0658_UVOT}).

\begin{figure*}
\centering
\subfloat[IceCube-171106A/87GB 223537.9+070825 (Section~\ref{sec:results_ehe171106A})]{%
  \includegraphics[width=0.45\textwidth]{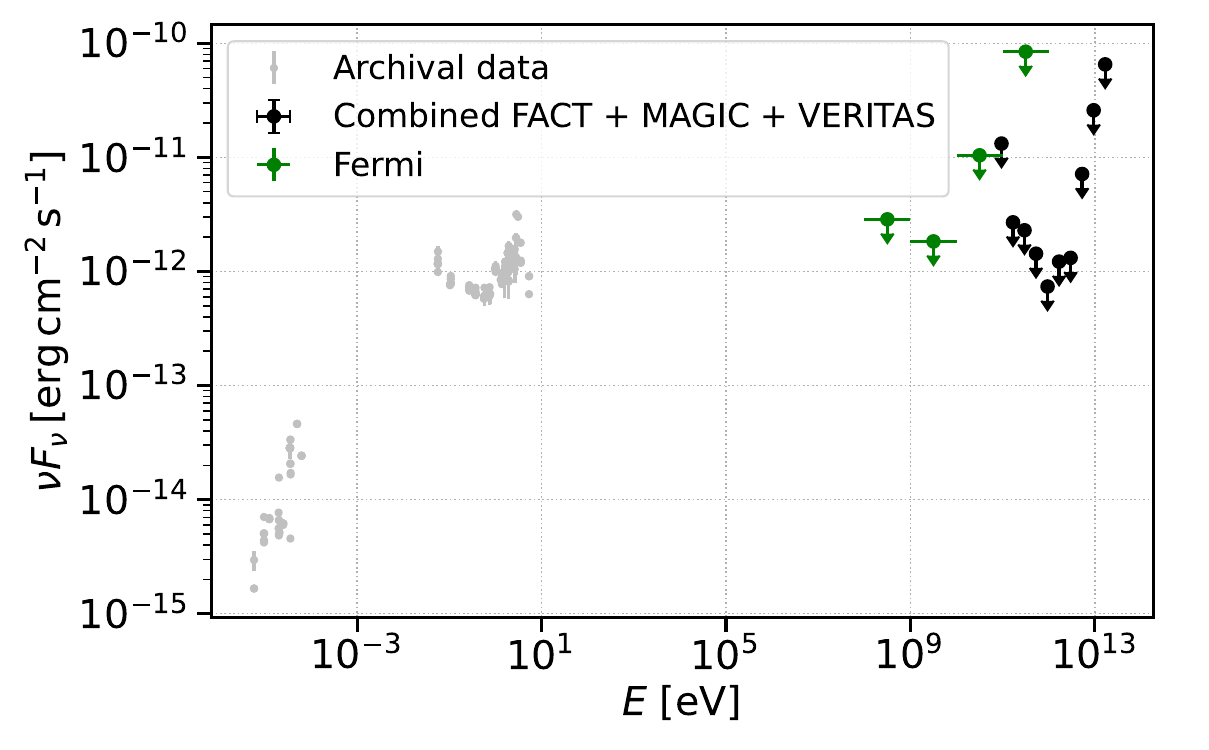}%
  \label{fig:IC171106A}%
} 
\subfloat[IceCube-190730A/PKS 1502+106 (Section~\ref{sec:results_ic190730})]{%
  \includegraphics[width=0.45\textwidth]{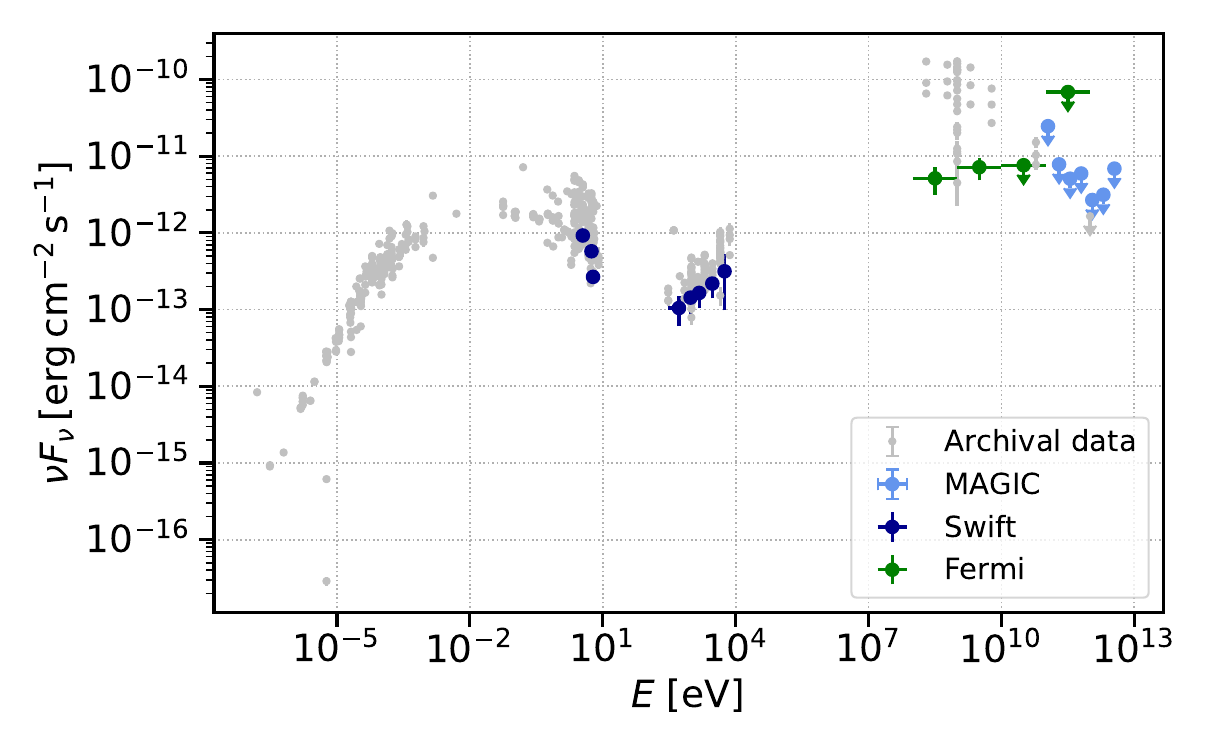}%
  \label{fig:IC190730A}%
}\\
\subfloat[IceCube-190922B/AT2019pqh (Section~\ref{sec:results_ic190922})]{%
  \includegraphics[width=0.45\textwidth]{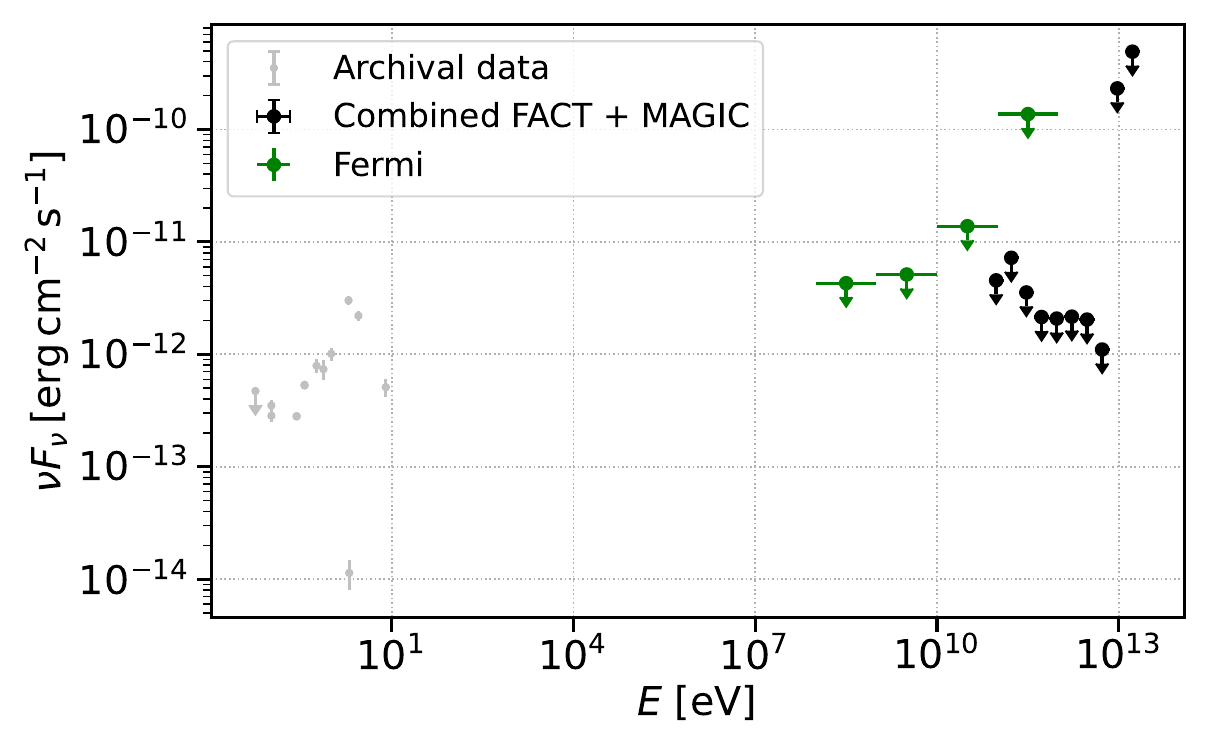}%
  \label{fig:IC190922B}%
} 
\subfloat[IceCube-191001A/AT2019dsg (Section~\ref{sec:results_ic191001})]{%
  \includegraphics[width=0.45\textwidth]{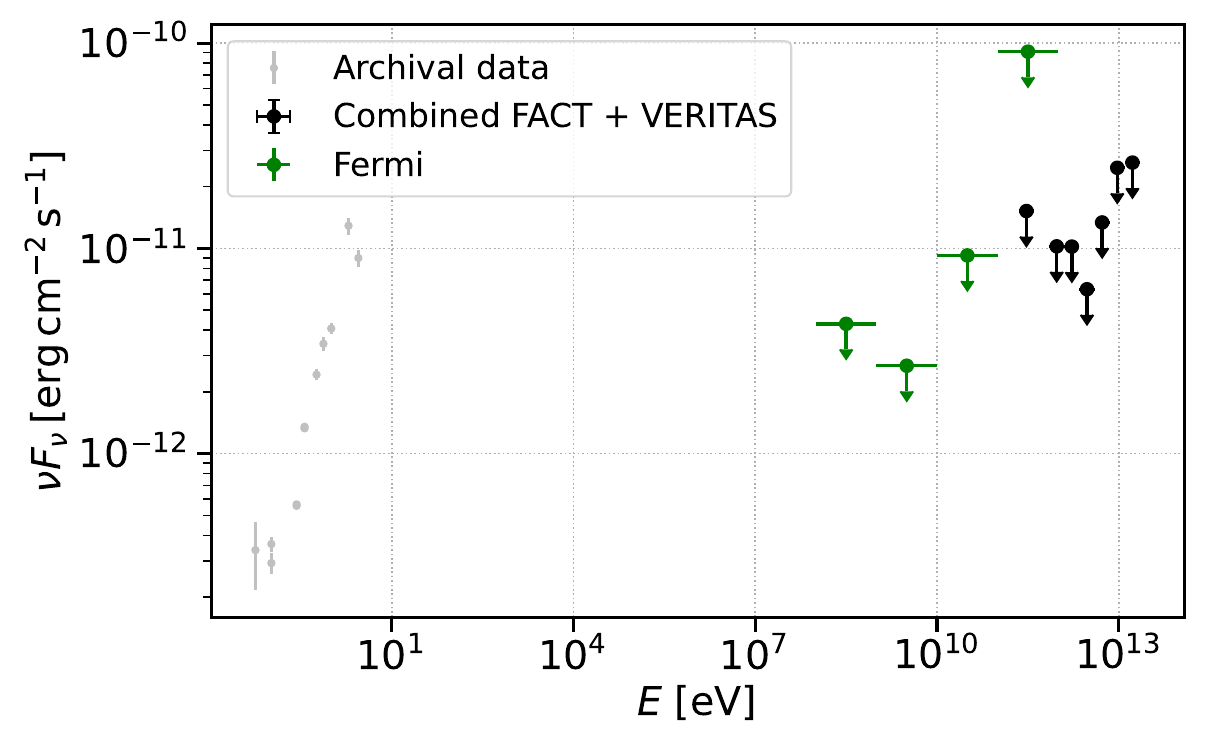}%
  \label{fig:IC191001A}%
}\\
\subfloat[IceCube-200107A/4FGL J0955.1+3551 (Section~\ref{sec:results_ic200107})]{%
  \includegraphics[width=0.45\textwidth]{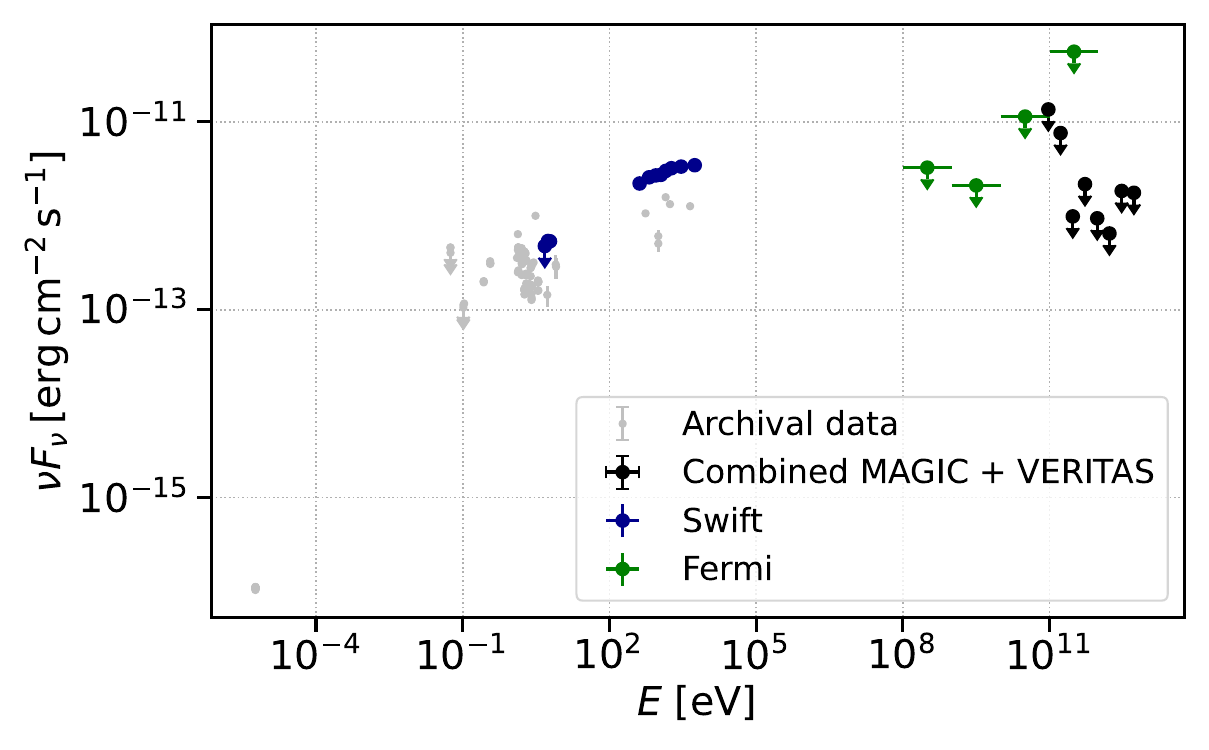}%
  \label{fig:IC200107A}%
} 
\subfloat[IceCube-201114A/4FGL J0658.6+0636 (Section~\ref{sec:results_ic201114})]{%
  \includegraphics[width=0.45\textwidth]{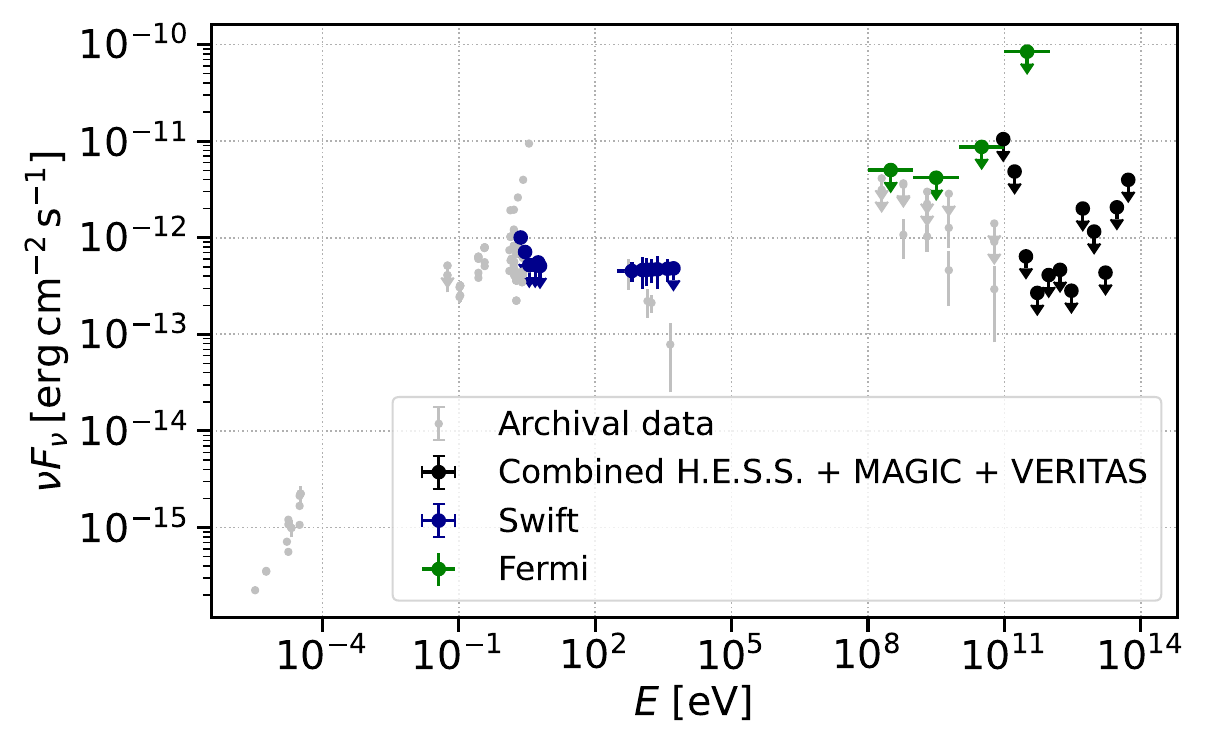}%
  \label{fig:IC201114A}%
}
\caption{SEDs for the potential counterparts of the single high-energy neutrino alerts. They comprise IACT ULs and simultaneous MWL data, together with archival data~\citep[from ASI ASDC,][]{Stratta2011Mar} provided for comparison.}
\label{fig:sed_tracks}
\end{figure*}

The MWL SED is shown in Fig.~\ref{fig:IC201114A}.

\subsection{IceCube-201222A}\label{sec:app_ic201222}
 
VERITAS followed up the alert on December 22, 2020, collecting 1~hour of quality-selected data taken at an average zenith angle of 39.2$^{\circ}$. Observations were performed using the standard wobble observation mode with a 0.7$^{\circ}$ offset in each of four cardinal directions. The upper limit on the integral flux above an energy threshold of 200\,GeV at the IceCube neutrino best-fit location is $1.09\times10^{-11}\mathrm{cm^{-2}}\,\mathrm{s^{-1}}$. An integral-flux upper-limit map and IceCube localization region are shown in Fig.~\ref{fig:IC201222A_VERITAS}.

\section{Summary and Outlook} 
\label{sec:discussion}

In this work, we have presented results from the neutrino ToO programs of the FACT, H.E.S.S., MAGIC, and VERITAS experiments.
No associations were found between $\gamma$-ray sources and observed neutrino events. Neutrino ToO programs have evolved significantly over the years as more has been learned about the properties of the astrophysical neutrino flux and its potential sources, and as new alert streams have been made available by IceCube (e.g., new GOLD and BRONZE single high-energy neutrino alerts and the upgraded GFU program), offering a unique opportunity to probe the possible hadronic nature of gamma-ray sources, which cannot be conclusively achieved with electromagnetic observations alone. Within this context, each IACT collaboration implemented its own observing priorities, and the resulting observations turn out to be complementary to each other (see Section~\ref{vhe-obs}). 
These efforts are an integral part of the current long-term science programs of the IACT observatories. The significant observation times allocated to neutrino ToO follow-ups enable the coverage of a large parameter range in terms of observed events, delay times between the neutrino event(s), and observation durations.

The IACTs conducted follow-up observations of six GFU-cluster alerts and one all-sky cluster alert from 2019 to 2020. 
The differential-flux upper limits given by the IACTs in the TeV $\gamma$-ray regime together with the X-ray observations can be used to constrain maximum contributions from photo-hadronic interactions. The combined upper limits from all IACTs increase our constraints on SED models.  

Furthermore, the IACTs performed follow-up observations of eleven IceCube single high-energy neutrino events between September 2017 and January 2021. These events have a relatively high signalness (compared to low energy events) and small uncertainty regions $\mathcal{O}(1^{\circ}$). In addition, potential counterpart sources are found within the uncertainty region of six events. It is an ideal data sample for IACTs to search for TeV $\gamma$-ray signals from photo-hadronic interactions. Integral-flux upper-limit maps are produced to cover the entire uncertainty regions of seven IceCube single high-energy events. 

The main purpose of this work is to serve as a legacy data set. While detailed SED modeling for all sources is beyond the scope of this paper, PKS~1502+106 was used as an example to discuss the potential effects as well as the limitation of our results on the current modeling work. The VHE $\gamma$-ray data detected from the source and presented here do not provide strong constraints on either of these models, as discussed in~Section \ref{sec:results_ic190730}. These results are a possible starting point for further modeling.

Looking to the future, the geographical distribution of the observatories in latitude has enabled full-sky coverage across the Northern and Southern hemispheres, and their location in longitude expands the field of regard (total sky area that can be observed by the telescope) of the combined IACT network, increasing the likelihood of prompt follow-up observations in cases where the visibility from one observatory site is constrained by weather, sunlight, bright moonlight, or technical issues. This aspect is critically important to enable VHE $\gamma$-ray observations of rare transient neutrino candidate events or in the search for other time-domain or multi-messenger triggers, as has been demonstrated recently by follow-up observations of gravitational wave events \citep[e.g., GW170817,][]{LIGOScientific:2017ync} and $\gamma$-ray bursts detectable in the VHE range (see \citealt{galaxies10030067} for an overview of recent detections). This underscores the value of conducting analyses combining all available IACT data as presented in this paper.

Follow-up programs are ongoing, and further analyses will prove highly beneficial in the search for VHE counterparts to neutrino events. The full adoption of common high-level data formats \citep[e.g., ``GADF,''][]{GADF_paper} and analysis tools \citep[e.g., ``Gammapy,''][]{2023arXiv230813584D} will simplify access and enable joint analyses that combine data from multiple IACTs. 

Given the absence of any clear association between multi-wavelength signatures and neutrino alerts in this study, these broad searches need to continue in the future. The upcoming Cherenkov Telescope Array Observatory (CTAO) has included neutrino follow-up observations as part of the high-priority Key Science Projects to be conducted in the early years of the observatory~\citep{Science_with_CTA}. With telescopes in both Northern and Southern Hemispheres, CTAO’s significant advancement in sensitivity and full-sky coverage promise exciting breakthroughs in these searches~\citep{Konstancja_ICRC2021}.

Further improvements are expected in the near future with the onset of science operations of the KM3NeT neutrino telescope  in the Mediterranean Sea \citep{Margiotta:2022kid}, further extensions to the Gigaton Volume Detector (GVD) installation in Lake Baikal~\citep{Baikal-GVD:2022fis}, and other proposed neutrino telescopes such as P-ONE~\citep{Resconi:2021ezb} and  TRIDENT~\citep{Ye:2022vbk}. These facilities will be able to identify more astrophysical neutrino candidate events, improve the angular resolution of the overall dataset, and, due to their location in the Northern Hemisphere, will offer a view of the Galactic Plane complementary to that provided by IceCube. 
The increased volume of the next generation IceCube detector, IceCube-Gen2~\citep{IceCube-Gen2:2020qha}, will significantly improve
its sensitivity to high-energy neutrinos in the next decade, thereby improving multi-messenger searches like the ones presented here.

\section{Data availability}
Data products presented in this paper can be downloaded via Zenodo (\href{https://doi.org/10.5281/zenodo.17238703}{DOI:10.5281/zenodo.17238703}). The repository contains the following data products:

- 1ES1312-423 MWL SED (Fig.~\ref{fig:1es1312_sed}).

- SEDs for the counterparts of the GFU-cluster alerts including IACT ULs, simultaneous MWL data and archival data provided for comparison (Fig.~\ref{fig:sed_multiplets}). 

- Differential-flux upper limits of each IACT and combined limits (Fig.~\ref{fig:ul_combined}).

- Fits files for the integral VHE $\gamma$-ray flux upper-limit maps (Fig.~\ref{fig:UL_maps}).

- SEDs for the potential counterparts of the single high-energy neutrino alerts (Fig.~\ref{fig:sed_tracks}).

\section*{acknowledgments}
\subsection*{FACT}
The important contributions from ETH Zurich grants ETH-10.08-2 and ETH-27.12-1 as well as the funding by the Swiss SNF and the German BMBF (Verbundforschung Astro- und Astroteilchenphysik) and HAP (Helmoltz Alliance for Astroparticle Physics) are gratefully acknowledged. Part of this work is supported by Deutsche Forschungsgemeinschaft (DFG) within the Collaborative Research Center SFB 876 "Providing Information by Resource-Constrained Analysis", project C3. We are thankful for the very valuable contributions from E.~Lorenz, D.~Renker and G.~Viertel during the early phase of the project. We thank the Instituto de Astrofísica de Canarias for allowing us to operate the telescope at the Observatorio del Roque de los Muchachos in La Palma, the Max-Planck-Institut für Physik for providing us with the mount of the former HEGRA CT3 telescope, and the MAGIC collaboration for their support.

\emph{Author contributions}:  
Within FACT, this project was led and coordinated by D.~Dorner, the analysis was carried out by B.~Schleicher and both discussed the results and contributed to the manuscript. T.~Bretz designed and implemented the automatic follow-up procedure in close collaboration with D.~Dorner. The rest of the collaboration contributed by other means to the design, construction and/or operation of the FACT project (telescope and everything ensuring the scientific output).

\subsection*{H.E.S.S.}
The support of the Namibian authorities and of the University of Namibia in facilitating the construction and operation of H.E.S.S. is gratefully acknowledged, as is the support by the German Ministry for Education and Research (BMBF), the Max Planck Society, the Helmholtz Association, the French Ministry of Higher Education, Research and Innovation, the Centre National de la Recherche Scientifique (CNRS/IN2P3 and CNRS/INSU), the Commissariat à l’énergie atomique et aux énergies alternatives (CEA), the U.K. Science and Technology Facilities Council (STFC), the Polish Ministry of Education and Science, agreement no. 2021/WK/06, the South African Department of Science and Innovation and National Research Foundation, the University of Namibia, the National Commission on Research, Science \& Technology of Namibia (NCRST), the Austrian Federal Ministry of Education, Science and Research and the Austrian Science Fund (FWF), the Australian Research Council (ARC), the Japan Society for the Promotion of Science, the University of Amsterdam and the Science Committee of Armenia grant 21AG-1C085. We appreciate the excellent work of the technical support staff in Berlin, Zeuthen, Heidelberg, Palaiseau, Paris, Saclay, Tübingen and in Namibia in the construction and operation of the equipment. This work benefited from services provided by the H.E.S.S. Virtual Organisation, supported by the national resource providers of the EGI Federation.

\subsection*{MAGIC}
We would like to thank the Instituto de Astrof\'{\i}sica de Canarias for the excellent working conditions at the Observatorio del Roque de los Muchachos in La Palma. The financial support of the German BMFTR, MPG and HGF; the Italian INFN and INAF; the Swiss National Fund SNF; the grants PID2019-107988GB-C22, PID2022-136828NB-C41, PID2022-137810NB-C22, PID2022-138172NB-C41, PID2022-138172NB-C42, PID2022-138172NB-C43, PID2022-139117NB-C41, PID2022-139117NB-C42, PID2022-139117NB-C43, PID2022-139117NB-C44, CNS2023-144504 funded by the Spanish MCIN/AEI/ 10.13039/501100011033 and "ERDF A way of making Europe; the Indian Department of Atomic Energy; the Japanese ICRR, the University of Tokyo, JSPS, and MEXT; the Bulgarian Ministry of Education and Science, National RI Roadmap Project DO1-400/18.12.2020 and the Academy of Finland grant nr. 320045 is gratefully acknowledged. This work was also been supported by Centros de Excelencia ``Severo Ochoa'' y Unidades ``Mar\'{\i}a de Maeztu'' program of the Spanish MCIN/AEI/ 10.13039/501100011033 (CEX2019-000920-S, CEX2019-000918-M, CEX2021-001131-S) and by the CERCA institution and grants 2021SGR00426 and 2021SGR00773 of the Generalitat de Catalunya; by the Croatian Science Foundation (HrZZ) Project IP-2022-10-4595 and the University of Rijeka Project uniri-prirod-18-48; by the Deutsche Forschungsgemeinschaft (SFB1491) and by the Lamarr-Institute for Machine Learning and Artificial Intelligence; by the Polish Ministry Of Education and Science grant No. 2021/WK/08; and by the Brazilian MCTIC, the CNPq Productivity Grant 309053/2022-6 and FAPERJ Grants E-26/200.532/2023 and E-26/211.342/2021.

\smallskip
\emph{Author contributions}:  
E. Bernardini: PI of MAGIC Neutrino ToO proposal, supervision, paper drafting and editing.
D. Miceli: production of MAGIC UL skymap.
K. Satalecka: former project leader, paper drafting.
I. Viale: project management, MAGIC data analysis, coordination of multi-wavelength results, paper drafting and editing.
M. Artero, A. Berti, H. Bökenkamp, L. Di Venere, A. Fattorini, L. Heckmann, S. Mangano, H. A. Mondal, K. Noda, S. Sakurai, S. Yoo: MAGIC data analysis.
The rest of the authors have contributed in one or several of the following ways: design, construction, maintenance and operation of the instrument(s); preparation and/or evaluation of the observation proposals; data acquisition, processing, calibration and/or reduction; production of analysis tools and/or related Monte Carlo simulations; discussion and approval of the contents of the draft.

\subsection*{VERITAS}
VERITAS is supported by grants from the U.S. Department of Energy Office of Science, the U.S. National Science Foundation and the Smithsonian Institution, by NSERC in Canada, and by the Helmholtz Association in Germany. This research used resources provided by the Open Science Grid, which is supported by the National Science Foundation and the U.S. Department of Energy's Office of Science, and resources of the National Energy Research Scientific Computing Center (NERSC), a U.S. Department of Energy Office of Science User Facility operated under Contract No. DE-AC02-05CH11231. We acknowledge the excellent work of the technical support staff at the Fred Lawrence Whipple Observatory and at the collaborating institutions in the construction and operation of the instrument.

\subsection*{IceCube}
The authors gratefully acknowledge the support from the following agencies and institutions:
USA {\textendash} U.S. National Science Foundation-Office of Polar Programs,
U.S. National Science Foundation-Physics Division,
U.S. National Science Foundation-EPSCoR,
U.S. National Science Foundation-Office of Advanced Cyberinfrastructure,
Wisconsin Alumni Research Foundation,
Center for High Throughput Computing (CHTC) at the University of Wisconsin{\textendash}Madison,
Open Science Grid (OSG),
Partnership to Advance Throughput Computing (PATh),
Advanced Cyberinfrastructure Coordination Ecosystem: Services {\&} Support (ACCESS),
Frontera and Ranch computing project at the Texas Advanced Computing Center,
U.S. Department of Energy-National Energy Research Scientific Computing Center,
Particle astrophysics research computing center at the University of Maryland,
Institute for Cyber-Enabled Research at Michigan State University,
Astroparticle physics computational facility at Marquette University,
NVIDIA Corporation,
and Google Cloud Platform;
Belgium {\textendash} Funds for Scientific Research (FRS-FNRS and FWO),
FWO Odysseus and Big Science programmes,
and Belgian Federal Science Policy Office (Belspo);
Germany {\textendash} Bundesministerium f{\"u}r Bildung und Forschung (BMBF),
Deutsche Forschungsgemeinschaft (DFG),
Helmholtz Alliance for Astroparticle Physics (HAP),
Initiative and Networking Fund of the Helmholtz Association,
Deutsches Elektronen Synchrotron (DESY),
and High Performance Computing cluster of the RWTH Aachen;
Sweden {\textendash} Swedish Research Council,
Swedish Polar Research Secretariat,
Swedish National Infrastructure for Computing (SNIC),
and Knut and Alice Wallenberg Foundation;
European Union {\textendash} EGI Advanced Computing for research;
Australia {\textendash} Australian Research Council;
Canada {\textendash} Natural Sciences and Engineering Research Council of Canada,
Calcul Qu{\'e}bec, Compute Ontario, Canada Foundation for Innovation, WestGrid, and Digital Research Alliance of Canada;
Denmark {\textendash} Villum Fonden, Carlsberg Foundation, and European Commission;
New Zealand {\textendash} Marsden Fund;
Japan {\textendash} Japan Society for Promotion of Science (JSPS)
and Institute for Global Prominent Research (IGPR) of Chiba University;
Korea {\textendash} National Research Foundation of Korea (NRF);
Switzerland {\textendash} Swiss National Science Foundation (SNSF).

\subsection*{Fermi-LAT}
The {\em Fermi}-LAT Collaboration acknowledges generous ongoing support from a number of agencies and institutes that have supported both the development and the operation of the LAT as well as scientific data analysis. These include the National Aeronautics and Space Administration and the Department of Energy in the United States, the Commissariat \`a l'Energie Atomique and the Centre National de la Recherche Scientifique / Institut National de Physique Nucl\'eaire et de Physique des Particules in France, the Agenzia Spaziale Italiana and the Istituto Nazionale di Fisica Nucleare in Italy, the Ministry of Education,Culture, Sports, Science and Technology (MEXT), High Energy Accelerator Research Organization (KEK) and Japan Aerospace Exploration Agency (JAXA) in Japan, and the K.~A.~Wallenberg Foundation, the Swedish Research Council and the Swedish National Space Board in Sweden. Additional support for science analysis during the operations phase is gratefully acknowledged from the Istituto Nazionale di Astrofisica in Italy and the Centre National d'\'Etudes Spatiales in France.
\\
\newline
We acknowledge support by Institut Pascal at Université Paris-Saclay during the Paris-Saclay Astroparticle Symposium 2021 and 2022, with the support of the P2IO Laboratory of Excellence (program “Investissements d’avenir” ANR-11-IDEX-0003-01 Paris-Saclay and ANR-10-LABX-0038), the P2I axis of the Graduate School Physics of Université Paris-Saclay, as well as IJCLab, CEA, IPhT, IAS, OSUPS, the IN2P3 master projet UCMN, APPEC, and EuCAPT. 

Fabian Sch\"ussler acknowledges the support of the French Agence Nationale de la Recherche (ANR), under grant ANR-22-CE31-0012 (project MOTS) and support by the Programme National des Hautes Energies of CNRS/INSU with INP and IN2P3, co-funded by CEA and CNES. 

This work was supported by the German Science Foundation DFG, research grant “Relativistic Jets in Active Galaxies” (FOR 5195, grant No. 443220636), and by the European Research Council, ERC Starting grant \emph{MessMapp}, Sara Buson Principal Investigator, under contract no. 949555. 

This work performed in part under DOE Contract DE-AC02-76SF00515.

Stefano Marchesi acknowledges support by the Next Generation EU funds within the National Recovery and Resilience Plan (PNRR), Mission 4 - Education and Research, Component 2 - From Research to Business (M4C2), Investment Line 3.1 - Strengthening and creation of Research Infrastructures, Project IR0000012 – ``CTA+ - Cherenkov Telescope Array Plus''.

Ilaria Viale and Elisa Prandini acknowledge the project “SKYNET: Deep Learning for Astroparticle Physics”, PRIN 2022 (CUP: D53D23002610006).

\software{Astro-COLIBRI~\citep{2021ApJS..256....5R}; Astropy \citep{astropy:2013, astropy:2018, astropy:2022}; gammapy~\citep{2023arXiv230813584D}; fermipy \citep{Wood:2018};}

\bibliography{IC-IACT_NeutrinoToOs}{}
\bibliographystyle{aasjournal}

\newpage
\appendix
\label{appendix}

\section{{\em Swift} results}
\label{appendix:swift}

\begin{table}[ht]
\caption{Logs and fit results of {\em Swift}-XRT observations of 1ES\,1312-423 using a power-law model with $N_{\rm H}$ fixed to Galactic absorption.}
\begin{center}
\begin{tabular}{ccccc}
\hline \hline
\multicolumn{1}{c}{{\it Swift} ObsID} &
\multicolumn{1}{c}{Observation} &
\multicolumn{1}{c}{Net Exposure Time} &
\multicolumn{1}{c}{Photon index} &
\multicolumn{1}{c}{Flux 0.3-10 keV} \\
\multicolumn{1}{c}{} &
\multicolumn{1}{c}{Date} &
\multicolumn{1}{c}{second} &
\multicolumn{1}{c}{$\Gamma$} &
\multicolumn{1}{c}{$\times$10$^{-11}$ erg cm$^{-2}$ s$^{-1}$} \\
\hline
00031915001 & 2011-01-25 &  4772 & 1.77 $\pm$ 0.05 &  4.03 $\pm$ 0.14 \\
00031915002 & 2019-03-12 &  1753 & 1.97 $\pm$ 0.08 &  3.77 $\pm$ 0.21 \\
00031915003 & 2019-03-13 &  1753 & 1.91 $\pm$ 0.08 &  4.15 $\pm$ 0.25 \\
00031915005 & 2019-03-15 &  1806 & 2.15 $\pm$ 0.10 &  4.21 $\pm$ 0.31 \\ 
00031915007 & 2019-04-17 &  3142 & 1.89 $\pm$ 0.06 &  5.43 $\pm$ 0.22 \\
00031915009 & 2019-04-19 &  3841 & 2.01 $\pm$ 0.05 &  4.14 $\pm$ 0.15 \\
\hline                  
\end{tabular}           
\end{center}            
\label{1312_XRT}       
\end{table}

\begin{table}[ht]
\caption{Observed magnitudes of 1ES~1312-423 obtained by {\em Swift}/UVOT.}
\begin{center}
\begin{tabular}{ccccccc}
\hline \hline
\multicolumn{1}{c}{Observation} &
\multicolumn{1}{c}{$v$}  &
\multicolumn{1}{c}{$b$}  &
\multicolumn{1}{c}{$u$}  &
\multicolumn{1}{c}{$w1$} &
\multicolumn{1}{c}{$m2$} &
\multicolumn{1}{c}{$w2$} \\
\multicolumn{1}{c}{Date} &
\multicolumn{1}{c}{mag}  &
\multicolumn{1}{c}{mag}  &
\multicolumn{1}{c}{mag}  &
\multicolumn{1}{c}{mag}  &
\multicolumn{1}{c}{mag}  &
\multicolumn{1}{c}{mag}  \\
\hline
2011-01-25 & -                & -                & -                & -                & 17.41 $\pm$ 0.06  & - \\
2019-03-12 & -                & -                & -                & -                & 16.74 $\pm$ 0.06  & - \\
2019-03-13 & -                & -                & -                & 16.74 $\pm$ 0.06 & -                 & - \\
2019-03-15 & -                & -                & -                & -                & -                 & 16.78 $\pm$ 0.06 \\
2019-04-17 & 16.47 $\pm$ 0.06 & 17.15 $\pm$ 0.06 & 16.36 $\pm$ 0.06 & 16.43 $\pm$ 0.07 & 16.45 $\pm$ 0.07  & 16.57 $\pm$ 0.07 \\
2019-04-19 & 16.54 $\pm$ 0.06 & 17.23 $\pm$ 0.06 & 16.56 $\pm$ 0.06 & 16.58 $\pm$ 0.07 & 16.66 $\pm$ 0.06  & 16.73 $\pm$ 0.07 \\
\hline
\end{tabular}
\end{center}
\label{1312_UVOT}
\end{table}

\begin{table}[ht]
\caption{Logs and fit results of {\em Swift}-XRT observations of OP\,313 using a power-law model with $N_{\rm H}$ fixed to Galactic absorption.}
\begin{center}
\begin{tabular}{ccccc}
\hline \hline
\multicolumn{1}{c}{{\it Swift} ObsID} &
\multicolumn{1}{c}{Observation} &
\multicolumn{1}{c}{Net Exposure Time} &
\multicolumn{1}{c}{Photon index} &
\multicolumn{1}{c}{Flux 0.3-10 keV} \\
\multicolumn{1}{c}{} &
\multicolumn{1}{c}{Date} &
\multicolumn{1}{c}{second} &
\multicolumn{1}{c}{$\Gamma$} &
\multicolumn{1}{c}{$\times$10$^{-12}$ erg cm$^{-2}$ s$^{-1}$} \\
\hline
00036384001 & 2007-04-03    & 2130  & 1.47 $\pm$ 0.17   &    4.31  $\pm$ 0.66  \\
00036384002 & 2007-04-07    & 2793  & 1.52 $\pm$ 0.15   &    3.99  $\pm$ 0.50  \\
00036384003 & 2007-06-27    & 465   & 1.63 $\pm$ 0.34   &    4.59  $\pm$ 1.20  \\
00030976001	& 2007-07-31    & 852   & 1.82 $\pm$ 0.30   &    2.97  $\pm$ 0.65  \\
00030976002	& 2007-08-01    & 4409  & 1.53 $\pm$ 0.15   &    3.09  $\pm$ 0.36  \\
00036768001 & 2007-08-05    & 4181  & 1.49 $\pm$ 0.12   &    4.61  $\pm$ 0.43  \\
00036768002, 00036384006 & 2008-05-12    & 7911  & 1.58 $\pm$ 0.11   &    2.58  $\pm$ 0.23  \\         
00036384007 & 2008-08-20    & 4755  & 1.48 $\pm$ 0.13   &    3.76  $\pm$ 0.40  \\
00036384008 & 2009-12-12    & 2303  & 1.66 $\pm$ 0.26   &    1.70  $\pm$ 0.33  \\
00036384009 & 2010-04-15    & 1975  & 1.47 $\pm$ 0.33   &    2.40  $\pm$ 0.62  \\
00036384010, 0003638411 & 2011-04-17    & 3249  & 1.61 $\pm$ 0.28   &    1.16  $\pm$ 0.25  \\
00036384012 & 2011-07-03    & 4722  & 1.90 $\pm$ 0.21   &    1.21  $\pm$ 0.17  \\
00091894001 & 2014-05-04    & 3486  & 1.65 $\pm$ 0.16   &    2.72  $\pm$ 0.40  \\
00091894004	& 2014-06-29    & 3494  & 1.60 $\pm$ 0.18   &    2.22  $\pm$ 0.30  \\
00036384013	& 2019-06-20    & 1870  & 2.03 $\pm$ 0.31   &    1.35  $\pm$ 0.28  \\
00036384014 & 2019-12-27    & 2035  & 1.55 $\pm$ 0.24   &    2.82  $\pm$ 0.48  \\
00036384015 & 2019-12-29    & 1865  & 1.53 $\pm$ 0.23   &    2.79  $\pm$ 0.52  \\
00036384016 & 2019-12-31    & 1249  & 1.47 $\pm$ 0.28   &    3.09  $\pm$ 0.69  \\
00036384017 & 2020-03-11    & 1641  & 1.70 $\pm$ 0.22   &    3.41  $\pm$ 0.55  \\
00036384018 & 2020-03-16    & 2238  & 1.67 $\pm$ 0.23   &    2.56  $\pm$ 0.45  \\
00036384019 & 2020-03-21    & 2218  & 1.50 $\pm$ 0.22   &    2.85  $\pm$ 0.51  \\
00036384020 & 2020-05-21    & 2492  & 1.61 $\pm$ 0.22   &    2.25  $\pm$ 0.38  \\
00036384021 & 2020-05-25    & 2230  & 1.59 $\pm$ 0.25   &    2.22  $\pm$ 0.40  \\
00036384022 & 2020-05-29    & 2048  & 1.70 $\pm$ 0.23   &    2.44  $\pm$ 0.41  \\ 
00036384023 & 2020-06-02    & 2794  & 1.48 $\pm$ 0.23   &    2.36  $\pm$ 0.42  \\
00036384024 & 2020-06-06    & 2899  & 1.51 $\pm$ 0.20   &    2.44  $\pm$ 0.39  \\
00036384025 &  2020-06-10    & 3037  & 1.56 $\pm$ 0.22   &    2.02  $\pm$ 0.35  \\
00036384026 & 2020-06-14    & 2737  & 1.68 $\pm$ 0.20   &    2.29  $\pm$ 0.35  \\
\hline                  
\end{tabular}           
\end{center}            
\label{OP313_XRT}       
\end{table}

\begin{table}[ht]
\caption{Observed magnitudes of OP\,313 obtained by {\em Swift}/UVOT.}
\begin{center}
\begin{tabular}{ccccccc}
\hline \hline
\multicolumn{1}{c}{Observation} &
\multicolumn{1}{c}{$v$}  &
\multicolumn{1}{c}{$b$}  &
\multicolumn{1}{c}{$u$}  &
\multicolumn{1}{c}{$w1$} &
\multicolumn{1}{c}{$m2$} &
\multicolumn{1}{c}{$w2$} \\
\multicolumn{1}{c}{Date} &
\multicolumn{1}{c}{mag}  &
\multicolumn{1}{c}{mag}  &
\multicolumn{1}{c}{mag}  &
\multicolumn{1}{c}{mag}  &
\multicolumn{1}{c}{mag}  &
\multicolumn{1}{c}{mag}  \\
\hline
2007-04-03 & 16.91 $\pm$ 0.08 & 17.39 $\pm$ 0.07 & 16.56 $\pm$ 0.07 & 16.58 $\pm$ 0.08 & 16.66 $\pm$ 0.08 & 16.96 $\pm$ 0.07 \\
2007-04-07 & 17.58 $\pm$ 0.09 & 18.20 $\pm$ 0.08 & 17.25 $\pm$ 0.07 & 17.20 $\pm$ 0.08 & 17.22 $\pm$ 0.09 & 17.57 $\pm$ 0.07 \\
2007-06-27 &          -       &          -       &           -      &             -    & 17.46 $\pm$ 0.10 &         -        \\
2007-07-31 & 16.56 $\pm$ 0.16 & 16.98 $\pm$ 0.12 & 16.20 $\pm$ 0.07 & 16.28 $\pm$ 0.06 & 16.15 $\pm$ 0.14 & 16.66 $\pm$ 0.12 \\
2007-08-01 & 16.73 $\pm$ 0.06 & 17.24 $\pm$ 0.06 & 16.45 $\pm$ 0.06 & 16.55 $\pm$ 0.06 & 16.66 $\pm$ 0.04 & 16.93 $\pm$ 0.06 \\
2008-05-12 &          -       &          -       &           -      &             -    & 17.32 $\pm$ 0.06 &         -        \\
2008-08-20 &          -       &          -       &           -      &             -    & 16.70 $\pm$ 0.06 &         -        \\ 
2009-12-12 &          -       &          -       &           -      &             -    &         -        & 18.59 $\pm$ 0.08 \\
2010-04-15 &          -       &          -       &           -      &             -    &         -        & 18.39 $\pm$ 0.08 \\
2011-04-17 & 18.85 $\pm$ 0.31 & 19.82 $\pm$ 0.31 & 18.22 $\pm$ 0.14 & 18.12 $\pm$ 0.14 & 18.08 $\pm$ 0.14 & 18.68 $\pm$ 0.13 \\ 
2011-04-17 &        -         &          -       & 18.40 $\pm$ 0.07 &           -      &             -    & 18.53 $\pm$ 0.12 \\
2011-07-03 & 18.10 $\pm$ 0.15 & 18.71 $\pm$ 0.13 & 17.72 $\pm$ 0.11 & 17.64 $\pm$ 0.11 & 17.74 $\pm$ 0.11 & 18.21 $\pm$ 0.07 \\  
2014-05-04 & 18.27 $\pm$ 0.08 & 17.91 $\pm$ 0.10 & 18.03 $\pm$ 0.16 & 18.16 $\pm$ 0.17 & 19.31 $\pm$ 0.26 & 18.28 $\pm$ 0.25 \\
2014-06-05 &        -         & 19.48 $\pm$ 0.26 & 18.20 $\pm$ 0.15 & 18.09 $\pm$ 0.15 &         -        & 18.57 $\pm$ 0.08 \\
2014-06-10 & $>$ 19.09        & 19.50 $\pm$ 0.24 & 19.58 $\pm$ 0.18 & 18.29 $\pm$ 0.16 & 18.08 $\pm$ 0.08 & 18.65 $\pm$ 0.14 \\
2014-06-29 & 18.91 $\pm$ 0.33 & 19.79 $\pm$ 0.31 & 18.20 $\pm$ 0.15 & 18.22 $\pm$ 0.15 & 18.15 $\pm$ 0.17 & 18.50 $\pm$ 0.08 \\
2019-06-20 & 16.52 $\pm$ 0.07 & 17.05 $\pm$ 0.07 & 16.31 $\pm$ 0.07 & 16.35 $\pm$ 0.08 & 16.35 $\pm$ 0.08 & 16.70 $\pm$ 0.07 \\
2019-12-27 & 16.51 $\pm$ 0.07 & 16.99 $\pm$ 0.07 & 16.16 $\pm$ 0.06 & 16.35 $\pm$ 0.08 & 16.34 $\pm$ 0.08 & 16.57 $\pm$ 0.07 \\
2019-12-29 & 16.52 $\pm$ 0.08 & 16.89 $\pm$ 0.06 & 16.20 $\pm$ 0.06 & 16.27 $\pm$ 0.08 & 16.27 $\pm$ 0.08 & 16.58 $\pm$ 0.07 \\
2019-12-31 & 16.43 $\pm$ 0.07 & 16.80 $\pm$ 0.06 & 16.20 $\pm$ 0.06 & 16.16 $\pm$ 0.07 & 16.44 $\pm$ 0.08 & 16.87 $\pm$ 0.08 \\
2020-03-11 &          -       &          -       &          -       & 15.84 $\pm$ 0.06 &          -       &          -       \\
2020-03-16 &          -       &          -       & 16.08 $\pm$ 0.05 &         -        &          -       &          -       \\
2020-03-21 &          -       &          -       &          -       &         -        &          -       & 16.32 $\pm$ 0.06 \\
2020-05-21 & 17.18 $\pm$ 0.09 & 17.87 $\pm$ 0.08 & 17.07 $\pm$ 0.08 & 17.14 $\pm$ 0.09 & 17.17 $\pm$ 0.09 & 17.45 $\pm$ 0.08 \\
2020-05-25 & 17.16 $\pm$ 0.09 & 17.76 $\pm$ 0.08 & 16.99 $\pm$ 0.08 & 17.03 $\pm$ 0.09 & 17.24 $\pm$ 0.10 & 17.52 $\pm$ 0.09 \\
2020-05-29 & 17.14 $\pm$ 0.09 & 17.68 $\pm$ 0.08 & 16.94 $\pm$ 0.08 & 16.94 $\pm$ 0.09 & 16.96 $\pm$ 0.11 & 17.28 $\pm$ 0.08 \\
2020-06-02 & 17.29 $\pm$ 0.09 & 17.73 $\pm$ 0.07 & 16.88 $\pm$ 0.07 & 17.00 $\pm$ 0.08 & 17.04 $\pm$ 0.09 & 17.24 $\pm$ 0.08 \\
2020-06-06 & 17.28 $\pm$ 0.09 & 17.90 $\pm$ 0.08 & 17.03 $\pm$ 0.07 & 17.06 $\pm$ 0.09 & 17.16 $\pm$ 0.09 & 17.34 $\pm$ 0.08 \\
2020-06-10 & 17.64 $\pm$ 0.11 & 18.08 $\pm$ 0.09 & 17.21 $\pm$ 0.08 & 17.30 $\pm$ 0.09 & 17.43 $\pm$ 0.10 & 17.72 $\pm$ 0.08 \\
2020-06-14 & 17.56 $\pm$ 0.11 & 18.03 $\pm$ 0.09 & 17.37 $\pm$ 0.08 & 17.30 $\pm$ 0.09 & 17.37 $\pm$ 0.10 & 17.59 $\pm$ 0.08 \\
 \hline           
 \end{tabular}    
 \end{center}     
 \label{OP313_UVOT}
 \end{table}

\begin{table}[ht]
\caption{Logs and fit results of {\em Swift}-XRT observations of GB6\,J0316+0904 using a power-law model with $N_{\rm H}$ fixed to Galactic absorption.}
\begin{center}
\begin{tabular}{ccccc}
\hline \hline
\multicolumn{1}{c}{{\it Swift} ObsID} &
\multicolumn{1}{c}{Observation} &
\multicolumn{1}{c}{Net Exposure Time} &
\multicolumn{1}{c}{Photon index} &
\multicolumn{1}{c}{Flux 0.3-10 keV} \\
\multicolumn{1}{c}{} &
\multicolumn{1}{c}{Date} &
\multicolumn{1}{c}{second} &
\multicolumn{1}{c}{$\Gamma$} &
\multicolumn{1}{c}{$\times$10$^{-12}$ erg cm$^{-2}$ s$^{-1}$} \\
\hline
00038370001 & 2009-03-09    & 2487   & 1.94 $\pm$ 0.10 & 16.0 $\pm$ 1.2  \\
00041581001 & 2010-11-23    &  974   & 2.23 $\pm$ 0.85 & 1.17 $\pm$ 0.45 \\
00038370002, 00038370003 & 2011-07-03/04 & 3259   & 2.27 $\pm$ 0.33 & 1.05 $\pm$ 0.23 \\     
00083407001 & 2015-01-09    & 2645   & 1.97 $\pm$ 0.24 & 2.34 $\pm$ 0.39 \\
\hline
\end{tabular}
\end{center}
\label{0316_XRT}
\end{table}

\begin{table}[ht]
\caption{Observed magnitudes of GB6 J0316+0904 obtained by {\em Swift}/UVOT.}
\begin{center}
\begin{tabular}{ccccccc}
\hline \hline
\multicolumn{1}{c}{Observation} &
\multicolumn{1}{c}{$v$}  &
\multicolumn{1}{c}{$b$}  &
\multicolumn{1}{c}{$u$}  &
\multicolumn{1}{c}{$w1$} &
\multicolumn{1}{c}{$m2$} &
\multicolumn{1}{c}{$w2$} \\
\multicolumn{1}{c}{Date} &
\multicolumn{1}{c}{mag}  &
\multicolumn{1}{c}{mag}  &
\multicolumn{1}{c}{mag}  &
\multicolumn{1}{c}{mag}  &
\multicolumn{1}{c}{mag}  &
\multicolumn{1}{c}{mag}  \\
\hline
2009-03-09    & 15.91 $\pm$ 0.12 & 16.47 $\pm$ 0.05 & 15.79 $\pm$ 0.05 & 16.37 $\pm$ 0.06 & $>$ 16.57        &  16.89 $\pm$ 0.07  \\
2010-11-23    & 16.64 $\pm$ 0.11 & 17.54 $\pm$ 0.10 & 17.02 $\pm$ 0.11 & 17.59 $\pm$ 0.14 & 18.32 $\pm$ 0.20 &  18.10 $\pm$ 0.13  \\
2011-07-03    & 16.38 $\pm$ 0.12 & 17.32 $\pm$ 0.12 & 16.53 $\pm$ 0.10 & 17.08 $\pm$ 0.12 & 17.74 $\pm$ 0.16 &  17.78 $\pm$ 0.08  \\ 
2011-07-04    & 16.62 $\pm$ 0.13 & 17.10 $\pm$ 0.10 & 16.55 $\pm$ 0.09 & 17.35 $\pm$ 0.13 & 17.73 $\pm$ 0.11 &  17.73 $\pm$ 0.11  \\     
2015-01-09    & 17.63 $\pm$ 0.12 & 18.33 $\pm$ 0.12 & 17.81 $\pm$ 0.11 & 18.39 $\pm$ 0.16 & 18.95 $\pm$ 0.19 &  19.04 $\pm$ 0.15  \\
\hline
\end{tabular}
\end{center}
\label{0316_UVOT}
\end{table}

\begin{table}[ht]
\caption{Log and fitting results of {\em Swift}-XRT observations of OC 457 using a power-law model with $N_{\rm H}$ fixed to Galactic absorption.}
\begin{center}
\begin{tabular}{ccccc}
\hline \hline
\multicolumn{1}{c}{{\it Swift} ObsID} &
\multicolumn{1}{c}{Observation} &
\multicolumn{1}{c}{Net Exposure Time} &
\multicolumn{1}{c}{Photon index} &
\multicolumn{1}{c}{Flux 0.3-10 keV} \\
\multicolumn{1}{c}{} &
\multicolumn{1}{c}{Date} &
\multicolumn{1}{c}{second} &
\multicolumn{1}{c}{$\Gamma$} &
\multicolumn{1}{c}{$\times$10$^{-12}$ erg cm$^{-2}$ s$^{-1}$} \\
\hline
00036759001 & 2007-07-16  & 7342  & 1.39 $\pm$ 0.16  & 2.10 $\pm$ 0.32    \\
00036759002 & 2007-10-09  & 1306  & 1.12 $\pm$ 0.42  & 3.47 $\pm$ 1.31    \\
00036759003 & 2007-10-10  & 3621  & 1.50 $\pm$ 0.22  & 2.21 $\pm$ 0.40    \\
00036759004 & 2007-10-11  & 4935  & 1.25 $\pm$ 0.19  & 3.03 $\pm$ 0.51    \\
00036188002 & 2007-11-22  & 4630  & 1.38 $\pm$ 0.15  & 3.44 $\pm$ 0.46    \\
00036188003 & 2008-01-14  & 5472  & 1.30 $\pm$ 0.16  & 2.80 $\pm$ 0.41    \\
00036188005 & 2008-02-11  & 4660  & 1.35 $\pm$ 0.23  & 2.60 $\pm$ 0.53    \\
00031123001 & 2008-02-14  & 4280  & 1.36 $\pm$ 0.17  & 3.38 $\pm$ 0.49    \\
00036188004 & 2008-11-18  & 6470  & 1.36 $\pm$ 0.14  & 2.94 $\pm$ 0.37    \\
00036188006	& 2009-09-07  &  357  & 1.51 $\pm$ 0.61  & 3.47 $\pm$ 1.56    \\
00036188005	& 2009-09-08  & 5162  & 1.32 $\pm$ 0.14  & 3.77 $\pm$ 0.47    \\
00036188007 & 2010-02-05  & 4183  & 1.38 $\pm$ 0.18  & 2.71 $\pm$ 0.53    \\
00036188008 & 2010-02-05  & 1451  & 1.50 $\pm$ 0.31  & 2.86 $\pm$ 0.70    \\
00036188009 & 2011-01-28  & 1648  & 1.35 $\pm$ 0.33  & 2.32 $\pm$ 0.69    \\
00036188010 & 2011-01-28  & 3501  & 1.42 $\pm$ 0.23  & 2.19 $\pm$ 0.44    \\
00036188011 & 2020-08-05  & 1988  & 1.78 $\pm$ 0.44  & 1.12 $\pm$ 0.34    \\
00036188012 & 2020-08-07  & 2567  & 1.39 $\pm$ 0.31  & 2.08 $\pm$ 0.53    \\
00036188013 & 2020-08-11  & 2962  & 1.38 $\pm$ 0.30  & 1.90 $\pm$ 0.47    \\
00036188014 & 2020-08-14  & 2647  & 1.69 $\pm$ 0.26  & 1.91 $\pm$ 0.39    \\
00036188015 & 2020-08-17  & 2605  & 1.18 $\pm$ 0.38  & 1.35 $\pm$ 0.47    \\
\hline
\end{tabular}
\end{center}
\label{OC457_XRT}
\end{table}

\begin{table}[ht]
\caption{Observed magnitudes of OC 457 obtained by {\em Swift}/UVOT.}
\begin{center}
\begin{tabular}{ccccccc}
\hline \hline
\multicolumn{1}{c}{Observation} &
\multicolumn{1}{c}{$v$}  &
\multicolumn{1}{c}{$b$}  &
\multicolumn{1}{c}{$u$}  &
\multicolumn{1}{c}{$w1$} &
\multicolumn{1}{c}{$m2$} &
\multicolumn{1}{c}{$w2$} \\
\multicolumn{1}{c}{Date} &
\multicolumn{1}{c}{mag}  &
\multicolumn{1}{c}{mag}  &
\multicolumn{1}{c}{mag}  &
\multicolumn{1}{c}{mag}  &
\multicolumn{1}{c}{mag}  &
\multicolumn{1}{c}{mag}  \\
\hline
2007-07-16    &      -        &          -       &           -      &          -       &       -          & 17.21 $\pm$ 0.05 \\
2007-11-22  & 14.96 $\pm$ 0.04 & 15.53 $\pm$ 0.04 & 14.94 $\pm$ 0.05 & 15.28 $\pm$ 0.06 & 15.63 $\pm$ 0.06 & 15.77 $\pm$ 0.06 \\
2008-01-14  & 15.40 $\pm$ 0.04 & 16.04 $\pm$ 0.05 & 15.50 $\pm$ 0.05 & 15.85 $\pm$ 0.06 & 16.28 $\pm$ 0.06 & 16.46 $\pm$ 0.06 \\
2008-02-14  & 16.23 $\pm$ 0.04 &          -       &           -      &          -       &       -          &                  \\
2008-11-18  & 15.83 $\pm$ 0.04 & 16.43 $\pm$ 0.05 & 15.84 $\pm$ 0.05 & 16.22 $\pm$ 0.06 & 16.55 $\pm$ 0.07 & 16.73 $\pm$ 0.06 \\
2009-09-07  & 15.99 $\pm$ 0.11 & 16.96 $\pm$ 0.11 & 16.08 $\pm$ 0.02 & 16.53 $\pm$ 0.13 & 16.89 $\pm$ 0.20 & 16.94 $\pm$ 0.12 \\ 
2009-09-08  & 16.24 $\pm$ 0.05 & 16.79 $\pm$ 0.05 & 16.18 $\pm$ 0.05 & 16.57 $\pm$ 0.06 & 16.87 $\pm$ 0.07 & 17.06 $\pm$ 0.06 \\
2010-02-05   &       -        &         -        & 17.56 $\pm$ 0.05 &         -        &          -       & 18.31 $\pm$ 0.08 \\ 
2010-02-05  & 17.62 $\pm$ 0.15 & 17.91 $\pm$ 0.13 & 17.45 $\pm$ 0.13 & 17.81 $\pm$ 0.14 & 18.10 $\pm$ 0.15 & 18.29 $\pm$ 0.12 \\ 
2011-01-28  &        -        &         -        &         -        &         -        &         -        & 17.97 $\pm$ 0.07 \\
2011-01-28  & 16.92 $\pm$ 0.09 & 17.46 $\pm$ 0.09 & 17.07 $\pm$ 0.10 & 17.40 $\pm$ 0.11 & 17.72 $\pm$ 0.13 & 17.84 $\pm$ 0.10 \\
2020-08-05  & 18.83 $\pm$ 0.25 & 19.40 $\pm$ 0.22 & 19.20 $\pm$ 0.27 & 19.07 $\pm$ 0.23 & 19.54 $\pm$ 0.35 & 20.11 $\pm$ 0.33 \\
2020-08-07  & $>$ 19.04        & $>$ 20.10        & 19.48 $\pm$ 0.31 & 19.58 $\pm$ 0.29 & 19.87 $\pm$ 0.33 & $>$ 20.45        \\
2020-08-11  & 18.98 $\pm$ 0.35 & 19.81 $\pm$ 0.26 & 18.87 $\pm$ 0.18 & 19.50 $\pm$ 0.25 & 19.00 $\pm$ 0.20 & 19.98 $\pm$ 0.27 \\
2020-08-14  & 19.10 $\pm$ 0.33 & 19.30 $\pm$ 0.18 & 19.42 $\pm$ 0.27 & 18.98 $\pm$ 0.20 & 19.41 $\pm$ 0.25 & 19.73 $\pm$ 0.22 \\
2020-08-17  & 18.76 $\pm$ 0.30 & 19.59 $\pm$ 0.23 & 19.69 $\pm$ 0.35 & 19.07 $\pm$ 0.20 & $>$ 19.88        & 19.62 $\pm$ 0.22 \\
\hline
\end{tabular}
\end{center}
\label{OC457_UVOT}
\end{table}

\begin{table}[ht]
\caption{Log and fitting results of {\em Swift}-XRT observations of NVSS J065844+063711 using a power-law model with $N_{\rm H}$ fixed to Galactic absorption.}
\begin{center}
\begin{tabular}{ccccc}
\hline \hline
\multicolumn{1}{c}{{\it Swift} ObsID} &
\multicolumn{1}{c}{Observation} &
\multicolumn{1}{c}{Net Exposure Time} &
\multicolumn{1}{c}{Photon index} &
\multicolumn{1}{c}{Flux 0.3-10 keV} \\
\multicolumn{1}{c}{} &
\multicolumn{1}{c}{Date} &
\multicolumn{1}{c}{second} &
\multicolumn{1}{c}{$\Gamma$} &
\multicolumn{1}{c}{$\times$10$^{-12}$ erg cm$^{-2}$ s$^{-1}$} \\
\hline
00047168001-00047168006 & 2012-05-01/10  &  4146  &  2.95 $\pm$ 0.65 & 0.80 $\pm$ 0.25 \\
00013876001 & 2020-11-15     &  3279  &  1.97 $\pm$ 0.37 & 1.64 $\pm$ 0.35 \\
00013876002 & 2020-11-18     &  3674  &  2.02 $\pm$ 0.34 & 1.58 $\pm$ 0.38 \\
00013876003 & 2020-12-11     &  3062  &  2.41 $\pm$ 0.48 & 1.61 $\pm$ 0.41 \\
\hline
\end{tabular}
\end{center}
\label{0658_XRT}
\end{table}

\begin{table}[ht]
\caption{Observed magnitudes of NVSS J065844+06371 obtained by {\em Swift}/UVOT.}
\begin{center}
\begin{tabular}{ccccccc}
\hline \hline
\multicolumn{1}{c}{Observation} &
\multicolumn{1}{c}{$v$}  &
\multicolumn{1}{c}{$b$}  &
\multicolumn{1}{c}{$u$}  &
\multicolumn{1}{c}{$w1$} &
\multicolumn{1}{c}{$m2$} &
\multicolumn{1}{c}{$w2$} \\
\multicolumn{1}{c}{Date} &
\multicolumn{1}{c}{mag}  &
\multicolumn{1}{c}{mag}  &
\multicolumn{1}{c}{mag}  &
\multicolumn{1}{c}{mag}  &
\multicolumn{1}{c}{mag}  &
\multicolumn{1}{c}{mag}  \\
\hline
2012-05-01 & - & - & 18.85 $\pm$ 0.13 & - & - & - \\
2012-05-02 & - & - & - & - & - & $>$ 20.09 \\ 
2012-05-05 & - & - & 17.33 $\pm$ 0.08 & - & - & - \\
2012-05-10 & - & - & - & - & - & 19.30 $\pm$ 0.24 \\ 
2012-05-10 & - & - & - & - & 19.31 $\pm$ 0.17 & - \\ 
2020-11-15 & $>$ 18.94 & $>$ 19.90 & $>$ 19.53 & $>$ 19.71 & 20.26 $\pm$ 0.29 & $>$ 20.15 \\ 
2020-11-18 & 19.22 $\pm$ 0.36 & 19.93 $\pm$ 0.31 & 18.97 $\pm$ 0.20 & $>$ 19.97 & $>$ 20.20 & 20.22 $\pm$ 0.32 \\
2020-12-11 & 19.07 $\pm$ 0.34 & 19.94 $\pm$ 0.33 & 19.13 $\pm$ 0.23 & 18.96 $\pm$ 0.20 & 19.91 $\pm$ 0.33 & 19.71 $\pm$ 0.24 \\
\hline
\end{tabular}
\end{center}
\label{0658_UVOT}
\end{table}

\begin{table}[ht]
\caption{Log and fitting results of {\em Swift}-XRT observations of 4FGL J0955.1+3551 using a power-law model with $N_{\rm H}$ fixed to Galactic absorption.}
\begin{center}
\begin{tabular}{ccccc}
\hline \hline
\multicolumn{1}{c}{{\it Swift} ObsID} &
\multicolumn{1}{c}{Observation} &
\multicolumn{1}{c}{Net Exposure Time} &
\multicolumn{1}{c}{Photon index} &
\multicolumn{1}{c}{Flux 0.3-10 keV} \\
\multicolumn{1}{c}{} &
\multicolumn{1}{c}{Date} &
\multicolumn{1}{c}{second} &
\multicolumn{1}{c}{$\Gamma$} &
\multicolumn{1}{c}{$\times$10$^{-12}$ erg cm$^{-2}$ s$^{-1}$} \\
\hline
00091400002 & 2012-04-19 &  914 & 2.03 $\pm$ 0.24 &  4.51 $\pm$ 0.75 \\
00091400006 & 2012-10-08 & 4795 & 1.80 $\pm$ 0.11 &  4.89 $\pm$ 0.39 \\ 
00091400007 & 2012-10-10 & 4220 & 1.92 $\pm$ 0.11 &  4.79 $\pm$ 0.38 \\
00091400008 & 2012-10-11 &  280 & 2.26 $\pm$ 0.49 &  5.19 $\pm$ 1.55 \\
00091400009 & 2012-10-12 & 2382 & 1.96 $\pm$ 0.15 &  4.71 $\pm$ 0.50 \\
00091400010 & 2012-10-16 & 1189 & 1.89 $\pm$ 0.24 &  3.75 $\pm$ 0.65 \\
00091400011 & 2012-10-17 &  614 & 1.85 $\pm$ 0.36 &  3.10 $\pm$ 0.81 \\
00091400012 & 2012-10-27 & 1666 & 1.99 $\pm$ 0.20 &  4.22 $\pm$ 0.57 \\
00091400013 & 2012-10-30 &  529 & 2.00 $\pm$ 0.36 &  3.82 $\pm$ 0.96 \\
00091400014 & 2012-11-01 &  739 & 2.38 $\pm$ 0.29 &  4.04 $\pm$ 0.74 \\
00091400015 & 2012-11-22 &  969 & 2.28 $\pm$ 0.25 &  4.03 $\pm$ 0.65 \\
00091400016 & 2012-11-23 & 2225 & 2.06 $\pm$ 0.22 &  3.22 $\pm$ 0.49 \\
00091400018 & 2012-12-23 & 1611 & 2.13 $\pm$ 0.13 &  4.95 $\pm$ 0.60 \\
00091400019 & 2012-12-29 &  704 & 2.08 $\pm$ 0.29 &  5.27 $\pm$ 1.03 \\
00091400021 & 2013-01-05 & 1159 & 2.16 $\pm$ 0.18 &  6.64 $\pm$ 0.77 \\
00091400022 & 2013-01-13 &  814 & 2.10 $\pm$ 0.22 &  6.03 $\pm$ 0.89 \\
00091400023 & 2013-01-26 & 1161 & 1.98 $\pm$ 0.18 &  6.79 $\pm$ 0.84 \\
00091400024 & 2013-01-29 &  607 & 1.61 $\pm$ 0.25 &  7.34 $\pm$ 1.40 \\
00091400025 & 2013-02-01 &  397 & 1.85 $\pm$ 0.33 &  6.29 $\pm$ 1.43 \\
00091400026 & 2013-02-06 & 1154 & 1.96 $\pm$ 0.19 &  6.43 $\pm$ 0.85 \\
00091400027 & 2013-02-09 & 1216 & 1.84 $\pm$ 0.22 &  5.80 $\pm$ 0.87 \\
00091400028 & 2013-02-11 &  252 & 2.94 $\pm$ 0.47 &  6.67 $\pm$ 1.82 \\
00013051001 & 2020-01-08 & 2904 & 1.81 $\pm$ 0.10 & 10.81 $\pm$ 0.65 \\ 
00013051002 & 2020-01-10 & 2829 & 1.94 $\pm$ 0.10 &  8.73 $\pm$ 0.63 \\
00013051003 & 2020-01-11 & 2844 & 1.84 $\pm$ 0.10 &  9.32 $\pm$ 0.69 \\
00013051004 & 2020-01-16 &  412 & 2.04 $\pm$ 0.26 &  9.64 $\pm$ 1.62 \\
00013051005 & 2020-01-21 &  834 & 1.99 $\pm$ 0.22 &  7.40 $\pm$ 1.04 \\
00013051007 & 2020-01-23 & 2592 & 1.92 $\pm$ 0.11 & 10.06 $\pm$ 0.75 \\
00013051008 & 2020-01-25 & 1908 & 1.92 $\pm$ 0.13 &  9.35 $\pm$ 0.87 \\
00013051010 & 2020-01-29 & 1581 & 1.84 $\pm$ 0.15 &  7.31 $\pm$ 0.80 \\
00013051011 & 2020-01-30 & 1086 & 2.13 $\pm$ 0.20 &  6.87 $\pm$ 0.90 \\
00013051013 & 2020-02-11 & 774  & 1.91 $\pm$ 0.21 &  9.34 $\pm$ 1.36 \\
00013051014 & 2020-02-16 & 822  & 1.97 $\pm$ 0.21 &  9.08 $\pm$ 1.33 \\
00013051015 & 2020-02-21 & 2155 & 1.86 $\pm$ 0.11 & 10.12 $\pm$ 0.81 \\ 
\hline
\end{tabular}
\end{center}
\label{0955_XRT}
\end{table}

\begin{table}[ht]
\caption{Observed magnitudes of 4FGL J0955.1+3551 obtained by {\em Swift}/UVOT.}
\begin{center}
\begin{tabular}{ccccccc}
\hline \hline
\multicolumn{1}{c}{Observation} &
\multicolumn{1}{c}{$v$}  &
\multicolumn{1}{c}{$b$}  &
\multicolumn{1}{c}{$u$}  &
\multicolumn{1}{c}{$w1$} &
\multicolumn{1}{c}{$m2$} &
\multicolumn{1}{c}{$w2$} \\
\multicolumn{1}{c}{Date} &
\multicolumn{1}{c}{mag}  &
\multicolumn{1}{c}{mag}  &
\multicolumn{1}{c}{mag}  &
\multicolumn{1}{c}{mag}  &
\multicolumn{1}{c}{mag}  &
\multicolumn{1}{c}{mag}  \\
\hline
2012-04-19 &  - & - & 19.35 $\pm$ 0.17 & 19.07 $\pm$ 0.25 & - & -  \\
2012-10-08 & $>$ 18.50 & $>$ 19.60 & 18.96 $\pm$ 0.08 & 18.52 $\pm$ 0.19 & 18.68 $\pm$ 0.20 & 18.81 $\pm$ 0.16 \\
2012-10-10 & $>$ 18.37 & $>$ 19.38 & 18.55 $\pm$ 0.26 & 18.76 $\pm$ 0.26 & 18.60 $\pm$ 0.08 & 18.55 $\pm$ 0.13 \\
2012-10-11 & - & - & - & 18.68 $\pm$ 0.19 & - & - \\
2012-10-12 & $>$ 18.45 & $>$ 19.45 & 19.04 $\pm$ 0.10 & 18.88 $\pm$ 0.27 & 18.86 $\pm$ 0.24 & 19.08 $\pm$ 0.20 \\ 
2012-10-16 & $>$ 18.41 & $>$ 19.39 & 18.91 $\pm$ 0.16 & 18.98 $\pm$ 0.30 & 18.79 $\pm$ 0.27 & 18.75 $\pm$ 0.19 \\
2012-10-17 & $>$ 18.14 & $>$ 19.21 & $>$ 18.90 & 18.54 $\pm$ 0.27 & 18.60 $\pm$ 0.28 & 18.85 $\pm$ 0.22 \\
2012-10-27 & - & $>$ 19.72 & 19-08 $\pm$ 0.30 & 18.61 $\pm$ 0.10 & - & 18.52 $\pm$ 0.17 \\
2012-10-30 &  $>$ 18.16 & $>$ 19.16 & 18.74 $\pm$ 0.36 & 18.74 $\pm$ 0.32 & 18.56 $\pm$ 0.27 & 18.88 $\pm$ 0.24 \\
2012-11-01 & $>$ 18.52 & $>$ 19.51 & 18.74 $\pm$ 0.27 & 18.82 $\pm$ 0.26 & 18.38 $\pm$ 0.21 & 18.47 $\pm$ 0.16 \\ 
2012-11-22 & $>$ 17.98 & $>$ 18.99 & $>$ 18.58 & 18.97 $\pm$ 0.36 & 18.56 $\pm$ 0.27 & 18.67 $\pm$ 0.15 \\
2012-11-23 & $>$ 18.46 & $>$ 19.56 & $>$ 19.24 & 19.32 $\pm$ 0.30 & 18.59 $\pm$ 0.13 & 18.80 $\pm$ 0.15 \\
2012-12-23 & $>$ 18.81 & $>$ 19.89 & 19.22 $\pm$ 0.21 & 18.73 $\pm$ 0.21 & 18.38 $\pm$ 0.18 & 18.38 $\pm$ 0.11 \\
2012-12-29 & $>$ 18.48 & 19.25 $\pm$ 0.32 & 18.69 $\pm$ 0.28 & 18.32 $\pm$ 0.22 & 18.17 $\pm$ 0.21 & 18.36 $\pm$ 0.16 \\
2013-01-05 & $>$ 18.85 & 19.77 $\pm$ 0.36 & 18.65 $\pm$ 0.21 & 18.48 $\pm$ 0.18 & 18.15 $\pm$ 0.17 & 18.25 $\pm$ 0.13 \\
2013-01-13 & $>$ 18.51 &  $>$ 19.52 & 18.86 $\pm$ 0.30 & 18.26 $\pm$ 0.20 & 18.25 $\pm$ 0.21 & 18.31 $\pm$ 0.15 \\
2013-01-26 & $>$ 18.76 & $>$ 19.75 & 18.98 $\pm$ 0.28 & 18.40 $\pm$ 0.18 & 18.49 $\pm$ 0.19 & 18.53 $\pm$ 0.14 \\
2013-01-29 & $>$ 18.34 & $>$ 19.34 & 18.87 $\pm$ 0.35 & 19.06 $\pm$ 0.34 & 18.54 $\pm$ 0.26 & 18.61 $\pm$ 0.19 \\
2013-02-01 &  - & - & 18.91 $\pm$ 0.16 & - & - & 18.57 $\pm$ 0.27 \\
2013-02-06 & 18.64 $\pm$ 0.34 & $>$ 19.74 & 18.91 $\pm$ 0.27 & 18.66 $\pm$ 0.21 & 18.47 $\pm$ 0.19 & 18.54 $\pm$ 0.15 \\
2013-02-09 & $>$ 18.77 & 19.64 $\pm$ 0.35 & 18.78 $\pm$ 0.24 &  18.78 $\pm$ 0.23 & 18.78 $\pm$ 0.21 & 18.52 $\pm$ 0.14 \\
2013-02-11 & - & $>$ 19.50 & 18.68 $\pm$ 0.28 & 18.35 $\pm$ 0.22 & - & - \\
2020-01-08 & $>$ 18.48 & - & - & $>$ 18.42 & 18.32 $\pm$ 0.11 & 18.37 $\pm$ 0.09 \\
2020-01-10 & 18.81 $\pm$ 0.32 & 19.79 $\pm$ 0.31 & 18.69 $\pm$ 0.19 & 18.58 $\pm$ 0.15 & 18.39 $\pm$ 0.12 & 18.30 $\pm$ 0.12 \\
2020-01-11 & 18.98 $\pm$ 0.35 & 19.68 $\pm$ 0.28 & 18.80 $\pm$ 0.20 & 18.33 $\pm$ 0.15 & 18.43 $\pm$ 0.11 & 18.23 $\pm$ 0.11 \\
2020-01-16 & $>$ 17.99 & $>$ 18.99 & 18.62 $\pm$ 0.36 & 18.65 $\pm$ 0.37 & 18.44 $\pm$ 0.30 & 18.53 $\pm$ 0.25 \\
2020-01-21 & $>$ 18.48 & $>$ 19.50 & 18.50 $\pm$ 0.25 & 18.63 $\pm$ 0.25 & 18.91 $\pm$ 0.28 & 18.79 $\pm$ 0.20 \\
2020-01-23 & $>$ 19.19 & 20.02 $\pm$ 0.34 & 18.73 $\pm$ 0.17 & 18.61 $\pm$ 0.16 & 18.53 $\pm$ 0.15 & 18.61 $\pm$ 0.12 \\
2020-01-25 & 19.04 $\pm$ 0.37 & 19.30 $\pm$ 0.22 & 18.81 $\pm$ 0.20 & 18.46 $\pm$ 0.16 &  18.15 $\pm$ 0.16 & 18.40 $\pm$ 0.12 \\
2020-01-29 & 18.86 $\pm$ 0.35 &  19.24 $\pm$ 0.23 & 18.66 $\pm$ 0.20 & 18.83 $\pm$ 0.21 & 18.51 $\pm$ 0.21 & 18.51 $\pm$ 0.13 \\
2020-01-30 & $>$ 18.61 & $>$ 19.62 & 18.96 $\pm$0.31 & 18.28 $\pm$ 0.20 & 18.50 $\pm$ 0.20 & 18.49 $\pm$ 0.16 \\
2020-02-11 & - & - & - & - & 18.36 $\pm$ 0.13 & - \\
2020-02-16 & - & - & - & 18.49 $\pm$ 0.11 & - & - \\
2020-02-21 & - & - & 18.69 $\pm$ 0.07 & - & - & \\
\hline
\end{tabular}
\end{center}
\label{0955_UVOT}
\end{table}

\begin{table}[ht]
\caption{Log and fitting results of {\em Swift}-XRT observations of PKS 1502+106 using a power-law model with $N_{\rm H}$ fixed to Galactic absorption.}
\begin{center}
\begin{tabular}{ccccc}
\hline \hline
\multicolumn{1}{c}{{\it Swift} ObsID} &
\multicolumn{1}{c}{Observation} &
\multicolumn{1}{c}{Net Exposure Time} &
\multicolumn{1}{c}{Photon index} &
\multicolumn{1}{c}{Flux 0.3-10 keV} \\
\multicolumn{1}{c}{} &
\multicolumn{1}{c}{Date} &
\multicolumn{1}{c}{second} &
\multicolumn{1}{c}{$\Gamma$} &
\multicolumn{1}{c}{$\times$10$^{-12}$ erg cm$^{-2}$ s$^{-1}$} \\
\hline
00094003011-00094003015 & 2019-01-03/31 & 5442 & 1.46 $\pm$ 0.26 & 1.14 $\pm$ 0.23 \\
00095003001, 00095003002 & 2019-06-20/27 & 2103 & 1.09 $\pm$ 0.52 & 1.38 $\pm$ 0.65 \\
00095003003-006, 	00011493001& 2019-07-04/30 & 4308 & 1.11 $\pm$ 0.37 & 1.10 $\pm$ 0.36 \\
00095003007, 00095003008, 00095088001 & 2019-08-01/09-20 & 3396 & 1.01 $\pm$ 0.42 & 1.17 $\pm$ 0.43 \\
00095003009, 00095003010 & 2019-12-20/27 & 2415 & 1.36 $\pm$ 0.41 & 1.20 $\pm$ 0.40 \\
00095003011-00095003015 & 2020-01-03/31 & 4423 & 1.40 $\pm$ 0.34 & 0.89 $\pm$ 0.24 \\
00095656001, 00095656002 & 2020-06-20/27 & 1673 & 1.54 $\pm$ 0.42 & 1.37 $\pm$ 0.45 \\
00095656003-00095656006 & 2020-07-04/25 & 3888 & 1.61 $\pm$ 0.25 & 1.77 $\pm$ 0.33 \\
00095656007, 00095656008 & 2020-08-01/08 & 1988 & 1.42 $\pm$ 0.33 & 2.12 $\pm$ 0.54 \\
00095656009, 00095656010 & 2020-12-20/27 & 1778 & 1.07 $\pm$ 0.45 & 1.49 $\pm$ 0.64 \\
\hline
\end{tabular}
\end{center}
\label{PKS1502_XRT}
\end{table}

\clearpage
\section{IceCube alert properties}

\begin{table}[ht]
\small
\caption{GFU-cluster alerts 
that are relevant to the results of this work.
For each alert, we provide the candidate source, the date of the first event in the cluster (Start Date), the Trigger Date, the duration of the cluster time window ($\Delta$T), the corresponding alert pre-trials $p$-value and Significance ($\sigma$), and the false-alert rate (FAR).
}
\label{tab_gfu_alerts}

\begin{tabular}{l|c|c|c|c|c|c}
\hline\hline
Source & Start Date & Trigger Date & $\Delta$T & $p$-value &Significance& FAR \\
 & & & [days] & & $\sigma$ & [yr$^{-1}$] \\
\hline
1ES~1312-423     & 2019-03-12 & 2019-03-12 &  0.26 & 3.46$\times10^{-4}$ &3.4  &0.02   \\[2ex]
MG1~J181841+0903 & 2019-01-19 & 2019-05-27 & 127.4 & 3.34$\times10^{-4}$ &3.3 & 0.1  \\
                 & 2019-01-19 & 2019-05-27 & 127.7 & 3.39$\times10^{-4}$ &3.3 & 0.1  \\
                 & 2019-01-19 & 2019-05-27 & 127.9 & 4.20$\times10^{-4}$ &3.3 & 0.1  \\
                 & 2019-01-19 & 2019-06-05 & 136.8 & 1.62$\times10^{-4}$ &3.6 & 0.1  \\
                 & 2019-01-19 & 2019-06-05 & 137.0 & 1.60$\times10^{-4}$ &3.6 & 0.1  \\[2ex]
PMN~J2016-09     & 2019-11-29 & 2019-11-29 &  0.01 & 1.86$\times10^{-3}$ &3.6  & 0.004\\[2ex]
OP~313           & 2020-02-12 & 2020-04-10 &  57.5 & 1.07$\times10^{-3}$  &3.1  & 0.05 \\
                 & 2020-02-12 & 2020-05-01 &  78.8 & 1.18$\times10^{-3}$ &3.0 & 0.05\\
                 & 2020-02-12 & 2020-05-17 &  94.7 & 1.23$\times10^{-3}$ &3.0 & 0.05\\
                 & 2020-03-13 & 2020-08-27 & 166.7 & 1.08$\times10^{-3}$ &3.1 & 0.05\\[2ex]
OC~457           & 2020-08-04 & 2020-08-04 &   0.3 & 4.20$\times10^{-4}$ &3.3 & 0.02  \\[2ex]
GB6~J0316+0904   & 2021-01-13 & 2021-01-15 &   2.3 & 9.80$\times10^{-4}$ &3.1  & 0.04 \\[2ex]
\hline
All-sky &&&&&\\
($\alpha$=51.2$^\circ$, $\delta$=-18.6$^\circ$) 
                 & 2019-09-15 & 2019-09-19 &   3.7  & 7.81$\times10^{-6}$ &4.3 & 0.7 \\[1ex]
 \hline
 \end{tabular}
\end{table}



\begin{table}[ht]
\small
\caption{List of IceCube single high-energy neutrino alerts followed up by at least one IACT. A link to the corresponding GCN Circular with updated coordinates (including information regarding the initial localization) released by IceCube is provided in the alert name. The alert nature is provided: E = EHE, H = HESE, G = GOLD. For each alert, the event direction (RA and Dec), energy, pre-trials signalness and false alert rate (FAR) are provided. The values given in parenthesis as a second line correspond to the offline reconstruction of direction, energy and signalness as reported in~\citep{2023ApJS..269...25A}.
Energy and signalness estimates are not available for 200107A and 190529A and FAR was not provided before 2019 alert stream upgrade.} 

\label{tab_tracks_extended}
\begin{center}
\begin{tabular}{lcccccc}
\hline\hline
Alert& & RA & Dec & Energy  & Signalness & FAR \\
& &[deg]&[deg] & [TeV] & &[yr$^{-1}$]  \\
 \hline
\href{https://gcn.nasa.gov/circulars/22105}{IC-171106A}& E&  $340.00^{+0.70}_{-0.50}$ & $+7.40^{+0.35}_{-0.25}$ & 230 & 0.75 & \\
&& ($340.14^{+0.62}_{-0.62}$) & ($+7.44^{+0.30}_{-0.26}$) & (1573) & (0.97)&  \\[2ex]

\href{https://gcn.nasa.gov/circulars/23375}{IC-181023A}& E&  $270.18^{+2.00}_{-1.70}$ & $-8.57^{+1.25}_{-1.30}$& 120 & 0.28 & \\
& & ($270.18^{+1.89}_{-1.71}$) & ($-8.42^{+1.13}_{-1.55}$) &(237) &  (0.15) &  \\[2ex]

\href{https://gcn.nasa.gov/circulars/24378}{IC-190503A} &E& $120.28^{+0.57}_{-0.77}$ & $+6.35^{+0.76}_{-0.70}$ & 100  & 0.36 &  \\
& & ($120.19^{+0.66}_{-0.66}$) & ($+6.43^{+0.68}_{-0.75}$) &(142) &  (0.34) &  \\[2ex]

\href{https://gcn.nasa.gov/circulars/24674}{IC-190529A}$^\dagger$ &H& $287.32^{+8.9}_{-8.9}$ & $+78.14^{+8.9}_{-8.9}$ & --- & --- &  \\
&&   --- & --- & --- & --- &\\[2ex]

\href{https://gcn.nasa.gov/circulars/25225}{IC-190730A}& G &  $225.79^{+1.28}_{-1.43}$ & $+10.47^{+1.14}_{-0.89}$ & 299 & 0.67 &0.68 \\
&&  ($226.14^{+1.27}_{-1.98}$) & ($+10.77^{+1.03}_{-1.17}$) & (298) & (0.67) & \\[2ex]

\href{https://gcn.nasa.gov/circulars/25806}{IC-190922B} &G & $5.76^{+1.19}_{-1.37}$ & $-1.57^{+0.93}_{-0.82}$& 187 & 0.50 & 1.33 \\
& & ($5.71^{+1.19}_{-1.27}$) & ($-1.53^{+0.90}_{-0.78}$) & (187) & (0.50) &  \\[2ex]

\href{https://gcn.nasa.gov/circulars/25913}{IC-191001A} &G & $314.08^{+6.56}_{-2.26}$ & $+12.94^{+1.50}_{-1.47}$ & 217 & 0.59 &0.86 \\ 
& & ($313.99^{+6.94}_{-2.46}$) & ($12.79^{+1.65}_{-1.64}$) & (218) & (0.59) & \\[2ex]

\href{https://gcn.nasa.gov/circulars/26655}{IC-200107A}$^\ddag$ & & $148.18^{+2.20}_{-1.83}$ & $+35.46^{+1.10}_{-1.22}$& --- & --- & --- \\
& & --- & --- & --- & --- & \\[2ex]

\href{https://gcn.nasa.gov/circulars/28504}{IC-200926A} &G & $96.46^{+0.73}_{-0.55}$ & $-4.33^{+0.61}_{-0.76}$&  670 & 0.44 &0.54 \\
& & ($96.46^{+0.70}_{-0.53}$) & ($-4.33^{+0.60}_{-0.75}$) & (670) &  (0.44) & \\[2ex]

\href{https://gcn.nasa.gov/circulars/28575}{IC-201007A} &G & $265.17^{+0.52}_{-0.52}$ & $+5.34^{+0.32}_{-0.23}$& 683 & 0.88 & 0.26  \\
& & ($265.17^{+0.48}_{-0.48}$) & ($+5.34^{+0.30}_{-0.19}$)& (683) & (0.89) & \\[2ex]

\href{https://gcn.nasa.gov/circulars/28887}{IC-201114A} &G & $105.25^{+1.28}_{-1.12}$ & $+6.05^{+0.95}_{-0.95}$& 214 & 0.56 & 0.92 \\
& & ($105.73^{+0.92}_{-1.27}$) & ($+5.87^{+1.05}_{-1.01}$) & (214) & (0.56) & \\[2ex]

\href{https://gcn.nasa.gov/circulars/29120}{IC-201222A} &G & $206.37^{+0.90}_{-0.80}$ & $+13.44^{+0.55}_{-0.38}$& 186 & 0.53 & 1.01 \\
& & ($206.37^{+0.88}_{-0.75}$) & ($+13.44^{+0.54}_{-0.34}$)& (186) & (0.53) &\\[1ex]
\hline
\multicolumn{7}{l}{%
  \begin{minipage}{11.5cm}~\\%
$\dag$ Retracted \\
$\ddag$ The high energy starting track was not identified as either GOLD or BRONZE
  \end{minipage}%
}\\
 \end{tabular}
 \end{center}

\end{table}
\end{document}